\documentclass[useAMS,usenatbib,a4paper]{mn2e}
\usepackage{amssymb}
\usepackage{amsmath}
\usepackage{graphicx}
\usepackage{subfigure}
\usepackage{booktabs}
\usepackage{txfonts}
\usepackage{mathtools} 
\usepackage[pdftex,colorlinks=true,breaklinks=true,citecolor=blue]{hyperref}
\usepackage{color}

\usepackage{tikz}
\usetikzlibrary{shapes.misc}
\usetikzlibrary{matrix}
\usetikzlibrary{arrows,backgrounds,fit,calc,shapes,automata}
\usetikzlibrary{bayesnet}
\tikzstyle{every picture}+=[remember picture]
\tikzset{
    >=stealth',
    punkt/.style={
           rectangle,
           rounded corners,
           draw=black,
           thick,
           text width=3.5cm,
           minimum height=1.0cm,
           text centered},
    pil/.style={
           ->,
           thick,
           shorten <=2pt,
           shorten >=2pt,}
}
\usetikzlibrary{decorations.pathreplacing,shapes}

\newif\ifbw
\bwfalse

\newcommand{\eqn}[1]{(#1)}

\newcommand{\tbl}[1]{Table~#1}

\newcommand{\fig}[1]{Fig.~#1}

\newcommand{\sectn}[1]{Sec.~#1}

\newcommand{\etal}{\mbox{\it et al.}}
\newcommand{\eg}{\mbox{\it e.g.}}
\newcommand{\ie}{\mbox{\it i.e.}}

\newcommand{\cf}{\mbox{\it cf.}}

\newcommand{\planck}{\textit{Planck}}

\newcommand{\isw}{{ISW}}
\newcommand{\iswtext}{{integrated Sachs-Wolfe}}

\newcommand{\healpix}{{\tt HEALPix}}

\newcommand{\sshtcode}{{\tt SSHT}}
\newcommand{\sothreecode}{{\tt SO3}}
\newcommand{\stwoletcode}{{\tt S2LET}}

\newcommand{\spcend}{\ensuremath{\:}}

\newcommand{\cconj}{\ensuremath{\ast}} 
 
\newcommand{\reals}{\ensuremath{\mathbb{R}}}

\newcommand{\integers}{\ensuremath{\mathbb{Z}}}
\newcommand{\naturals}{\ensuremath{\mathbb{N}}}

\newcommand{\ltwo}{\ensuremath{\mathrm{L}^2}}
\newcommand{\sphere}{\ensuremath{{\mathbb{S}^2}}}
\newcommand{\sothree}{\ensuremath{{\mathrm{SO}(3)}}}

\newcommand{\vect}[1]{\ensuremath{\mbox{\boldmath ${#1}$}}}

\newcommand{\opnexpv}{\ensuremath{\langle}}
\newcommand{\clsexpv}{\ensuremath{\rangle}}

\newcommand{\dx}{\ensuremath{\mathrm{\,d}}}
\newcommand{\dmu}[1]{\ensuremath{\dx \Omega(#1)}}

\newcommand{\deul}[1]{\ensuremath{\dx \varrho(#1)}}

\newcommand{\innerp}[2]{\ensuremath{\langle {#1},\: {#2} \rangle}}

\newcommand{\sa}{\ensuremath{\omega}}
\newcommand{\saa}{\ensuremath{\theta}}
\newcommand{\sab}{\ensuremath{\varphi}}
\newcommand{\sas}{\ensuremath{\saa, \sab}}

\newcommand{\eul}{\ensuremath{\mathbf{\rho}}}
\newcommand{\euls}{\ensuremath{\eula, \eulb, \eulc}}
\newcommand{\eula}{\ensuremath{\alpha}}
\newcommand{\eulb}{\ensuremath{\beta}}
\newcommand{\eulc}{\ensuremath{\gamma}}

\newcommand{\el}{\ensuremath{\ell}}
\newcommand{\m}{\ensuremath{m}}
\newcommand{\n}{\ensuremath{n}}

\newcommand{\elmax}{\ensuremath{{L}}}

\newcommand{\nmax}{\ensuremath{{N}}}

\newcommand{\p}{\ensuremath{^\prime}}

\newcommand{\kron}[2]{\ensuremath{\delta_{{#1}{#2}}}}

\renewcommand{\exp}[1]{\ensuremath{{\rm e}^{#1}}}

\newcommand{\shf}[2]{\ensuremath{Y_{#1#2}}}

\newcommand{\shc}[3]{\ensuremath{{#1}_{{#2}{#3}}}}
\newcommand{\shcc}[3]{\ensuremath{{#1}_{{#2}{#3}}^\cconj}}

\newcommand{\gammafun}{\ensuremath{\Gamma}}

\newcommand{\dmatbig}{\ensuremath{D}}
\newcommand{\Dlmn}{\ensuremath{ \dmatbig_{\m\n}^{\el} }}

\newcommand{\rot}{\ensuremath{\mathcal{R}}}
\newcommand{\rotarg}[1]{\ensuremath{\mathcal{R}_{#1}}}
\newcommand{\rotmat}{\ensuremath{\mathbf{\mathsf{R}}}}
\newcommand{\rotmatarg}[1]{\ensuremath{\rotmat_{#1}}}

\newcommand{\f}{\ensuremath{f}}

\newcommand{\wav}{\ensuremath{\psi}}

\newcommand{\wavs}{\ensuremath{\Phi}}

\newcommand{\wcoeff}{\ensuremath{W}}
\newcommand{\scoeff}{\ensuremath{W}}
\newcommand{\wscale}{\ensuremath{j}}
\newcommand{\wscalemax}{\ensuremath{J}}
\newcommand{\wscalemin}{\ensuremath{J_0}}
\newcommand{\wposn}{\ensuremath{\eul}}
\newcommand{\wscaleposn}{\ensuremath{{\wscale\wposn}}}
\newcommand{\dilparam}{\ensuremath{\alpha}}
\newcommand{\wavker}{\ensuremath{\kappa}}
\newcommand{\wavsteer}{\ensuremath{\zeta}}

\newcommand{\sumlmn}{\ensuremath{\sum_{\el=0}^{\infty} \sum_{\m=-\el}^\el} \sum_{\n=-\el}^\el}

\newcommand{\summ}{\ensuremath{\sum_{\m=-\el}^\el}}

\newcommand{\sumulm}{\ensuremath{\sum_{\el\m}}}
\newcommand{\sumulmn}{\ensuremath{\sum_{\el\m\n}}}

\newcommand{\nside}{\ensuremath{{N_{\rm{side}}}}}

\newcommand{\kurtosis}{\ensuremath{\kappa}}

\newcommand{\prob}{\ensuremath{{\rm P}}}
\newcommand{\given}{\ensuremath{{\,|\,}}}
\newcommand{\evidence}{\ensuremath{{E}}}

\newcommand{\fitmodelsel}{\ensuremath{M}}

\newcommand{\gmu}{\ensuremath{G \mu}}
\newcommand{\fstring}{\ensuremath{s}}
\newcommand{\fcmb}{\ensuremath{c}}
\newcommand{\fdata}{\ensuremath{d}}
\newcommand{\fnoise}{\ensuremath{n}}
\newcommand{\fcmbnoise}{\ensuremath{{g}}}
\newcommand{\fstringrecov}{\ensuremath{\overline{\fstring}}}
\newcommand{\wcoeffstring}{\ensuremath{\wcoeff^\fstring}}
\newcommand{\wcoeffcmb}{\ensuremath{\wcoeff^\fcmb}}
\newcommand{\wcoeffdata}{\ensuremath{\wcoeff^\fdata}}
\newcommand{\wcoeffnoise}{\ensuremath{\wcoeff^\fnoise}}
\newcommand{\wcoeffcmbnoise}{\ensuremath{\wcoeff^{\fcmbnoise}}}
\newcommand{\wcoeffstringp}{\ensuremath{\wcoeffstring_{\wscaleposn}}}
\newcommand{\wcoeffcmbp}{\ensuremath{\wcoeffcmb_{\wscaleposn}}}
\newcommand{\wcoeffdatap}{\ensuremath{\wcoeffdata_{\wscaleposn}}}
\newcommand{\wcoeffnoisep}{\ensuremath{\wcoeffnoise_{\wscaleposn}}}
\newcommand{\wcoeffcmbnoisep}{\ensuremath{\wcoeffcmbnoise_{\wscaleposn}}}
\newcommand{\wcoeffstringprecov}{\ensuremath{\overline{\wcoeff}^\fstring_{\wscaleposn}}}
\newcommand{\wcoeffstringrecov}{\ensuremath{\overline{\wcoeff}^\fstring}}
\newcommand{\wvarcmb}{\ensuremath{(\sigma^\fcmb_\wscale)^2}}
\newcommand{\wstdcmb}{\ensuremath{\sigma^\fcmb_\wscale}}
\newcommand{\wvarstring}{(\sigma^\fstring_\wscale)^2}

\newcommand{\wkurstring}{\ensuremath{\kurtosis^\fstring_\wscale}}
\newcommand{\ggdshape}{\ensuremath{\changed{\xi}}}
\newcommand{\ggdscale}{\ensuremath{\changed{\zeta}}}
\newcommand{\ggdshapew}{\ensuremath{{\ggdshape_\wscale}}}
\newcommand{\ggdscalew}{\ensuremath{{\ggdscale_\wscale}}}
\newcommand{\modelstring}{\ensuremath{{{\rm M}^{\rm \fstring}}}}
\newcommand{\modelcmb}{\ensuremath{{{\rm M}^{\rm \fcmb}}}}

\DeclarePairedDelimiter{\ceil}{\lceil}{\rceil}

\renewcommand{\wav}{\ensuremath{\Psi}}
\renewcommand{\scoeff}{\ensuremath{S}}
\renewcommand{\dilparam}{\ensuremath{\lambda}}
\renewcommand{\opnexpv}{\ensuremath{\mathbb{E} \bigl [ }}
\renewcommand{\clsexpv}{\ensuremath{\bigr ]}}
\newcommand{\opnexpvb}{\ensuremath{\mathbb{E} \biggl [ }}
\newcommand{\clsexpvb}{\ensuremath{ \biggr ]}}
\renewcommand{\exp}[1]{\ensuremath{{\rm exp}{#1}}}
\renewcommand{\elmax}{\ensuremath{{\el_{\rm max}}}}
\renewcommand{\prob}{\ensuremath{{\rm P}}}
\renewcommand{\eqn}[1]{Eq.~(#1)}

\newcommand{\stwocode}{{\sc s2}}
\renewcommand{\sshtcode}{{\sc ssht}}
\renewcommand{\sothreecode}{{\sc so3}}
\renewcommand{\stwoletcode}{{\sc s2let}}
\newcommand{\stwodwcode}{{\sc s2dw}}
\renewcommand{\healpix}{{\sc healpix}}
\newcommand{\fftwcode}{{\sc fftw}}
\newcommand{\cambcode}{{\sc camb}}

\newcommand{\changed}[1]{\textcolor{red}{#1}}

\newlength{\plotwidth}

\title[Wavelet-Bayesian inference of cosmic strings]
   {Wavelet-Bayesian inference of
   cosmic strings embedded in the cosmic microwave background}

\author[McEwen \etal]
{%
  J.~D.~McEwen${}^{1}$\thanks{jason.mcewen@ucl.ac.uk}, S.~M.~Feeney${}^{2,3}$, H.~V.~Peiris${}^{4,5}$,
  Y.~Wiaux${}^{6}$, C.~Ringeval${}^{7}$\newauthor and F.~R.~Bouchet${}^{8}$\\
  ${}^1$Mullard Space Science Laboratory (MSSL), University College London, Surrey RH5 6NT, U.K.\\
  ${}^2$Astrophysics Group, Imperial College London, Blackett Laboratory, Prince Consort Road, London SW7 2AZ, U.K.\\
  ${}^3$Center for Computational Astrophysics, 160 5th Avenue, New York, NY 10010, USA\\
  ${}^4$Department of Physics and Astronomy, University College London, London WC1E 6BT, U.K.\\
  ${}^5$The Oskar Klein Centre for Cosmoparticle Physics, Stockholm University, Stockholm, Sweden\\
  ${}^6$Institute of Sensors, Signals, and Systems, Heriot-Watt University, Edinburgh EH14 4AS, U.K.\\
  ${}^7$Centre for Cosmology, Particle Physics and Phenomenology, Universit\'e Catholique de Louvain, Louvain-la-Neuve B-1348, Belgium\\
  ${}^8$Institut d'Astrophysique de Paris, Paris 75014, France
}

\date{Accepted ---. Received ---; in original form ---}
\pagerange{\pageref{sec:introduction}--\pageref{lastpage}}
\pubyear{2016}

\def\LaTeX{L\kern-.36em\raise.3ex\hbox{a}\kern-.15em
    T\kern-.1667em\lower.7ex\hbox{E}\kern-.125emX}

\begin{document}
\maketitle

\begin{abstract}
  Cosmic strings are a well-motivated extension to the standard cosmological model and could induce a subdominant component in the anisotropies of the cosmic microwave background (CMB), in addition to the standard inflationary component.  The detection of strings, while observationally challenging, would provide a direct probe of physics at very high energy scales.  We develop a framework for cosmic string inference from observations of the CMB made over the celestial sphere, performing a Bayesian analysis in wavelet space where the string-induced CMB component has distinct statistical properties to the standard inflationary component.  Our wavelet-Bayesian framework provides a principled approach to compute the posterior distribution of the string tension \gmu\ and the Bayesian evidence ratio comparing the string model to the standard inflationary model.  Furthermore, we present a technique to recover an estimate of any string-induced CMB map embedded in observational data.   Using \planck-like simulations we demonstrate the application of our framework and evaluate its performance.  The method is sensitive to $\gmu \sim 5 \times 10^{-7}$ for Nambu-Goto string simulations that include an \iswtext\ (\isw) contribution only and do not include any recombination effects, before any parameters of the analysis are optimised. The sensitivity of the method compares favourably with other techniques applied to the same simulations.
\end{abstract}

\begin{keywords}
  cosmology: cosmic background radiation -- cosmology: observations --
  methods: data analysis -- methods: statistical.
\end{keywords}

\section{Introduction}
\label{sec:introduction}

High-precision measurements of the anisotropies of the cosmic microwave background (CMB) strongly favour a standard cosmological model in which the large-scale structure of the Universe is seeded by nearly scale-invariant Gaussian density perturbations created during a phase of inflation~\citep{hinshaw:2013,planck2014-a15}. These measurements do, however, leave room for additional subdominant contributions to the CMB generated by processes beyond the standard inflationary paradigm.
Cosmic strings represent a particularly well motivated extension to the standard model (for reviews see \citealt{brandenberger:1994,vilenkin:1994,hindmarsh:1995,copeland:2009}). Arising in a range of attempts at Grand Unification, cosmic strings are linear topological defects produced when the Universe undergoes certain symmetry-breaking phase transitions. In an expanding Universe, the existence of causally separate regions prevents the symmetry from being broken in the same way throughout space, with a network of cosmic strings inevitably forming as a result~\citep{kibble:1976}. Such a string network cannot be solely responsible for producing the anisotropies of the CMB --- cosmic strings cannot explain the acoustic peaks of the CMB power spectrum~\citep{pen:1997}.  However, cosmic strings could induce a subdominant contribution to the CMB through the Kaiser-Stebbins effect~\citep{kaiser:1984}, which induces a step-like (\ie, highly non-Gaussian) temperature change between photons passing either side of a moving string. The magnitude of the contribution to the relative CMB temperature anisotropies from a \emph{single}, \emph{straight} string is given by
\begin{equation}
  \frac{\Delta T}{T_0} = 8\pi G \mu v \gamma_{\rm s},
\end{equation}
where $v$ is the transverse string velocity, $\gamma_{\rm s}$ is the corresponding relativistic gamma factor, $G$ is the gravitational constant and $\mu$ is the string tension (throughout we use natural units $c=1$).  More generally, the situation is complicated by the existence of an evolving \emph{network} of \emph{wiggly} strings in an expanding universe.  Calculating accurate observable effects of a network of cosmic strings is a rich and computationally demanding area of research \citep{albrecht:1989, bennett:1989, bennett:1990, allen:1990, hindmarsh:1993, bouchet:1988, vincent:1998, moore:2001, landriau:2002, ringeval:2005, fraisse:2007, landriau:2010, blancopillado:2011, ringeval:2012}, requiring the numerical evolution of the network in the presence of photons, matter and dark energy. Tools to simulate full-sky, high-resolution maps of string-induced CMB anisotropies incorporating all physical effects are not yet available. The current state-of-the-art methods produce matter-free simulations, which faithfully represent the small-scale structure imparted by the string network via the integrated Sachs-Wolfe (ISW) effect~\citep{ringeval:2012} but do not include recombination effects. These simulations nevertheless remain computationally intensive, requiring hundreds of thousands of CPU hours to simulate a single full-sky map at \planck\ resolution.

\begin{figure}
 \centering
 \includegraphics[width=\columnwidth]{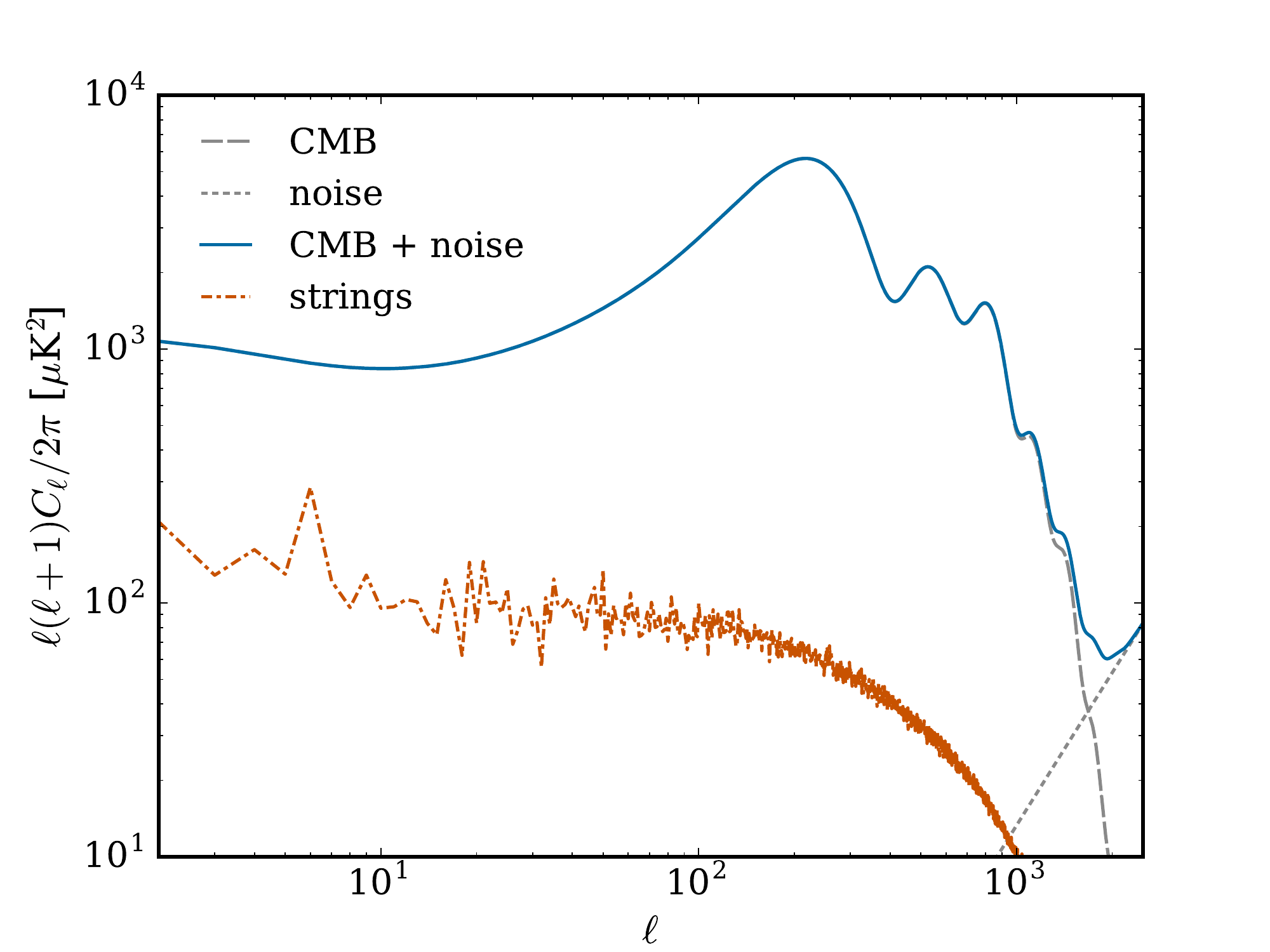}
 \caption{Power spectra of the fiducial CMB, smoothed by the \planck\ beam and the \healpix\ pixel window function corresponding to  $\nside=2048$  (grey, long dash), \planck\ instrumental noise (grey, short dash) and their sum (blue, solid).  For comparison, a power spectrum estimated from a simulated full-sky cosmic string-induced CMB component \citep{ringeval:2012} is plotted, corresponding to $\gmu = 5\times10^{-7}$ (orange, dot-dash). The string contribution is clearly subdominant, highlighting the challenge in constraining cosmic string models.}
 \label{fig:input_power_spectra}
\end{figure}

The energy scale of the string-inducing phase transition $\eta$ is directly related to the string tension $\mu$ by $\mu \sim \eta^2$. Detecting the signatures of cosmic strings would therefore provide a direct probe of physics at extremely high energy scales.  However, since any string signature must be subdominant, detecting strings is a significant observational challenge. The magnitude of the task is demonstrated in \fig{\ref{fig:input_power_spectra}}, in which we compare the power spectrum of a simulated string-induced CMB contribution~\citep{ringeval:2012} (with amplitude close to current observational limits) and a standard inflationary component, as would be observed by \planck. Hereafter, we refer to string-induced CMB anisotropy maps with the shorthand ``string maps''.

Various methods have been developed to search for string-induced contributions to the CMB, from power-spectrum constraints~\citep{lizarraga:2014,lizarraga:2014b,lizarraga:2016,charnock:2016}, to higher-order statistics such as the bispectrum~\citep{planck2013-p20,regan:2015} and trispectrum~\citep{fergusson:2010}, and tools such as edge detection~\citep{lo:2005,amsel:2008,stewart:2009,danos:2010}, Minkowski functionals~\citep{gott:1990,ducout:2013,planck2013-p20}, wavelets and curvelets~\citep{starck:2004,hammond:2009,wiaux:2010:csstring,planck2013-p20,hergt:2016}, level crossings~\citep{movahed:2011}, and peak-peak correlations~\citep{movahed:2013}.  
Current constraints on the string tension depend on the string model and simulation technique adopted.
For Nambu-Goto strings, power spectrum analyses based on simulations computed by the unconnected segment model (USM; \citealt{albrecht:1997,albrecht:1999,pogosian:1999}) constrain the string tension to $\gmu < 1.3 \times 10^{-7}$ \citep{planck2013-p20} using \planck\ temperature data and to $\gmu < 1.1 \times 10^{-7}$ \citep{charnock:2016} when \planck\ polarisation data are also included.
Recombination effects have been considered by \citet{regan:2015} but were found not to have a significant effect on the bispectrum.
Beyond spectra, non-Gaussian analyses for Nambu-Goto strings---based on high-resolution simulations of stringy CMB maps including only the \isw\ (\iswtext) contribution and no recombination effects~\citep{ringeval:2012}---constrain the string tension to $\gmu < 7.8 \times 10^{-7}$ \citep{planck2013-p20} using \planck\ temperature data. Considering only the ISW effect enables the production of high-resolution full-sky string maps, but these maps are necessarily conservative, and the resulting constraints are hence weaker. Furthermore, effects of recombination physics would increase the string anisotropy signal considerably \citep{planck2013-p20}.  While power spectrum statistics are inherently lossy, map-based analyses have the potential to better discriminate cosmic strings from other potential subdominant CMB signals.

\begin{figure}
  \centering
  \subfigure[Pixel space\label{fig:wav_vs_pix_separability:pixel}]{\includegraphics[width=.7\columnwidth]{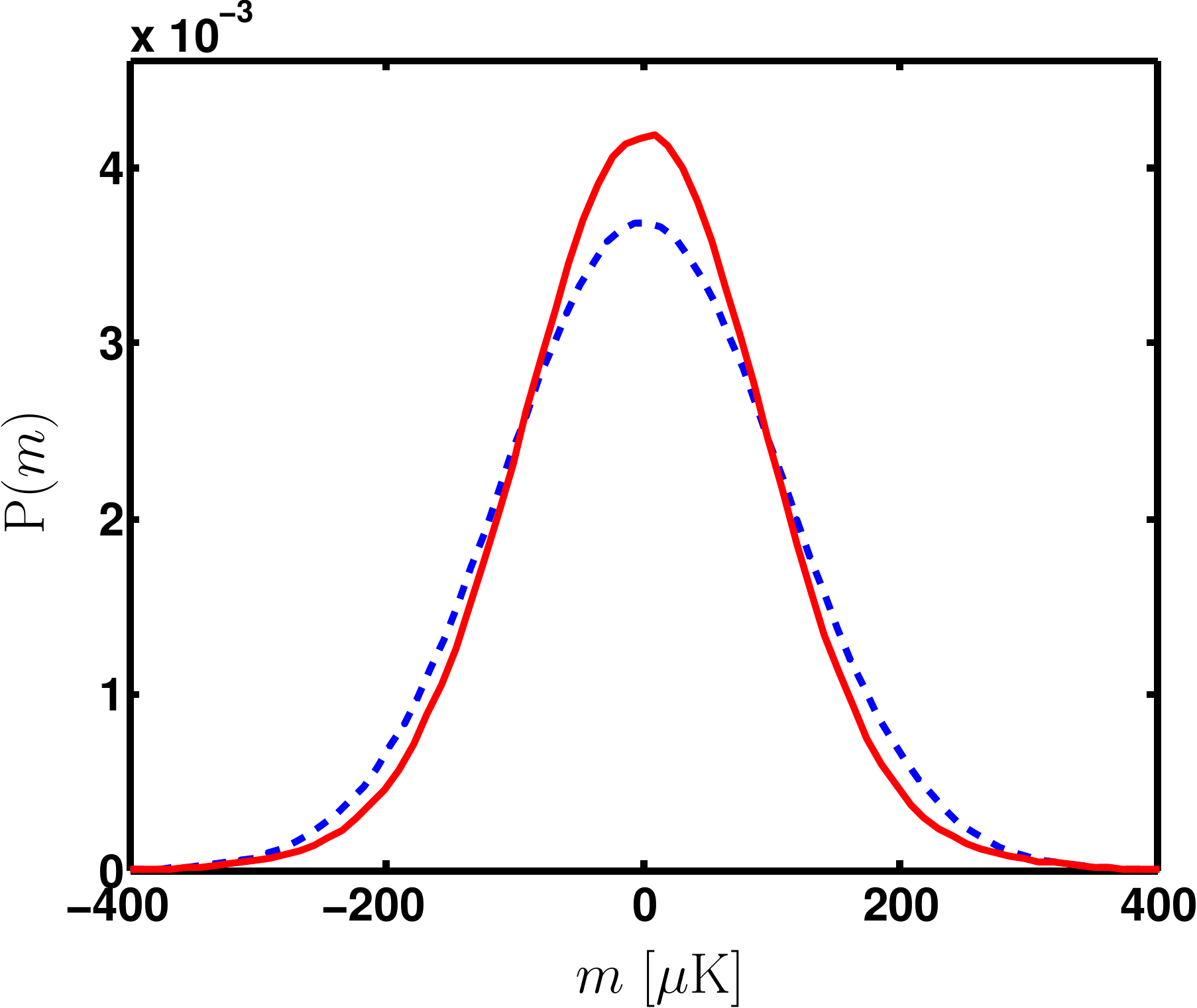}} \hfill
  \subfigure[Wavelet space ($\wscale=0$)\label{fig:wav_vs_pix_separability:wavelet}]{\includegraphics[width=.7\columnwidth]{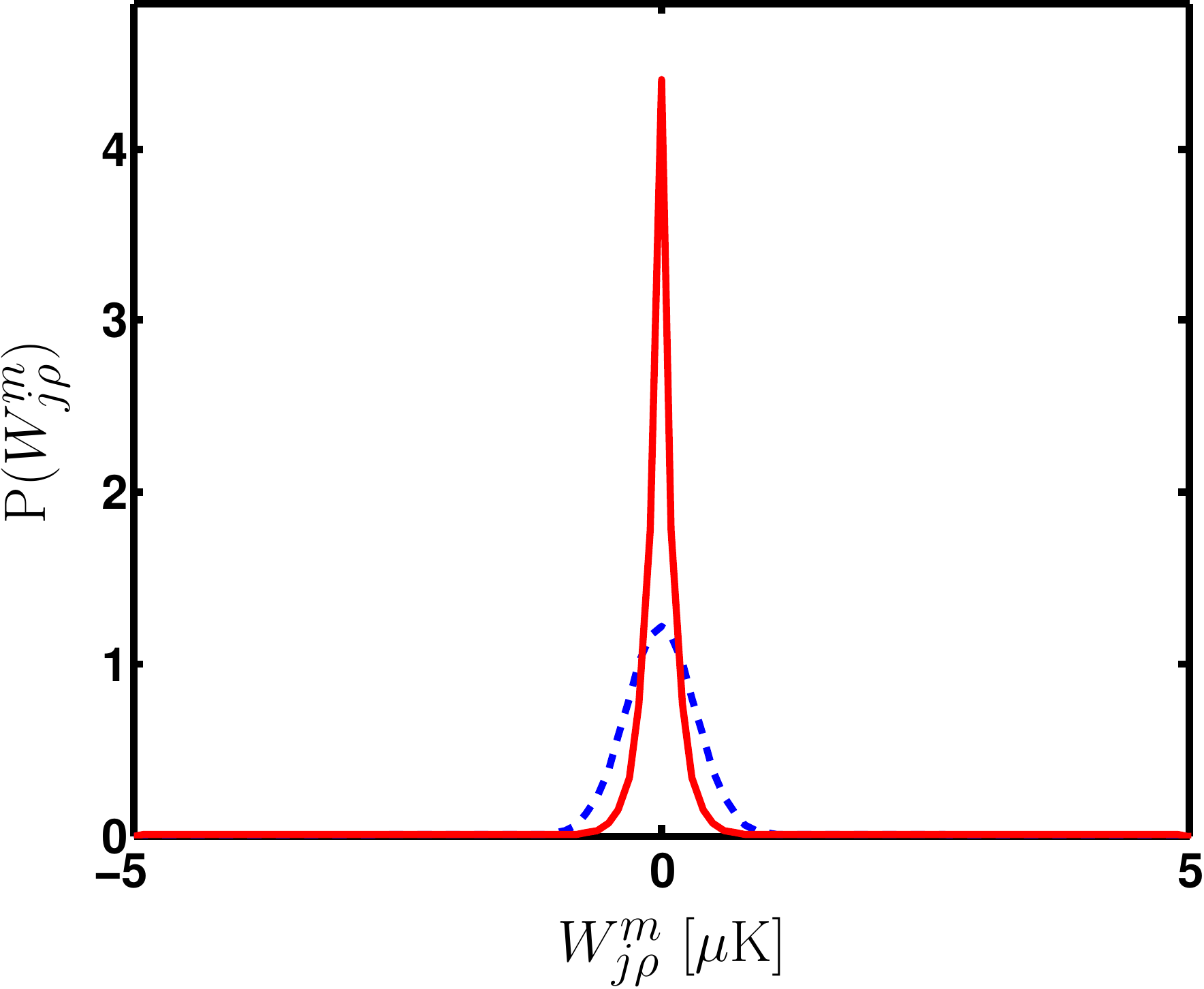}}
  \caption{Distributions of map $m \in \{c, s\}$, comprising either a CMB simulated inflationary component $c$ (blue, dashed) or string component $s$ (red, solid) CMB components, in both pixel space and scale-discretised wavelet space (for $\gmu=2\times10^{-6}$). The string-induced component is simulated by the method of \citet{ringeval:2012}. The shape of the distributions is reasonably similar in pixel space, whereas in wavelet space the distributions are markedly different.}
  \label{fig:wav_vs_pix_separability}
\end{figure}

As constraints on the amplitude of any string-induced component tighten, analysis techniques must become more sensitive to improve on the status quo. Wavelets are a particularly powerful tool for searching for cosmic strings due to their ability to simultaneously characterise signal structure in both scale and position. Furthermore, wavelets that are well-matched to the expected structure of string maps can be adopted, facilitating extraction of the string signal from the CMB and instrumental noise.  Although string-induced CMB anisotropies are non-Gaussian, the statistical distribution of the pixels of a cosmic string map nevertheless remains close to Gaussian. In \fig{\ref{fig:wav_vs_pix_separability}} histograms of simulated inflationary and cosmic string components are plotted, in both pixel (\fig{\ref{fig:wav_vs_pix_separability:pixel}}) and wavelet (\fig{\ref{fig:wav_vs_pix_separability:wavelet}}) space. The shape of the distributions is reasonably similar in pixel space, whereas in wavelet space the distributions are markedly different.  The distribution of the string component in wavelet space is highly peaked (\ie\ leptokurtic) due to the sparse representation of the string component in wavelet space (\ie\ due to the property that many of the wavelet coefficients of the string component are near zero).  The inflationary CMB component, however, remains Gaussian distributed in wavelet space, since the wavelet transform is linear.  The difference in the statistical properties of the string and inflationary CMB components in wavelet space can be exploited to isolate and estimate the parameters of any string component.  This is the approach taken in the current work.\footnote{For very small scales, the underlying string distribution in pixel space becomes increasingly different to a Gaussian distribution in its tails, however these features can be washed out observationally by instrumental beams.  In any case, for small instrumental beams that preserve these features the wavelet approach presented in the current work would characterise such structure, improving the sensitivity of the analysis.}

\begin{figure*}
  \centering
  \includegraphics[width=1.6\columnwidth]{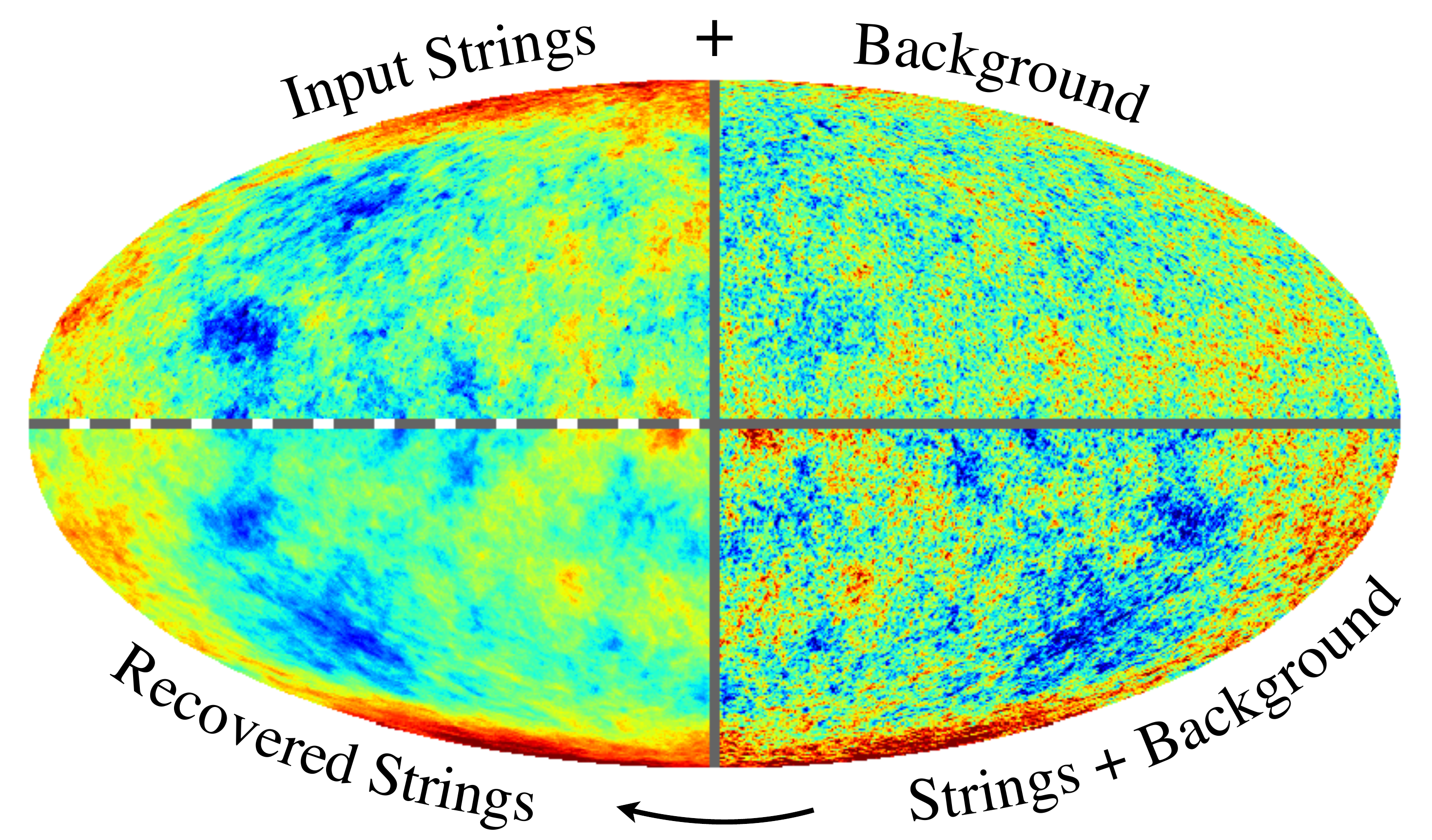}
  \caption{Illustration of string map recovery.  Under the string model, the CMB is comprised of a string-induced component (top left quadrant) and a Gaussian inflationary background component (top right quadrant), yielding the observed CMB map (bottom right quadrant).  For visualisation purposes, in this simulation the string component is generated for a relatively large value of the string tension ($\gmu=2\times10^{-6}$), which controls the amplitude of the string map.  We invert this process to recover an estimate of the input string component (bottom left quadrant).  This is achieved by our hybrid wavelet-Bayesian analysis pipeline, which estimates the posterior distribution of the string tension, the Bayesian evidence for strings, and the mean-posterior string map.  The same quadrant of each map is displayed in each panel, rotated and reflected as required.}
  \label{fig:denoising_illustration}
\end{figure*}

In this article we develop a hybrid wavelet-Bayesian approach to infer the presence and parameters of any cosmic string component in the CMB.  We do not consider (insufficient) summary statistics like many alternative methods (\eg\ the kurtosis), for which the origin of any non-Gaussian component cannot be rigorously determined.  Instead, we learn and exploit the complex non-Gaussian structure of string-induced CMB contributions.  We follow the approach of \citet{hammond:2009}, generalising from the planar setting to the celestial sphere.  In \citet{hammond:2009}, techniques using planar wavelets are presented to learn the statistical structure of string-induced CMB contributions and to exploit this structure to recover an estimate of a planar map of the string component. We generalise these techniques to the full-sky setting using scale-discretised wavelets defined on the sphere \citep{mcewen:s2let_spin,leistedt:s2let_axisym,wiaux:2007:sdw}, adopting directional wavelets with parameters selected to match the characteristic step-like temperature changes induced by strings in the CMB \citep{kaiser:1984}.  While \citet{hammond:2009} adopt a power spectrum approach to estimate the string tension, we recover the posterior distribution of the string tension in our wavelet formalism.  Moreover, we also compute the Bayesian evidence to distinguish between the cosmic string model and the standard inflationary model.  
In summary, we present a principled and robust statistical framework based on Bayesian inference for parameter estimation and model selection, performing a Bayesian analysis in wavelet space where the inflationary and string induced CMB components have very different statistical properties.
An overview of the string model and the recovery of any string-induced component is illustrated in \fig{\ref{fig:denoising_illustration}}, while an example of the scale-discretised wavelets considered is shown in \fig{\ref{fig:wavelet}}.

We restrict our attention here to simulated observations modelling idealised \planck\ observations.  An optimisation of the parameters of the method and the application to \planck\ data will be presented in a subsequent study.
We employ the matter-free simulations of~\citet{ringeval:2012}, which faithfully represent the small-scale string anisotropies produced by the ISW effect. As previously stated, these simulations come at a cost of hundreds of thousands of CPU hours per full-sky \planck-resolution map. Fortunately, our approach requires only two realisations of string maps: one to train our method and one to test it (see \fig{\ref{fig:input_string_sims}}).

The outline of the paper is as follows.  In \sectn{\ref{sec:wavelets}}, we review the wavelets used in this analysis.  In \sectn{\ref{sec:inference}} we describe in detail our hybrid wavelet-Bayesian framework for inferring the string tension and determining the Bayesian evidence for the string model relative to the standard inflationary model.  We present in \sectn{\ref{sec:denoising}} the approach to recovering an estimate of the string-induced CMB component at the map level in the full-sky setting, which can be viewed as a Bayesian thresholding approach to denoising the observed CMB signal.  In \sectn{\ref{sec:results}} we apply our framework to simulated observations and discuss the results.  Concluding remarks are made in \sectn{\ref{sec:conclusions}}.

\section{Scale-discretised wavelets on the sphere}
\label{sec:wavelets}

Wavelets on the sphere have found widespread use in analyses of the CMB \citep[\eg][]{vielva:2004, mcewen:2005:ng, mcewen:2006:ng, mcewen:2008:ng, vielva:2005, mcewen:2006:isw, mcewen:2007:isw2, feeney:2011b, feeney:2011a, feeney:2012, planck2013-p06, planck2013-p09, planck2013-p20} due to their ability to localise signal content in scale and space simultaneously (for a review see \citealt{mcewen:2006:review}).

Initial stable wavelet constructions on the sphere were based largely on continuous methodologies \citep[\eg][]{antoine:1998,antoine:1999,wiaux:2005,sanz:2006,mcewen:2006:cswt2}, which do not support the exact synthesis of a sampled signal from its wavelet coefficients in practice.  Consequently, cosmological analyses based on these constructions were limited to the analyses of wavelet coefficients; sampled signals on the sphere could not be accurately recovered from processed wavelet coefficients.  Alternative discrete constructions based on the lifting scheme \citep{sweldens:1997} were developed \citep{schroder:1995, barreiro:2000, mcewen:2008:fsi, mcewen:szip}, however these do not necessarily lead to a stable basis \citep{sweldens:1997}.

More recently, a number of exact discrete wavelet frameworks on the sphere have
been developed, with underlying continuous representations and fast
implementations that have been made available publicly, including needlets
\citep{narcowich:2006, baldi:2009, marinucci:2008}, directional scale-discretised
wavelets \citep{wiaux:2007:sdw, leistedt:s2let_axisym, mcewen:2013:waveletsxv},
and the isotropic undecimated and pyramidal wavelet transforms
\citep{starck:2006}.  Each approach has also been extended to analyse spin
functions on the sphere \citep{geller:2008, geller:2010:sw, geller:2010,
geller:2009_bis, mcewen:s2let_spin, mcewen:s2let_spin_sccc21_2014,
starck:2009} and functions defined on the three-dimensional ball formed by
augmenting the sphere with the radial line \citep{durastanti:2014,
leistedt:flaglets,   mcewen:flaglets_sampta, leistedt:flaglets_spin, lanusse:2012}.  Ridgelet and curvelet wavelets on the sphere have also been constructed \citep{starck:2006, mcewen:s2let_ridgelets, chan:s2let_curvelets}.

In this work we adopt directional scale-discretised wavelets
\citep{wiaux:2007:sdw, leistedt:s2let_axisym, mcewen:2013:waveletsxv,mcewen:s2let_spin, mcewen:s2let_localisation}, which are essentially the generalisation of needlets \linebreak \citep{narcowich:2006, baldi:2009, marinucci:2008} to directional wavelets \citep{mcewen:s2let_localisation}.  Directional scale-discretised wavelets have recently been shown to satisfy quasi-exponential localisation and asymptotic uncorrelation properties similar to needlets \citep{mcewen:s2let_localisation} and consequently have excellent spatial localisation properties.

In the remainder of this section we review directional scale-discretised wavelets concisely; for further details please see the related literature \citep{wiaux:2007:sdw, leistedt:s2let_axisym, mcewen:2013:waveletsxv,mcewen:s2let_spin, mcewen:s2let_localisation}.  The reader not interested in the details may safely skip the following subsections and simply note the notation used to denote wavelet coefficients specified in \eqn{\ref{eqn:analysis}}.

\subsection{Wavelet transform and inversion}

The scale-discretised wavelet transform of a function $\f \in
\ltwo(\sphere)$ on the sphere \sphere\ is defined by the directional
convolution of \f\ with the wavelet $\wav^{(\wscale)} \in
\ltwo(\sphere)$.  In order to perform directional, spherical convolutions it is necessary to rotate functions on the sphere.
The rotation operator $\rotarg{\eul}$ is defined by
\begin{equation}
  \wav_{\wscaleposn}(\sa)
  \equiv (\rotarg{\eul} \wav_{\wscale})(\sa)
  \equiv \wav_{\wscale}(\rotmatarg{\eul}^{-1} \vect{\hat{\sa}})
  \spcend ,
\end{equation}
where $\rotmatarg{\eul}$ is the three-dimensional rotation matrix
corresponding to $\rotarg{\eul}$.
Spherical coordinates are denoted \mbox{$\sa=(\sas) \in \sphere$} with
colatitude $\saa \in [0,\pi]$ and longitude $\sab \in [0,2\pi)$, where $\vect{\hat{\sa}}$ denotes the
Cartesian vector corresponding to $\sa$.  Rotations are specified by elements
of the rotation group $\sothree$, parameterised by the Euler angles
$\eul=(\euls) \in \sothree$, with $\eula \in [0,2\pi)$, $\eulb \in
[0,\pi]$ and $\eulc \in [0,2\pi)$.
The scale-discretised wavelet transform on the sphere then reads
\begin{align}
  \wcoeff^f_{\wscaleposn} \equiv
  \wcoeff^{f}_{\wscale}(\eul) &\equiv ( \f \circledast \wav_{\wscale}) (\eul)
  \equiv \innerp{\f}{\wav_{\wscaleposn}} \nonumber \\
  &= \int_\sphere \dmu{\sa} \f(\sa) \wav_{\wscaleposn}^\cconj(\sa)
  \spcend ,
  \label{eqn:analysis}
\end{align}
where $\wscale$ denotes the wavelet scale, which encodes the angular
localisation of the wavelet, $\dmu{\sa} = \sin\saa \dx\saa \dx\sab$
is the usual rotation-invariant
measure on the sphere, and $\cdot^\cconj$ denotes complex conjugation.
The inner product of functions on the sphere is denoted
$\innerp{\cdot}{\cdot}$, while the operator $\circledast$ denotes
directional convolution on the sphere.

The wavelet transform of \eqn{\ref{eqn:analysis}} thus probes directional
structure in the signal of interest \f, where \eulc\ can be viewed as
the orientation about each point on the sphere $(\sas) = (\eulb,
\eula)$.  Wavelet coefficient at scale $\wscale$ therefore live on the rotation group, \ie\ $\wcoeff^{f}_{\wscale} \in \ltwo(\sothree)$.
We adopt the shorthand notation $\wcoeff^f_{\wscaleposn}$ to denote the wavelet coefficients of the signal $\f$ at scale $\wscale$ and position and orientation $\eul$, in order to simplify subsequent statistical calculations.

The wavelet coefficients do not encode the low-frequency content of the signal \f; a scaling function is introduced for this purpose.  The scaling coefficients $\scoeff^\f \in \ltwo(\sphere) $
are given by the convolution of \f\ with the axisymmetric scaling
function $\wavs \in \ltwo(\sphere)$ and read
\begin{align}
  \scoeff^{f}_\sa \equiv \scoeff^{f}(\sa) &\equiv ( \f \odot \wavs) (\sa)
  \equiv \innerp{\f}{\wavs_\sa} \nonumber \\
  &= \int_\sphere \dmu{\sa\p} \f(\sa\p) \wavs_\sa^\cconj(\sa\p)
  \spcend ,
\end{align}
where the rotated scaling function is defined by
\begin{equation}
  \wavs_{\sa}(\sa\p)
  \equiv (\rotarg{\sa} \wavs)(\sa\p)
  \equiv \wavs(\rotmatarg{\sa}^{-1} \vect{\hat{\sa}}\p)
  \spcend ,
\end{equation}
with $\rotarg{\sa} = \rot_{(\sab,\saa, 0)}$.  The operator $\odot$
denotes axisymmetric convolution on the sphere.  Note that the scaling
coefficients live on the sphere, and not the rotation group \sothree,
since directional structure of the low-frequency content of \f\ is
not typically of interest.  We adopt the shorthand notation $\scoeff^{f}_\sa $ to denote the scaling coefficients of the signal $\f$ at position $\sa$.
In addition, we introduce the shorthand notation $\wcoeff^f = \mathcal{W}(f)$ to represent the overall wavelet analysis of \f, \ie\ including both wavelet and scaling coefficients.

Provided the wavelets and scaling function satisfy an admissibility condition (see \sectn{\ref{sec:wavelets:construction}}), the original signal \f\ can be synthesised exactly from its wavelet and scaling coefficients by
\begin{align}
  \f(\sa)
  = &\int_\sphere \dmu{\sa\p}
  \scoeff^f_{\sa\p} \wavs_{\sa\p}(\sa)
  +
  \sum_{\wscale=0}^\wscalemax \int_\sothree \deul{\eul}
  \wcoeff^f_{\wscaleposn} \wav_\wscaleposn(\sa)
  \spcend ,
  \label{eqn:synthesis}
\end{align}
where $\deul{\eul} = \sin\eulb \dx\eula \dx\eulb \dx\eulc$ is the usual
invariant measure on \sothree.
We introduce the shorthand notation \mbox{$f = \mathcal{W}^{-1} (\wcoeff^f)$} to represent the synthesis of a signal from its wavelet and scaling coefficients.

We adopt the same convention as \citet{wiaux:2007:sdw} and \citet{mcewen:s2let_localisation} for the wavelet
scales \wscale, with increasing \wscale\ corresponding to larger
angular scales, \ie\ lower frequency content.\footnote{Note that this
  differs to the convention adopted in
  \citet{leistedt:s2let_axisym} and \citet{mcewen:s2let_spin} where increasing
  \wscale\ corresponds to smaller angular scales and higher frequency
  content.}
The maximum possible wavelet scale $\wscale$ is denoted by
 $\wscalemax_{\rm max}$ and is set
to ensure the wavelets probe the entire scale (frequency) range (except zero) of the signal of interest,
yielding $\wscalemax_{\rm max} = \ceil{\log_\dilparam(\elmax)}$, where \dilparam\ is a dilation parameter (see \citealt{wiaux:2007:sdw, leistedt:s2let_axisym,mcewen:s2let_spin, mcewen:s2let_localisation}).  The maximum wavelet
scale considered in a given analysis $\wscalemax$ may be freely chosen, provided $0 \leq \wscalemax <
\wscalemax_{\rm max}$.  For \mbox{$\wscalemax=\wscalemax_{\rm max}$}, the wavelets probe the entire frequency
content of the signal of interest except its mean, which is incorporated in
the scaling coefficients.

\subsection{Wavelet construction}
\label{sec:wavelets:construction}

For the original signal to be synthesised perfectly from its wavelet and scaling coefficients through \eqn{\ref{eqn:synthesis}} the wavelets and scaling function must satisfy the following admissibility property:
\begin{equation}
  \label{eqn:admissibility}
  \frac{4\pi}{2\el+1}
  \vert \shc{\wavs}{\el}{0} \vert^2 +
  \frac{8\pi^2}{2\el+1}
  \sum_{\wscale=\wscalemin}^\wscalemax
  \summ \vert \shc{(\wav_{\wscale})}{\el}{\m}\vert^2 = 1
  \spcend ,
  \quad \forall\el
  \spcend ,
\end{equation}
where $\shc{\wavs}{\el}{0} \kron{\m}{0} =
\innerp{\wavs}{\shf{\el}{\m}}$ and
$\shc{(\wav_\wscale)}{\el}{\m} = \innerp{\wav_\wscale}{\shf{\el}{\m}}$ are the spherical harmonic coefficients
of $\wavs$ and $\wav_{\wscale}$, respectively, where $\kron{i}{j}$
for $i,j \in \integers$ denotes the Kronecker delta.  The spherical harmonic functions are denoted by $\shf{\el}{\m} \in \ltwo(\sphere)$, with $\el\in\naturals$ and $\m \in \integers$, $\vert\m\vert\leq \el$.

\begin{figure}
  \centering
  \includegraphics[width=\columnwidth, trim=4mm 8mm 4mm 8mm, clip=true]{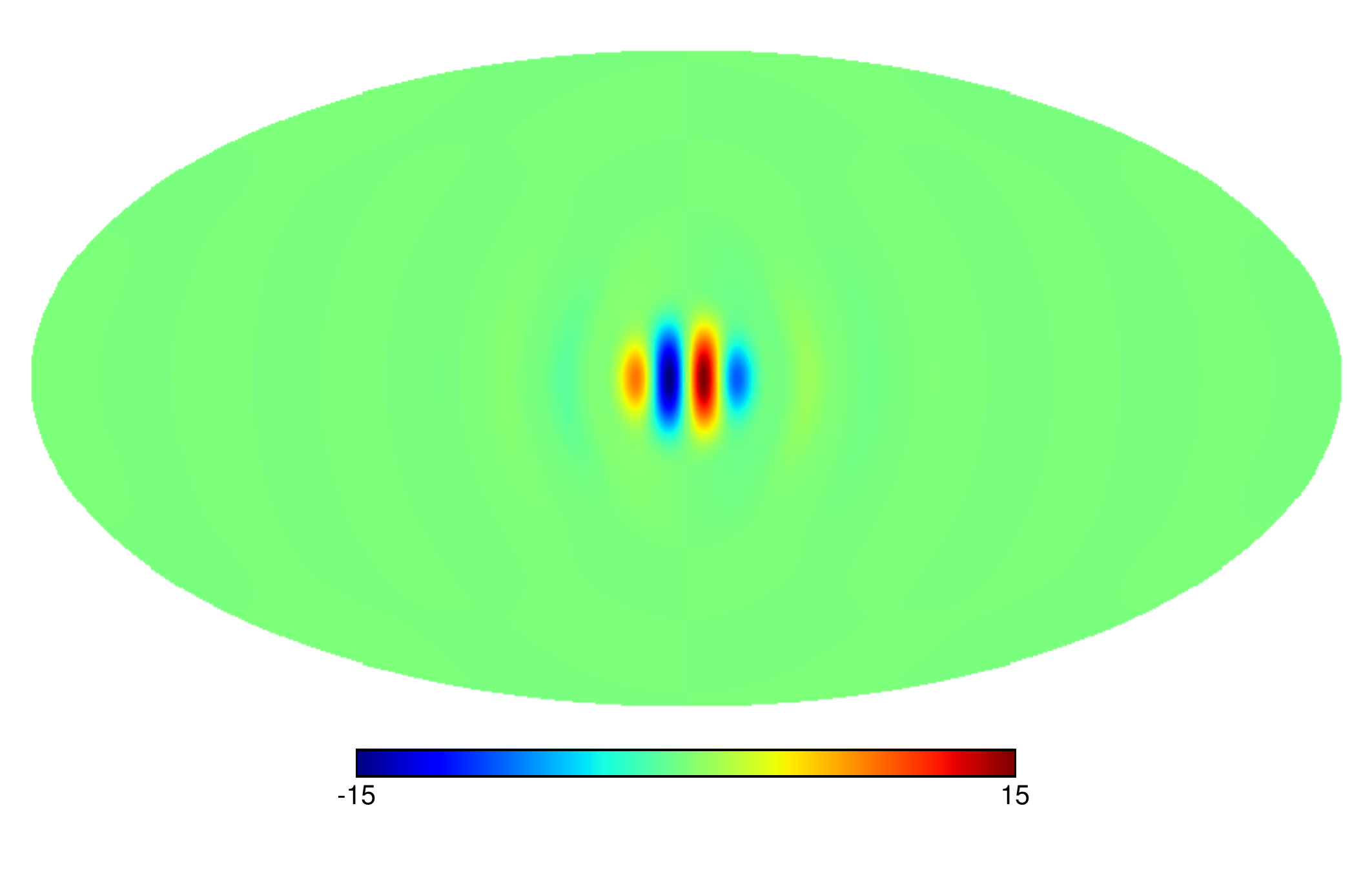}
  \caption{Directional scale-discretised wavelet with odd azimuthal symmetry for $\nmax=4$, $\wscale=7$, $\elmax=2048$, and $\dilparam=2$. The wavelet is rotated from the North pole to the equator for visualisation purposes. The same wavelet parameters are assumed when analysing cosmic string maps, although lower $\wscale$ (\ie\ smaller scales) are also considered. The wavelet is selected to match the step-like structure of contributions to the CMB due to cosmic strings, in order to yield a sparse representation of the string component in wavelet space.}
  \label{fig:wavelet}
\end{figure}

Wavelets are defined in harmonic space in the separable form
\begin{equation}
  \label{eqn:wav_factorized}
  \shc{(\wav_\wscale)}{\el}{\m} \equiv
  \sqrt{\frac{2\el+1}{8\pi^2}} \:
  \wavker_{\wscale}(\el) \:
  \shc{\wavsteer}{\el}{\m}
  \spcend,
\end{equation}
in order to control their angular and directional localisation
separately, respectively through the kernel
$\wavker_{\wscale} \in \ltwo(\reals^{+})$ and directionality
component $\wavsteer \in \ltwo(\sphere)$, with harmonic coefficients
$\shc{\wavsteer}{\el}{\m} = \innerp{\wavsteer}{\shf{\el}{\m}}$.
Without loss of generality, the directionality component is
normalised to impose
\begin{equation}
  \label{eqn:directionality_normalisation}
  \summ \vert \shc{\wavsteer}{\el}{\m} \vert^2 = 1,\quad \forall \el
  \spcend.
\end{equation}
An azimuthal band-limit \nmax\ is imposed on the
directionality component such that $\shc{\wavsteer}{\el}{\m}=0$,
$\forall \el,\m$ with $\vert \m \vert \geq \nmax$, which controls the directional selectivity of the wavelet.  Moreover, the wavelets are constructed to
exhibit odd (even) azimuthal symmetry for $\nmax-1$ odd (even). For further detail regarding the explicit construction of the wavelet kernel and directionality component see, \eg, \citet{mcewen:s2let_localisation}.
An example of a scale-discretised wavelet on the sphere is plotted in \fig{\ref{fig:wavelet}}

\subsection{Computation}

By appealing to sampling theorems on the sphere \citep[\eg][]{mcewen:fssht} and rotation group \citep[\eg][]{mcewen:so3}, the forward and inverse wavelet transforms of sampled signals can be computed exactly in theory for band-limited signals on the sphere, \ie\ signals with spherical harmonic coefficients $\shc{\f}{\el}{\m} = 0$,
$\forall \el \geq \elmax$, where
$\shc{\f}{\el}{\m} = \innerp{\f}{\shf{\el}{\m}}$. The only error arising in numerical computations is that due to the finite representation of floating point numbers (indeed, numerical errors are found to be on the order of machine precision; see \eg\ \citealt{mcewen:s2let_spin}).  In practice, many real-world signals can be approximated accurately by band-limited signals. Furthermore, fast algorithms to compute the harmonic transforms associated with sampling theorems on the sphere and rotation group \citep[\eg][respectively]{mcewen:fssht,mcewen:so3} can be exploited to render forward and inverse scale-discretised wavelet transforms computationally feasible for large cosmological data-sets (\eg\ \planck\ maps).

The scale-discretised wavelet transform on the sphere is implemented in the \stwoletcode\ code
\citep{leistedt:s2let_axisym, mcewen:s2let_spin}.  The core algorithms of \stwoletcode\ are
implemented in C, while Matlab, Python and IDL interfaces are
also provided.  Consequently, \stwoletcode\ is able to handle very
large harmonic band-limits, corresponding to data-sets containing tens
of millions of pixels.
\stwoletcode\footnote{\url{http://www.s2let.org}} is publicly
available, and relies on the
\sshtcode\footnote{\url{http://www.spinsht.org}} code
\citep{mcewen:fssht} to compute spherical harmonic transforms, the
\sothreecode\footnote{\url{http://www.sothree.org}} code
\citep{mcewen:so3} to compute Wigner transforms and the
\fftwcode\footnote{\url{http://www.fftw.org}} code to compute Fourier
transforms.  Note that it also supports the analysis of data on the
sphere defined in the common
\healpix\footnote{\url{http://healpix.jpl.nasa.gov}}
\citep{gorski:2005} format.  \stwoletcode\ provides the most recent and feature-rich implementation of scale-discretised wavelets, however, development on this project and \stwoletcode\ was concurrent, and here we therefore use the previous \stwodwcode\footnote{\url{http://www.s2dw.org}} code \citep{wiaux:2007:sdw} (which is functionally identical for the setting considered). \stwodwcode\ is implemented in Fortran, and relies on the \stwocode\footnote{\url{http://www.jasonmcewen.org/codes.html}} code \citep{mcewen:2006:fcswt,mcewen:2006:filters} to handle data defined on the sphere and \fftwcode\ to perform Fourier transforms.

\section{Inference of cosmic string model}
\label{sec:inference}

While a cosmic string-induced component embedded in the CMB will not be Gaussian, the statistical distribution of the pixels of a cosmic string map nevertheless remains close to Gaussian.  In \fig{\ref{fig:wav_vs_pix_separability}} histograms of simulated inflationary and string-induced CMB components are plotted, in both pixel (\fig{\ref{fig:wav_vs_pix_separability:pixel}}) and wavelet (\fig{\ref{fig:wav_vs_pix_separability:wavelet}}) space.  In pixel space the distributions are similar.  In wavelet space, however, while the distribution of the inflationary CMB component remains Gaussian (since the wavelet transform is linear), the distribution of the string-induced component is highly non-Gaussian. The latter distribution is peaked sharply about zero, illustrating the sparsifying nature of the wavelet transform for strings: the wavelet coefficients of the string-induced CMB component are sparsely distributed in wavelet space, while the coefficients of the inflationary CMB component are not.

We construct a hybrid wavelet-Bayesian framework to infer the presence of cosmic strings from CMB temperature observations.  By constructing the statistical framework in wavelet space, where the inflationary and string-induced components have quite different statistical properties, we exploit the sparseness of the wavelet representation of the string signal to effectively determine the presence and parameters of any such component.

In this section we first describe the various models considered.  We then define the statistical distributions of the inflationary and string-induced CMB components and noise, before presenting the framework for estimating the posterior distribution of the string tension \gmu\ and for estimating the Bayesian evidence in order to perform model selection.

\begin{figure}
  \begin{center}


\begin{tikzpicture}
  [squarednode/.style={rectangle, draw=black, minimum size=8mm},
  latent/.style={circle, draw=black, minimum size=10mm}]


  \node[obs]                               (wd) {\large $W^d$};
  \node[latent, above=0.3cm of wd, xshift=-1.3cm]  (ws) {\large $W^s$};
  \node[latent, left=0.8cm of ws, xshift=0cm]  (s) {\large $s$};
    \node[const, above=-0.25cm of s, xshift=0.9cm] () {\large $\mathcal{W}^{-1}$};
  \node[latent, above=0.3cm of wd, xshift=1.3cm]  (wg) {\large $W^g$};
  \node[latent, right=0.8cm of wg, xshift=0cm]  (g) {\large $g$};
    \node[const, above=-0.25cm of g, xshift=-0.9cm] () {\large $\mathcal{W}$};
  \node[latent, above=1.5cm of g, xshift=-0.7cm]  (cl) {\large $C_\ell$};
    \node[latent, above=1.5cm of g, xshift=0.7cm]  (nl) {\large $N_\ell$};
  \node[factor, above=0.5cm of g, xshift=0cm]  (G) {G};
    \node[const, right=0.1cm of G] () {\normalsize Gaussian};

  \node[squarednode, above=1.6cm of ws, xshift=0.5cm] (nu) {\large $\zeta_j$};
  \node[squarednode, above=1.6cm of ws, xshift=1.5cm] (upsilon) {\large $\xi_j$};
  \node[latent, above=1.5cm of ws, xshift=-0.8cm] (gmu) {\large $G\mu$};

  \node[factor, above=0.5cm of ws, xshift=0cm]  (GGD) {G};
    \node[const, right=0.1cm of GGD] () {\normalsize Generalised Gaussian};


  \draw [->, dashed] (ws) -- (wd);
  \draw [->, dashed] (wg) -- (wd);
  \draw [->, dashed] (ws) -- (s);
  \draw [->, dashed] (g) -- (wg);

  \factoredge{cl,nl}{G}{g}
  \factoredge{gmu,nu,upsilon}{GGD}{ws}

  \platenode {} {(nu)(upsilon)} {\scriptsize  $j \in \{0, ..., J \}$}; %

\end{tikzpicture}

  \caption{Graphical Bayesian model \modelstring\ of the observed  inflationary (Gaussian) and string induced (non-Gaussian) CMB components, represented in both wavelet and pixel spaces.  Solid lines represent stochastic dependencies, while dashed lines represent deterministic dependencies. The string component is modelled by a generalised Gaussian distribution (GGD) in wavelet space, while the inflationary and noise components are modelled by Gaussian distributions in pixel space.  The string component, Gaussian component and observed data are denoted by \fstring, \fcmbnoise, and \fdata, respectively, while wavelet coefficients are denoted by $\wcoeff$ with superscript representing the relevant signal.  The forward and inverse wavelet transforms are represented by the shorthand notation $\mathcal{W}$ and $\mathcal{W}^{-1}$, respectively.  The GGDs modelling the wavelet coefficients of the string component are defined by the string tension $\gmu$ and the scale and shape parameters \ggdscalew\ and \ggdshapew, respectively.  The Gaussian component is defined by the inflationary CMB and noise power spectra, $C_\el$ and $N_\el$, respectively.}
  \label{fig:hbm}
  \end{center}
\end{figure}
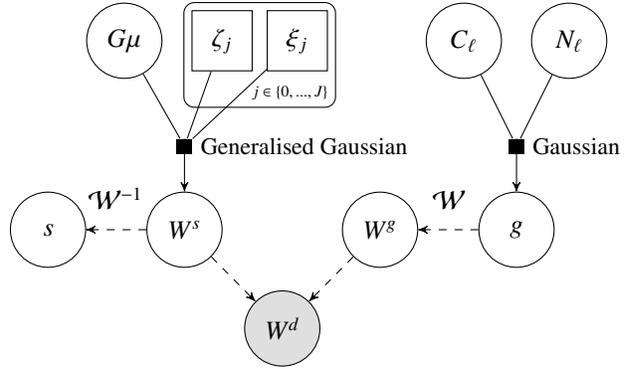

\subsection{Models}

In the presence of a subdominant contribution due to cosmic strings, we model full-sky observational CMB data \fdata\ as the sum of a string component \fstring, an inflationary Gaussian component \fcmb, and noise \fnoise:
\begin{equation}
  \modelstring: \fdata = \fstring + \fcmb + \fnoise
  \spcend .
\end{equation}
All signals are assumed to be zero-mean since we study the perturbations of cosmological signals about their mean. We denote this string model by $\modelstring$.  The alternative (standard) model is denoted by $\modelcmb$ and consists of an inflationary CMB component and noise only, absent of any string component:
\begin{equation}
  \modelcmb: \fdata = \fcmb + \fnoise
  \spcend ,
\end{equation}

We work predominantly in wavelet space, where the inflationary and string-induced CMB components exhibit very different statistical distributions.  Since the wavelet transform is linear, the models considered can be recast in wavelet space by, respectively,
\begin{equation}
  \modelstring: \wcoeffdatap = \wcoeffstringp + \wcoeffcmbp + \wcoeffnoisep
\end{equation}
and
\begin{equation}
  \modelcmb: \wcoeffdatap = \wcoeffcmbp + \wcoeffnoisep
  \spcend ,
\end{equation}
where $\wcoeffdata = \mathcal{W}(\fdata)$, $\wcoeffstring = \mathcal{W}(\fstring)$, $\wcoeffcmb = \mathcal{W}(\fcmb)$  and $\wcoeffnoise = \mathcal{W}(\fnoise)$ are the wavelet coefficients of the observed CMB data, string component, inflationary component and noise, respectively.  Here we denote the wavelet coefficients for each scale \wscale\ and rotation \eul\ separately (hence the subscripts).  Similar expressions hold for the scaling coefficients.

A graphical representation of the string model \modelstring\ in both wavelet and pixel space is shown in \fig{\ref{fig:hbm}}.  The distributions modelling the string component, inflationary component and noise are defined subsequently.
While some of the variables used in the model shown in \fig{\ref{fig:hbm}} have yet to be defined, it is nevertheless useful to present the general model now, which can then be used as a reference as the details of the model and distributions are specified in the subsequent subsections.

\subsection{Statistical distributions}
\label{sec:inference:distributions}

We determine the statistical distributions of the inflationary, noise and string-induced CMB components in wavelet space.  The first two can be calculated analytically from an assumed power spectrum (since they are Gaussian), whereas the latter must be learnt from a string training simulation.

Since we determine these distributions in wavelet space, it is necessary in the following derivations to relate the wavelet coefficients of a signal to its spherical harmonic representation \citep[\eg][]{mcewen:2006:fcswt}:
\begin{equation}
  \wcoeffdatap =
  \sumlmn
  \shc{\fdata}{\el}{\m} \:
  \shc{(\wav_{\wscale})}{\el}{\n}^\cconj \:
  \Dlmn(\eul)
  \spcend ,
  \label{eqn:wavelet_transform_harmonic}
\end{equation}
where $\shc{d}{\el}{\m} = \innerp{d}{\shf{\el}{\m}}$ and $\Dlmn \in \ltwo(\sothree)$ are the Wigner D-functions.  We adopt the shorthand notation $\sumlmn=\sumulmn$ henceforth and, assuming signals band-limited at $\elmax$, truncate sums over $\el$ to $\elmax$.

\subsubsection{CMB}

The inflationary CMB component is assumed to be a homogeneous and isotropic Gaussian random field on the sphere defined by its power spectrum $C_\el$:
\begin{equation}
  \opnexpv
  \shc{\fcmb}{\el}{\m}
  \shc{\fcmb}{\el\p}{\m\p}^\cconj
  \clsexpv
  =
  C_\el \:
  \kron{\el}{\el\p} \:
  \kron{\m}{\m\p}
  \spcend ,
\end{equation}
where $\opnexpv \cdot \clsexpv$ denotes expectation.
The cosmological parameters defining the power spectrum $C_\el$ are assumed fixed at concordance values  \citep{planck2014-a15} since the string contribution is subdominant. Since the wavelet transform is linear, the wavelet coefficients of the inflationary CMB component are also Gaussian.  Their variance for scale $\wscale$ is
\begin{align}
  \wvarcmb
  &=
  \opnexpv
  \wcoeffcmbp \:
  \wcoeffcmbp{}^\cconj
  \clsexpv \nonumber \\
  &=
  \opnexpvb
  \sumulmn
  \Dlmn(\eul) \:
  \shc{\fcmb}{\el}{\m} \:
  \shc{(\wav_{\wscale}^\cconj)}{\el}{\n}
  \sum_{\el\p\m\p\n\p}
  \dmatbig_{\m\p\n\p}^{\el\p\cconj}(\eul) \:
  \shcc{\fcmb}{\el\p}{\m\p} \:
  \shc{(\wav_{\wscale})}{\el\p}{\n\p}
  \clsexpvb \nonumber \\
  &=
  \sumulm
  C_\el \:
  | \shc{(\wav_\wscale)}{\el}{\m} |^2
  \label{eqn:cmb_wavelet_variance}
  \spcend ,
\end{align}
where we have used \eqn{\ref{eqn:wavelet_transform_harmonic}} and the Wigner property \citep{varshalovich:1989}
\begin{equation}
  \sum_{\m}
  \dmatbig_{\m\n}^{\el}(\eul)
  \dmatbig_{\m\n\p}^{\el\cconj}(\eul)
  =  \kron{\n}{\n\p}
\end{equation}
(for an alternative proof of \eqn{\ref{eqn:cmb_wavelet_variance}} see \citealt{mcewen:2006:isw}).
Consequently, the probability distribution of the wavelet coefficients of the inflationary CMB component on scale \wscale\ read:
\begin{equation}
  \prob_\wscale^\fcmb(\wcoeffcmbp) =
  \frac{1}{\sqrt{2\pi\wvarcmb}} \:
  \exp{\Biggl[ -\frac{1}{2} \biggl(\frac{\wcoeffcmbp}{\wstdcmb}\biggr)^2 \Biggr]}
  \spcend .
\end{equation}
We use $\prob(\cdot)$ to denote generic probability distributions; however, when referring to a particular distribution we add appropriate superscripts and subscripts.  Although this notation is not strictly necessary it improves the readability of the Bayesian analysis that follows.
As typically considered in statistical wavelet analyses we assume wavelet coefficients are independent and do not include their full covariance structure.  We revisit the assumption of independence later and introduce measures to account for this approximation.

\subsubsection{Noise}

Assuming Gaussian noise, we can include noise by simply modifying the Gaussian inflationary component to include the inflationary signal \fcmb\ and noise \fnoise:
\begin{equation}
  \fcmbnoise = \fcmb + \fnoise
  \spcend .
\end{equation}
The resulting term \fcmbnoise\ is Gaussian distributed since both \fcmb\ and \fnoise\ are Gaussian distributed. For modelling simplicity, we assume homogeneous and isotropic noise defined by power spectrum $N_\el$ such that
\begin{equation}
  \opnexpv
  \shc{\fcmbnoise}{\el}{\m}
  \shc{\fcmbnoise}{\el\p}{\m\p}^\cconj
  \clsexpv
  =
  (C_\el
  + N_\el) \:
  \kron{\el}{\el\p} \:
  \kron{\m}{\m\p}
  \spcend .
\end{equation}
In practice a beam $b_\el$ and pixel window function $p_\el$ may also be incorporated, yielding
\begin{equation}
  \opnexpv
  \shc{\fcmbnoise}{\el}{\m}
  \shc{\fcmbnoise}{\el\p}{\m\p}^\cconj
  \clsexpv = ( b_\el^2 p_\el^2 C_\el + N_\el ) \:
  \kron{\el}{\el\p} \:
  \kron{\m}{\m\p}
  \spcend .
\end{equation}

\subsubsection{Cosmic strings}
\label{sec:inference:distributions:strings}

Since the cosmic string-induced CMB component is not Gaussian and its map space distribution is not known \emph{a priori}, it is not possible to analytically determine its distribution in wavelet space.  Hence, we learn its distribution from a \emph{training} simulated string map and test the distribution on a separate \emph{testing} simulated string map.

\begin{figure}
  \centering
  \subfigure[Training\label{fig:input_string_sims:training}]{\includegraphics[width=\columnwidth, trim=4mm 8mm 4mm 8mm, clip=true]{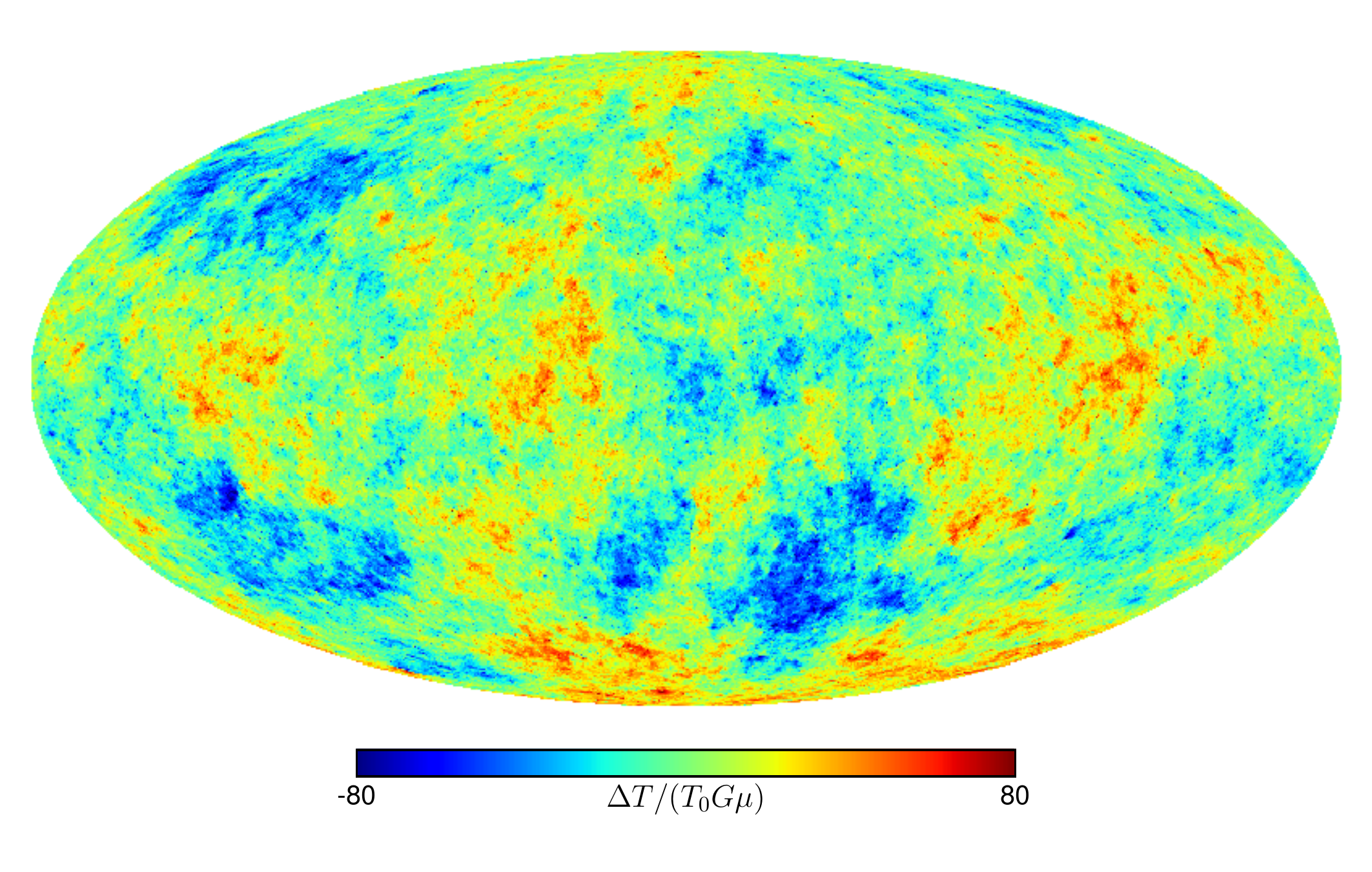}}
  \subfigure[Testing\label{fig:input_string_sims:testing}]{\includegraphics[width=\columnwidth, trim=4mm 8mm 4mm 8mm, clip=true]{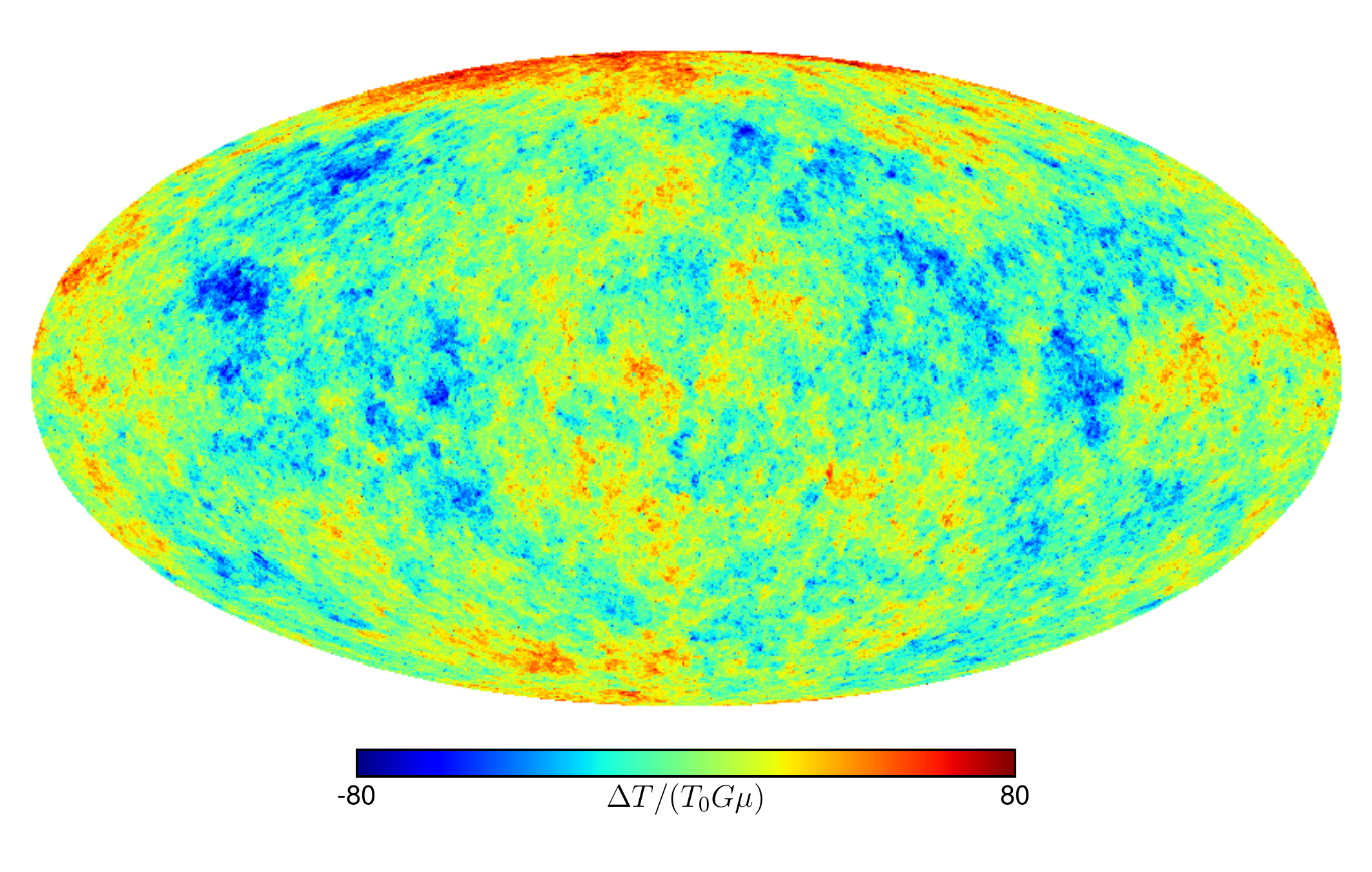}}
  \caption{Cosmic string-induced CMB anisotropies simulated by the method of \citet{ringeval:2012}.}
  \label{fig:input_string_sims}
\end{figure}

String maps simulated by the method of \citet{ringeval:2012} are shown in \fig{\ref{fig:input_string_sims}}.  As previously stated, simulating these full-sky string maps at high resolution is extremely computationally demanding, requiring hundreds of thousands of CPU hours.  Thankfully, we require only two simulated string maps: one for training, \ie\ learning the statistical properties of string-induced CMB components; and one for testing our framework.

We adopt a generalised Gaussian distribution (GGD) to model the string-induced component in wavelet space and fit its parameters from the training simulation.  The GGD of wavelet coefficients given a string tension \gmu\ is defined by
\begin{equation}
  \prob_\wscale^\fstring(\wcoeffstringp \given \gmu) =
  \frac{\ggdshapew}{2 \gmu  \ggdscalew \gammafun(\ggdshapew^{-1})} \:
  \exp{\Biggl( - \biggl|\frac{\wcoeffstringp}{\gmu \ggdscalew}\biggr|^\ggdshapew \Biggr)}
  \spcend ,
\end{equation}
where \ggdscalew\ and \ggdshapew\ are scale and shape parameters respectively (note that the overall scale of the distribution is dependent on the string tension and is given by $\gmu \ggdscalew$ at each scale) and $\gammafun(\cdot)$ denotes the Gamma function.  The GGD reduces to many common distributions for various shape parameters \ggdshape.  Gaussian and Laplacian distributions are recovered for $\ggdshape=2$ and $\ggdshape=1$ respectively, and in the limit $\ggdshape \to \infty$ the uniform distribution is recovered.  The shape parameter can thus be considered as a measure of sparsity of the underlying signal.  Note that GGDs have been used to model wavelet coefficients previously \citep[\eg][]{simoncelli:1996}.
Due to statistical isotropy, the parameters of the GGD modelling the string contribution depend on wavelet scale \wscale\ only and not the position or orientation of wavelet coefficients \eul.
For small scales we expect the distribution of wavelet coefficients of the string map to be sparse in wavelet space, which we check by testing whether the shape of the distribution is leptokurtic, \ie\ if $\ggdshapew<2$.

We learn the shape and scale parameters of the GGD for the wavelet coefficients of a string-induced CMB component by the method of moments approach outlined in \citet{hammond:2009}.
The variance and kurtosis of the GGD distributed wavelet coefficients of the string signal are given by, respectively,
\begin{equation}
  \wvarstring =
  \frac{ (\gmu)^2 \ggdscalew^2 \gammafun(3\ggdshapew^{-1}) }
  {\gammafun(\ggdshapew^{-1})}
\end{equation}
and
\begin{equation}
  \wkurstring =
  \frac{ \gammafun(5\ggdshapew^{-1}) \gammafun(\ggdshapew^{-1})}
  {\bigl(\gammafun(3\ggdshapew^{-1})\bigr)^2}
  \spcend .
\end{equation}
We compute the variance and kurtosis of the string training map and then solve these equations numerically to recover the scale and shape parameters of the GGD describing the wavelet coefficients of the string signal at each scale \wscale.
In practice, we train on the training string map with a beam and pixel windowed function applied.

\begin{figure*}
  \centering
  \subfigure[$\wscale = 0$]{\includegraphics[width=.24\textwidth]{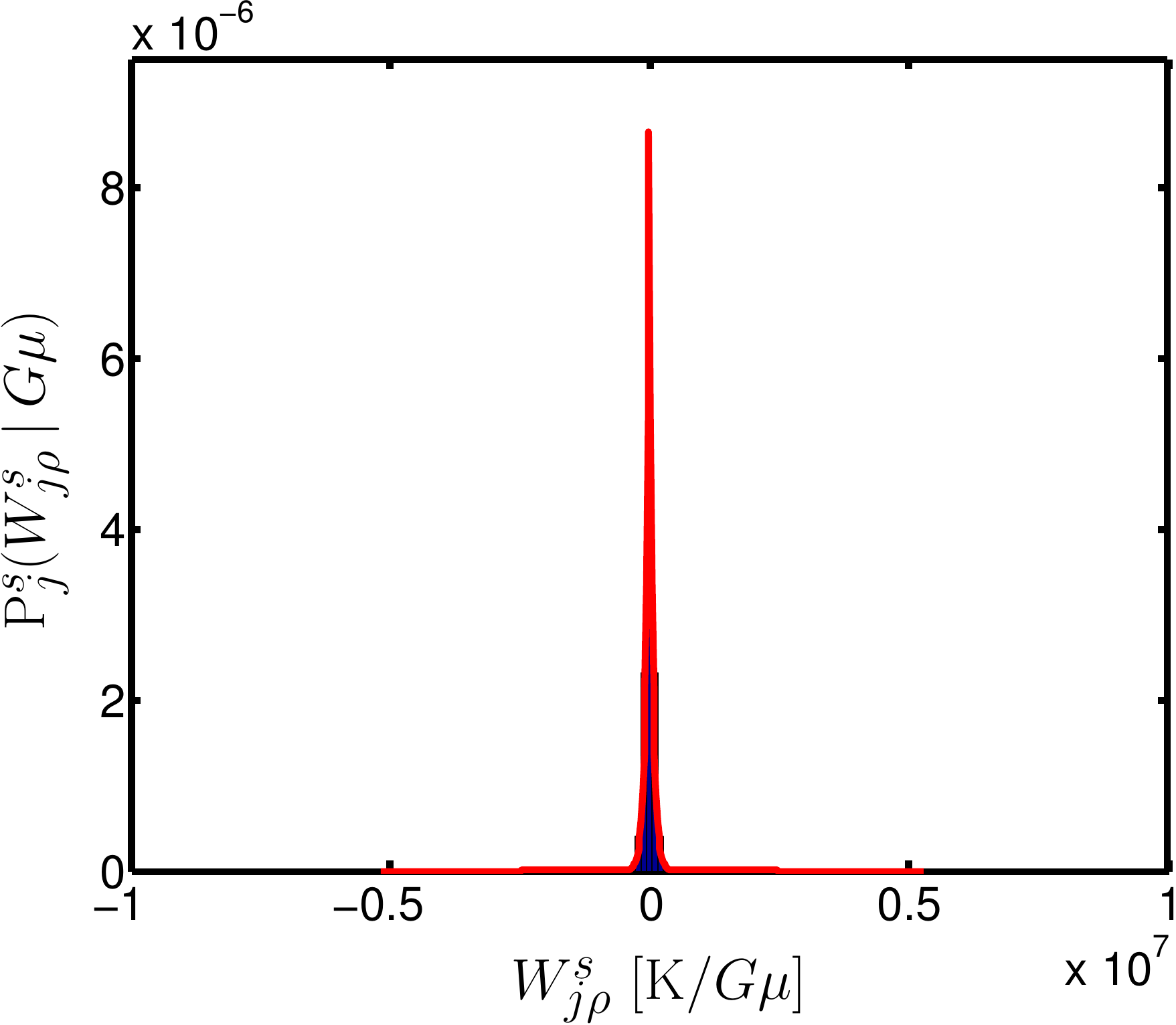}}\hfill
  \subfigure[$\wscale = 1$]{\includegraphics[width=.24\textwidth]{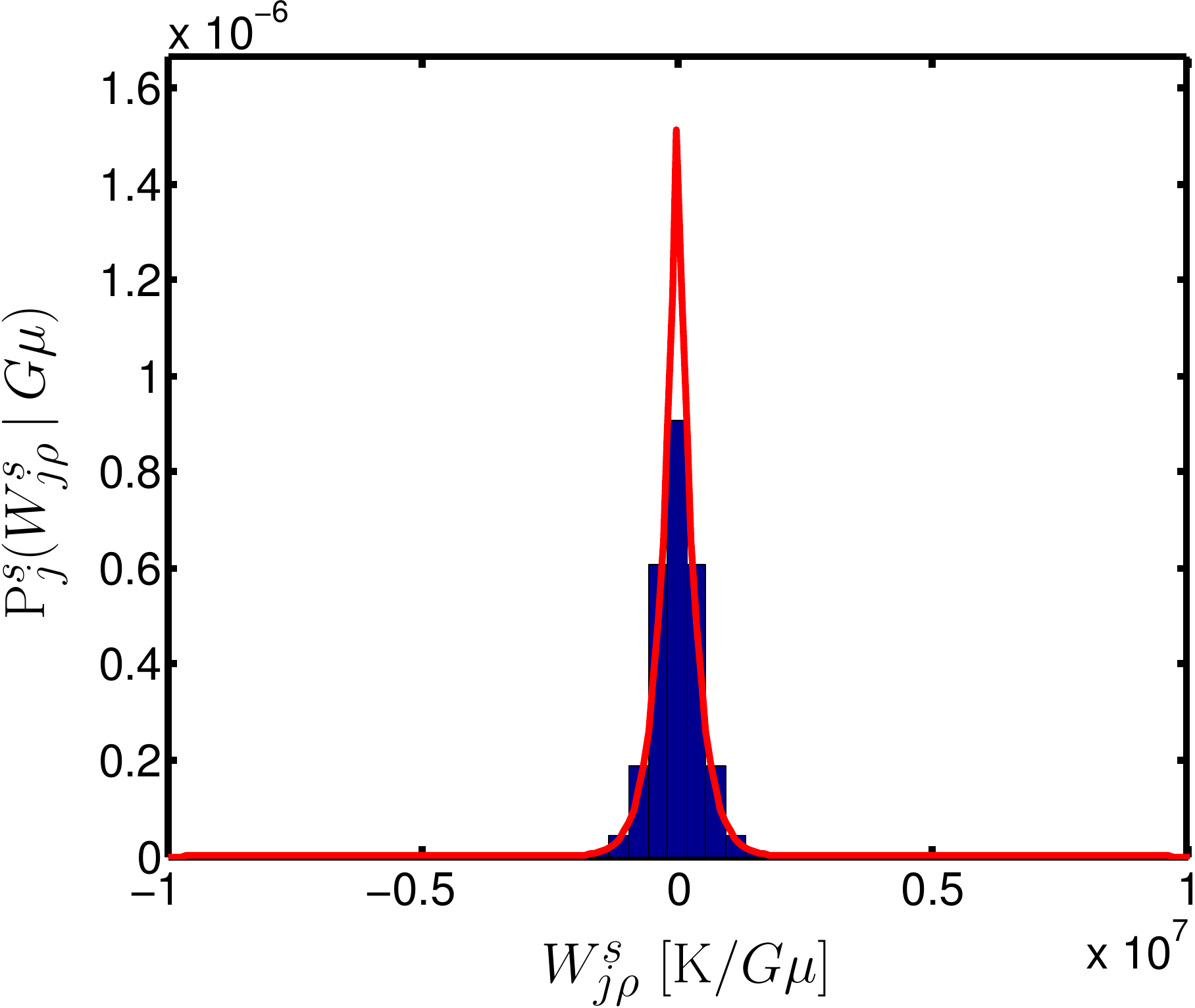}}\hfill
  \subfigure[$\wscale = 2$]{\includegraphics[width=.24\textwidth]{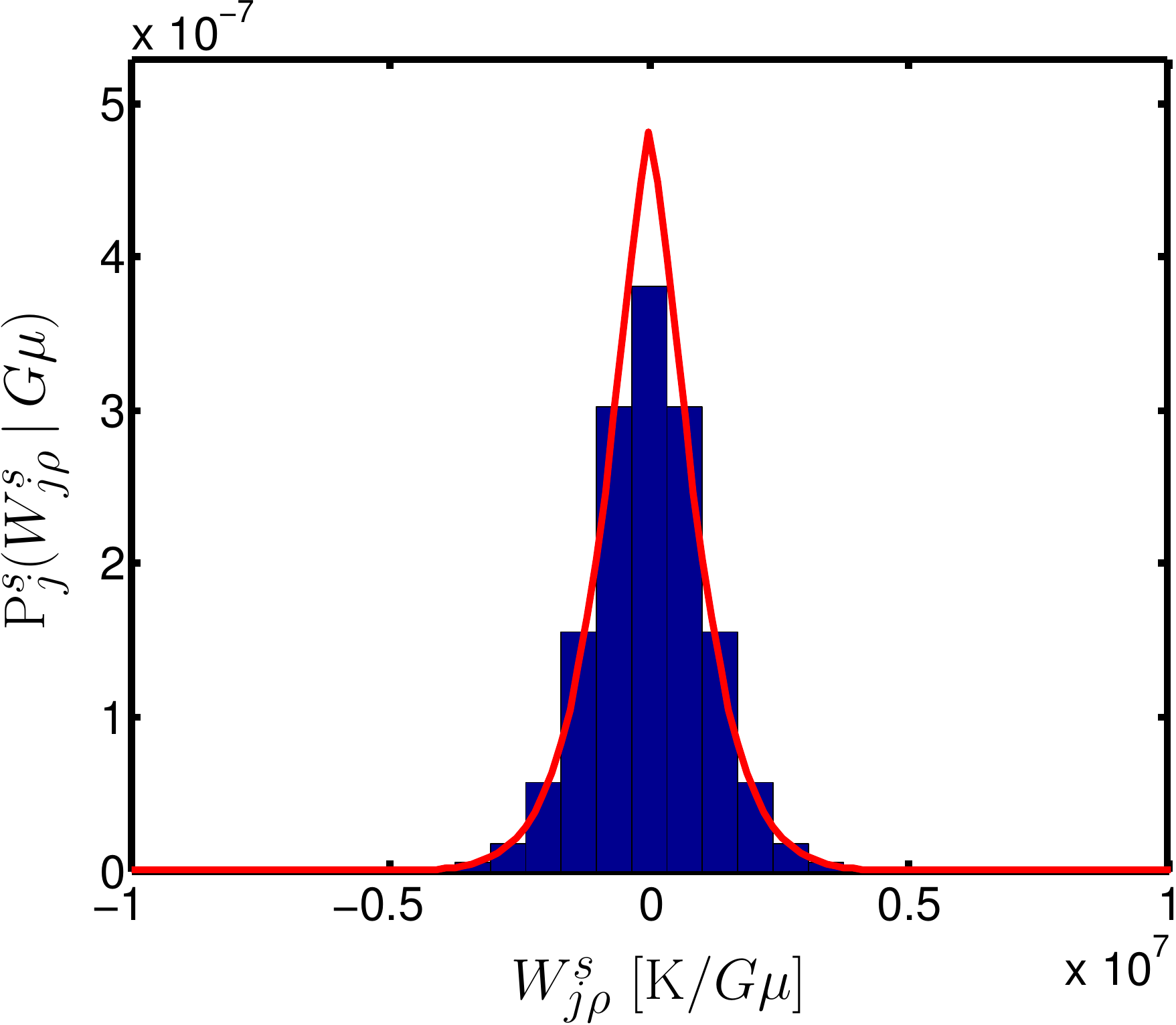}}\hfill
  \subfigure[$\wscale = 3$]{\includegraphics[width=.24\textwidth]{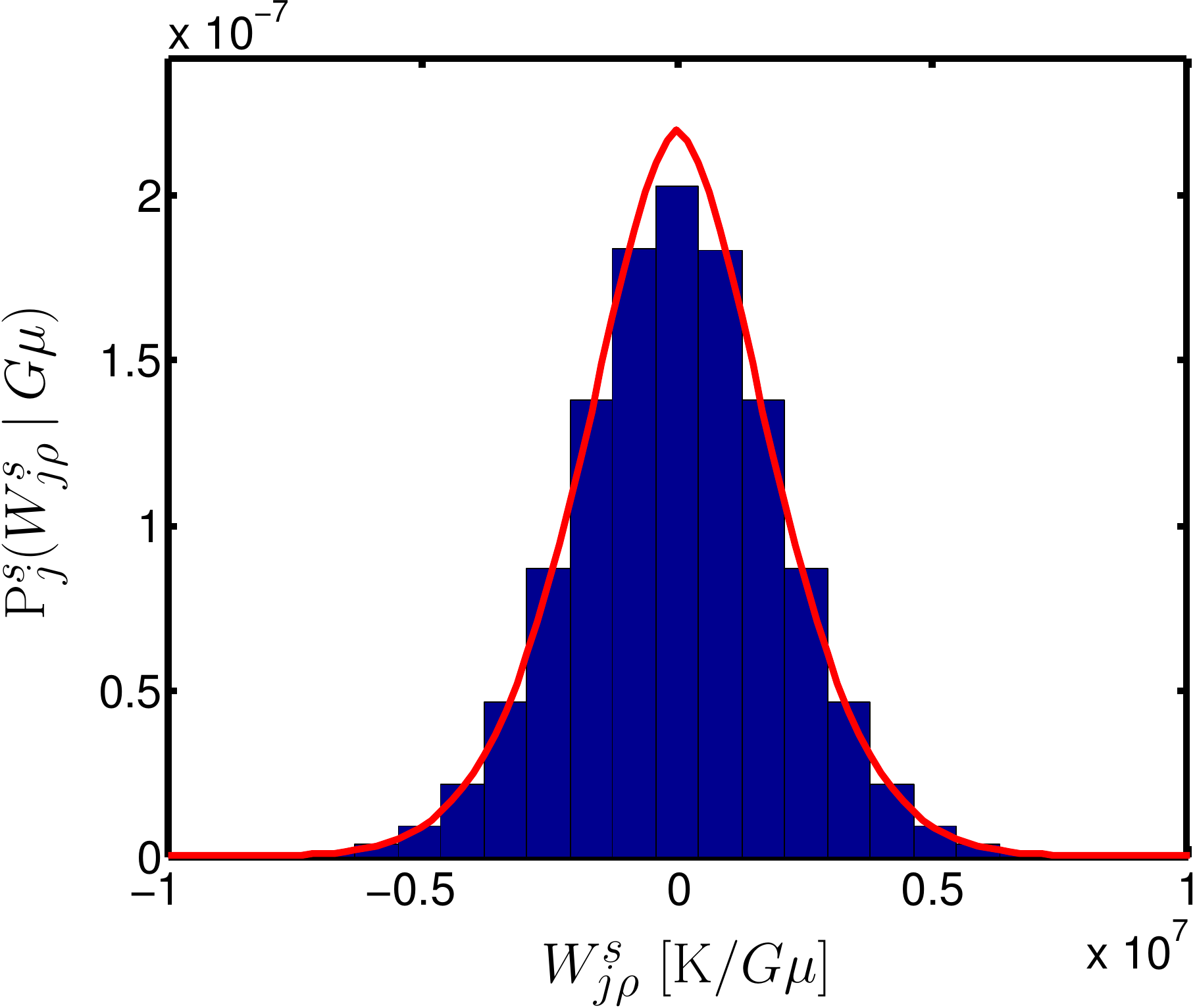}} \\
  \subfigure[$\wscale = 4$]{\includegraphics[width=.24\textwidth]{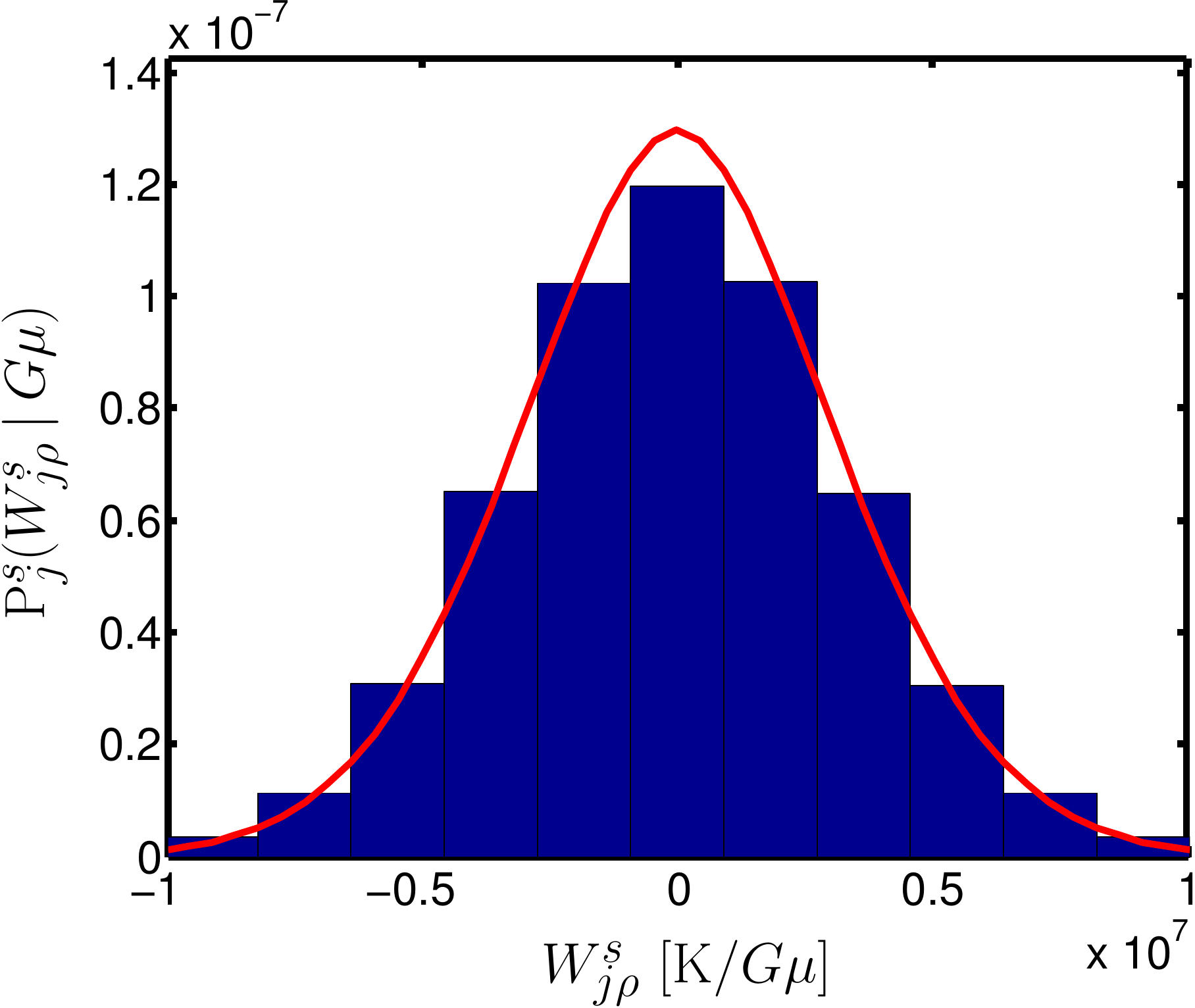}}\hfill
  \subfigure[$\wscale = 5$]{\includegraphics[width=.24\textwidth]{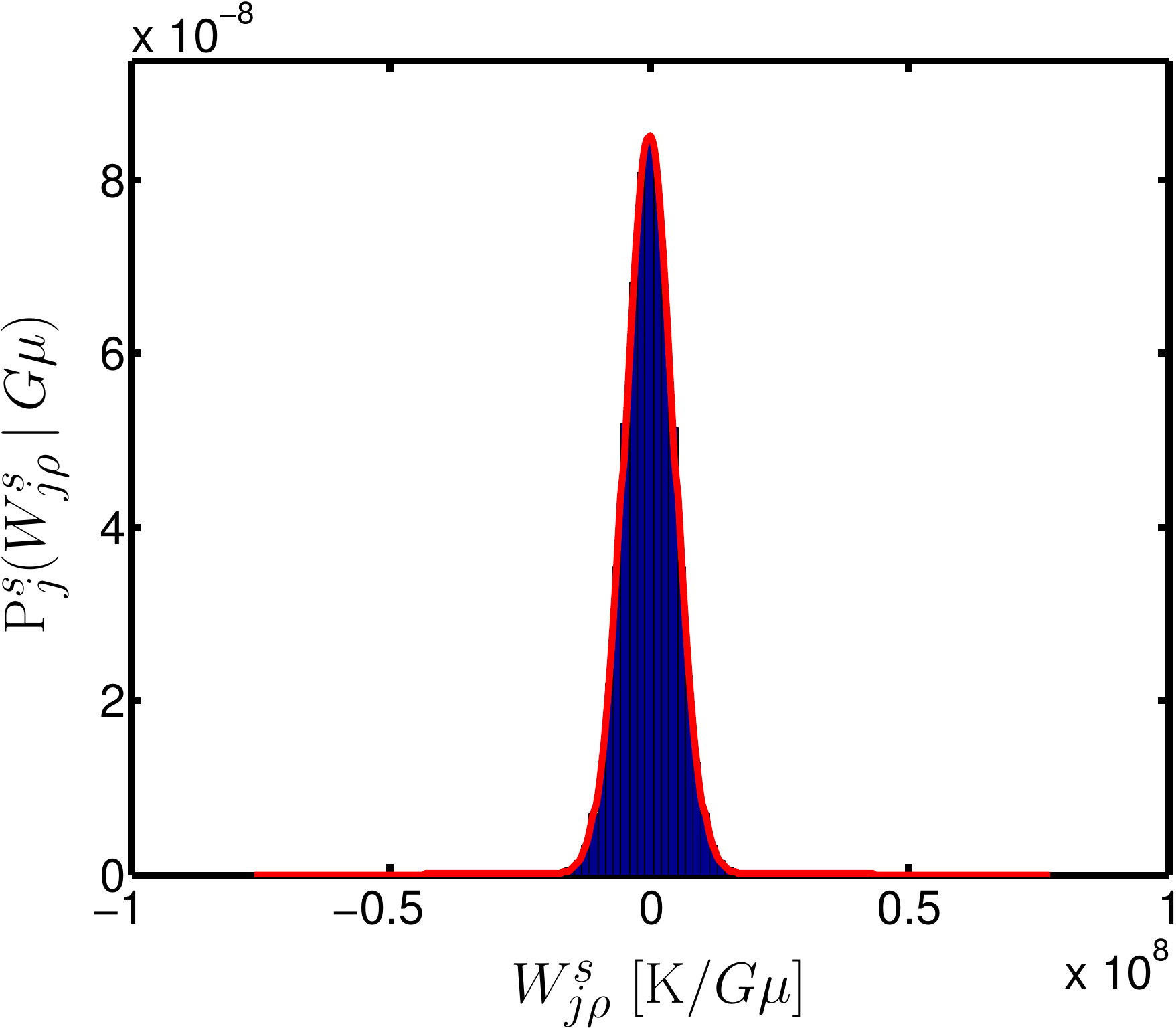}}\hfill
  \subfigure[$\wscale = 6$]{\includegraphics[width=.24\textwidth]{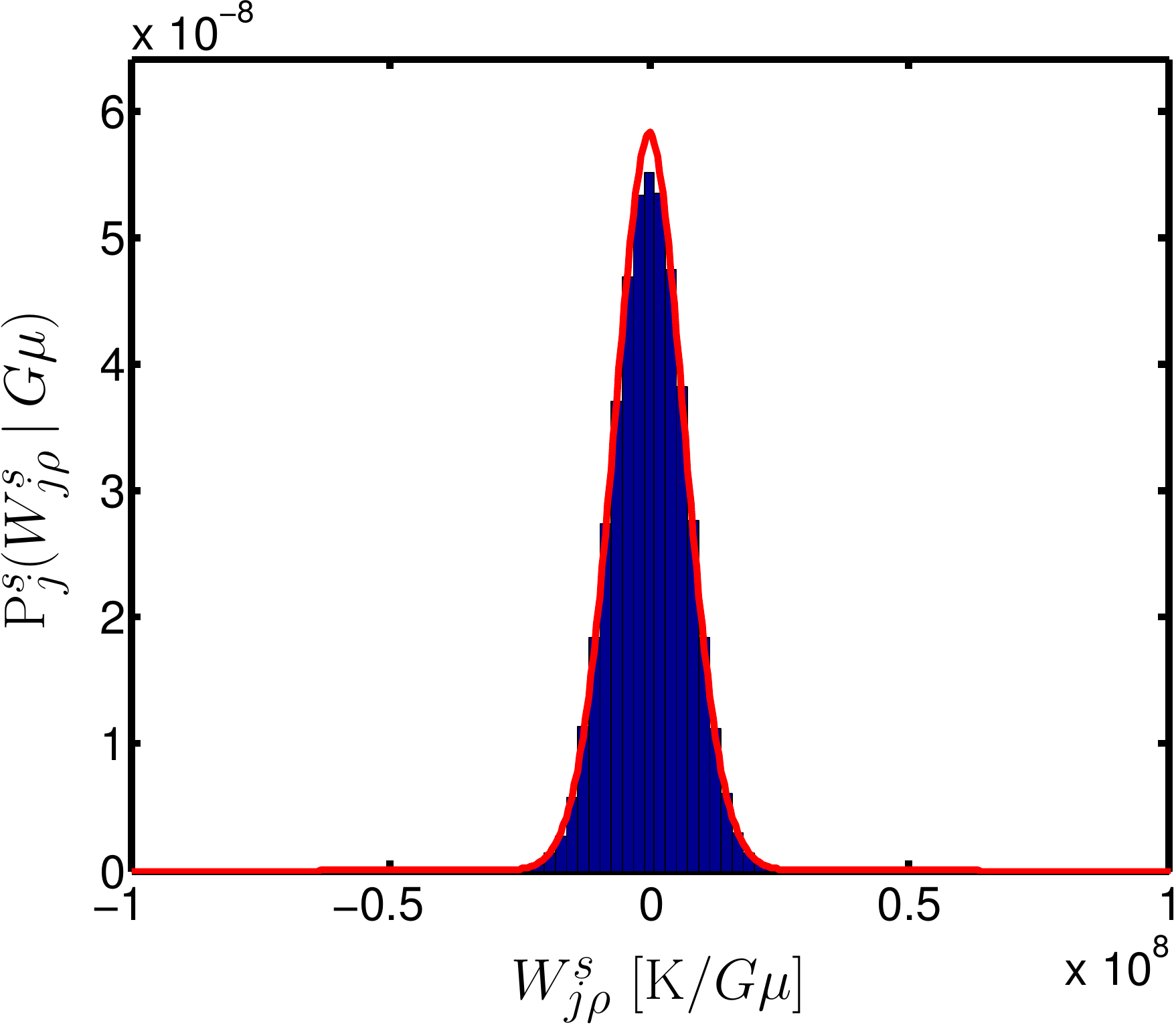}}\hfill
  \subfigure[$\wscale = 7$]{\includegraphics[width=.24\textwidth]{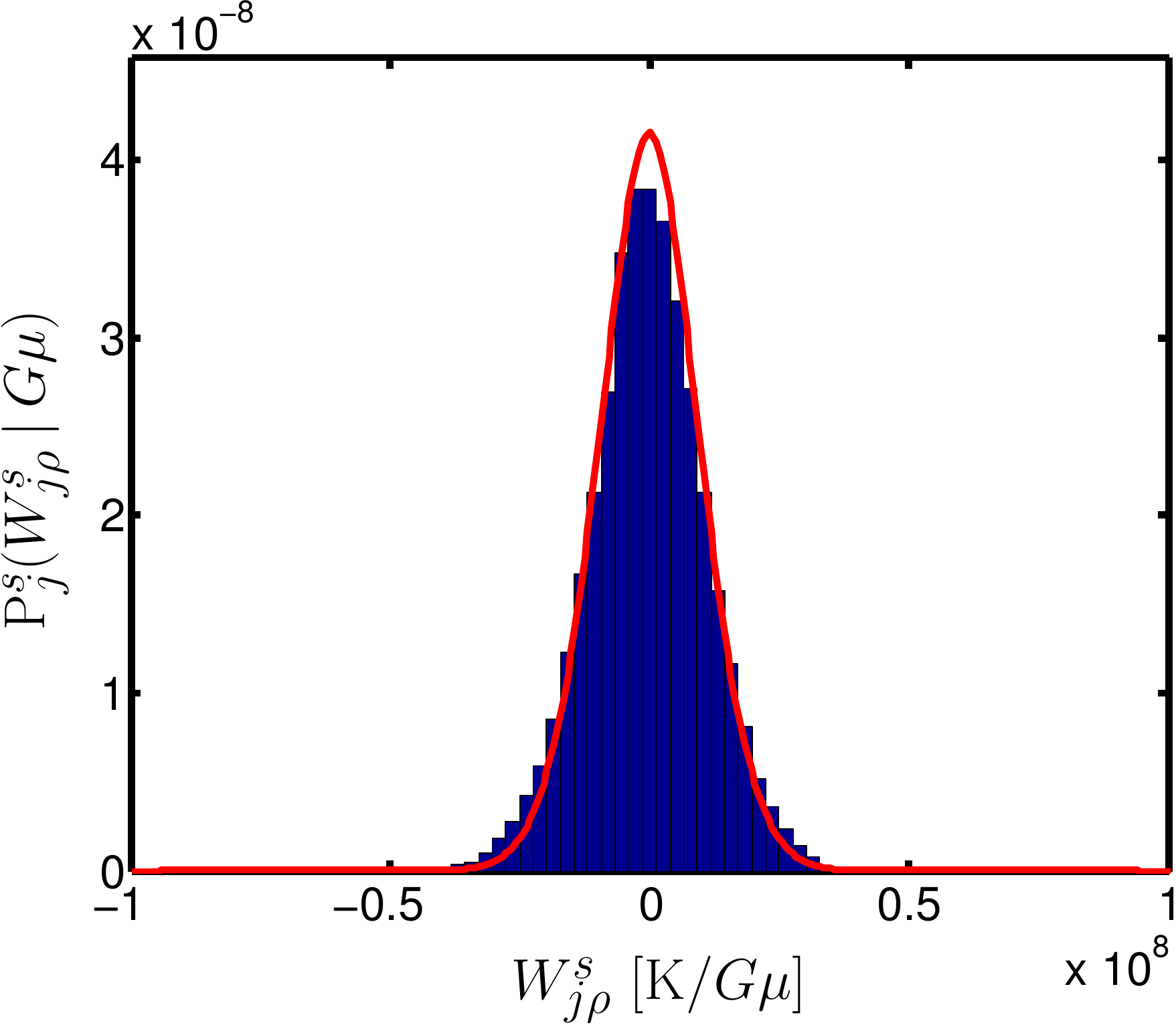}} \\
  \subfigure[$\wscale = 8$]{\includegraphics[width=.24\textwidth]{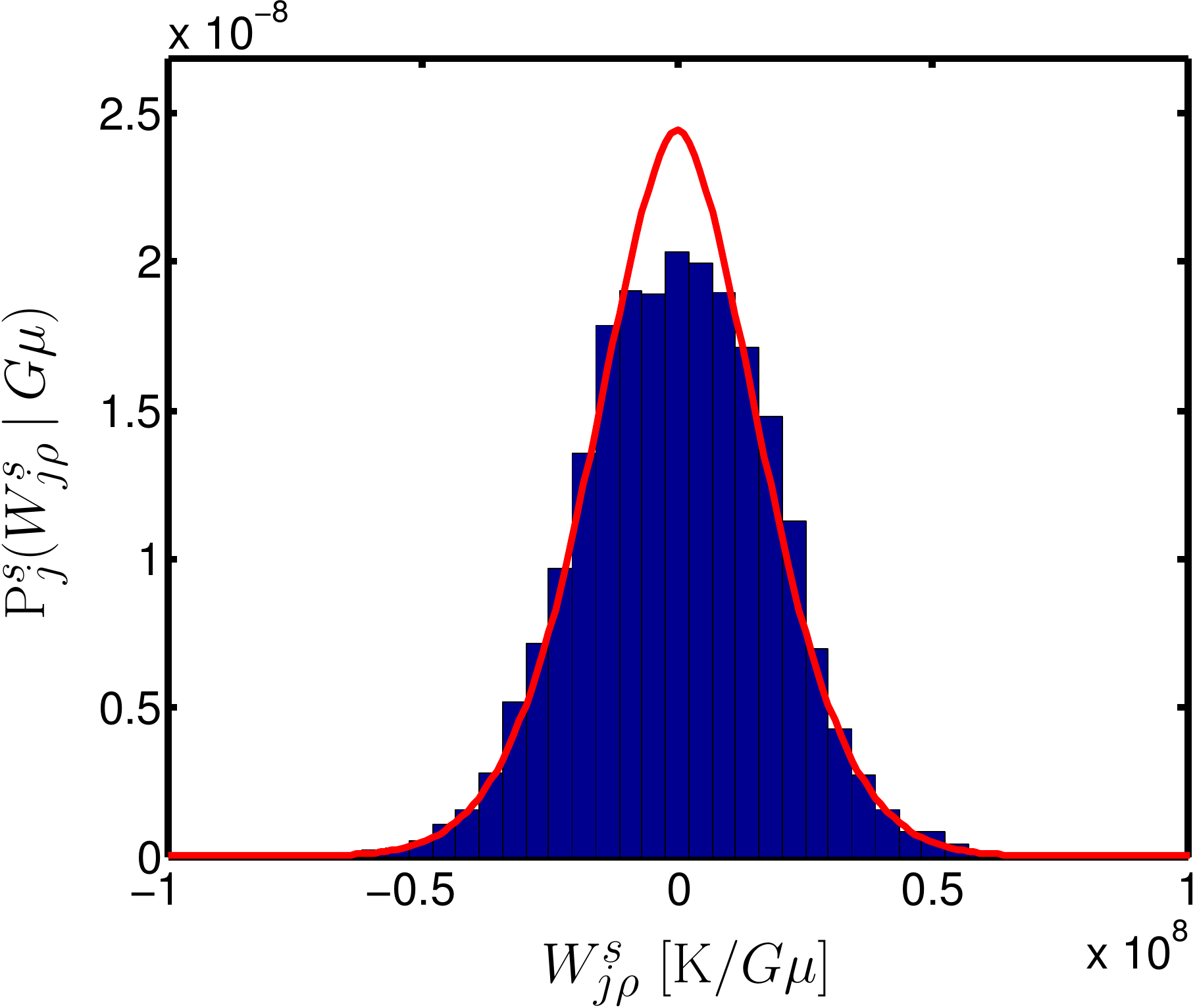}}\hfill
  \subfigure[$\wscale = 9$]{\includegraphics[width=.24\textwidth]{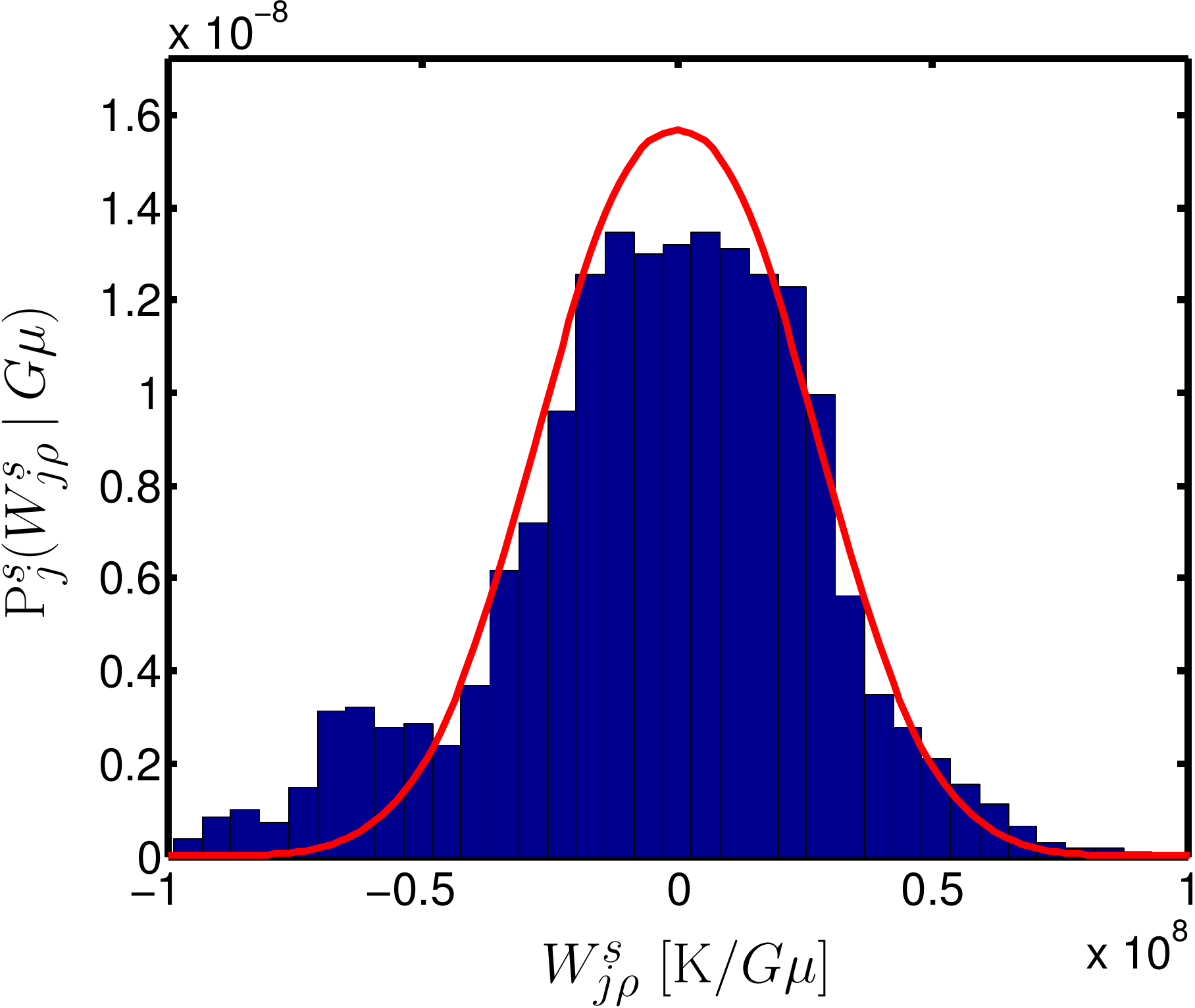}}\hfill
  \subfigure[$\wscale = 10$]{\includegraphics[width=.24\textwidth]{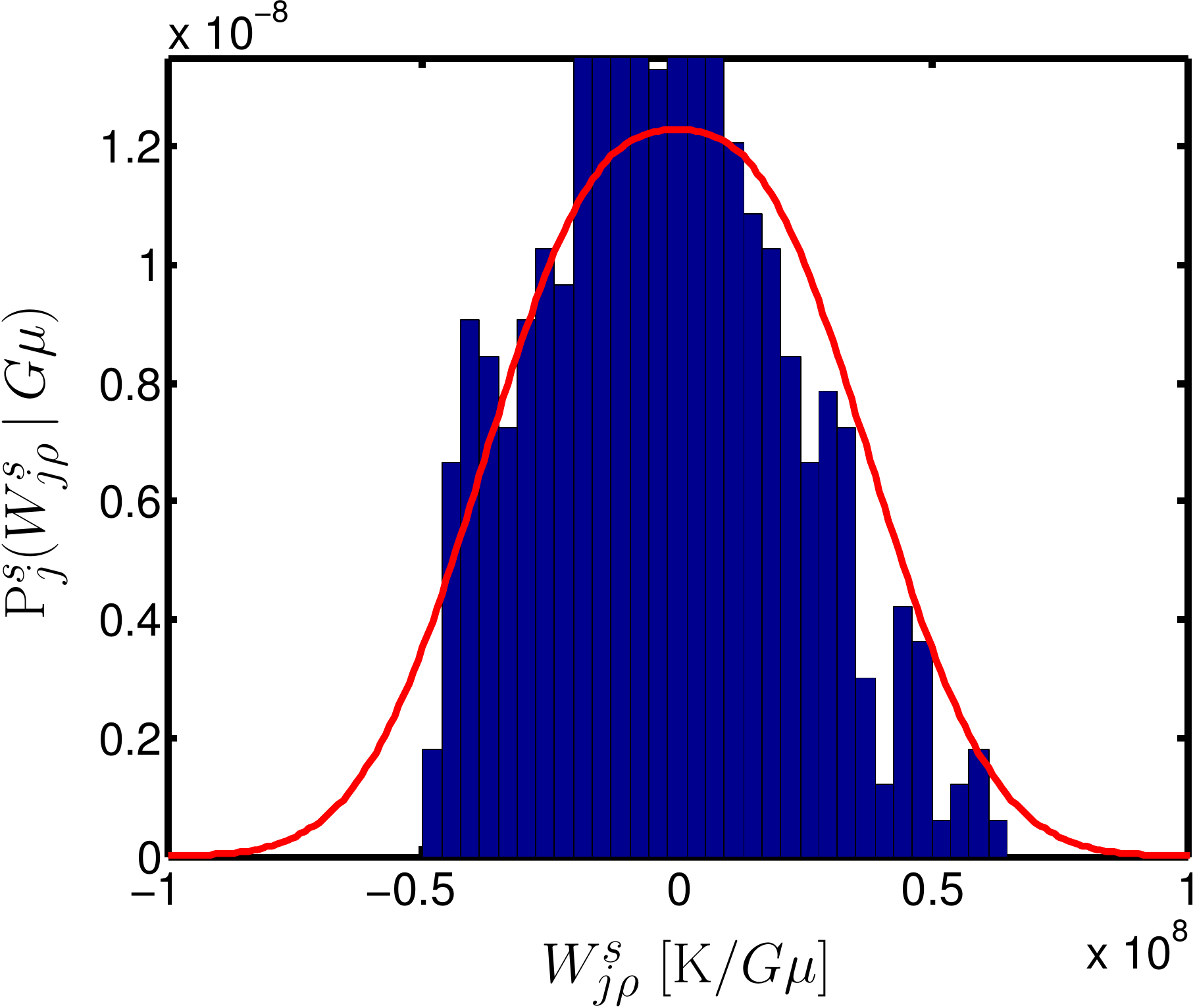}}\hfill
  \phantom{\includegraphics[width=.24\textwidth]{figures/psfrag/dist_j10_psfrag}}
  \caption{Distribution of the cosmic string-induced CMB component in wavelet space, for each wavelet scale $\wscale$ (for parameters $\elmax=2048$, $\nmax=4$, and $\dilparam=2$).  The GGD distribution fitted to the training map is shown by the solid red curve, while the raw distribution of the testing map is shown by the solid blue histogram. The fitted GGD distribution of the training map matches the histogram of the testing map well for small scales (low \wscale), indicating that the learnt GGD accurately models the general statistical properties of cosmic string included CMB maps.  As the scale becomes larger (higher \wscale) the match becomes less accurate due to cosmic variance.  The distributions are also highly leptokurtic for small scales (low \wscale), indicating that the string map is indeed sparse in wavelet space, as expected. As the scale increases the distribution becomes less leptokurtic, also as expected. For these reasons we consider wavelet coefficients up to and including scale $\wscale=7$ only in the subsequent analysis.}
  \label{fig:string_distributions_linear}
\end{figure*}

\begin{figure*}
  \centering
  \subfigure[$\wscale = 0$]{\includegraphics[width=.24\textwidth]{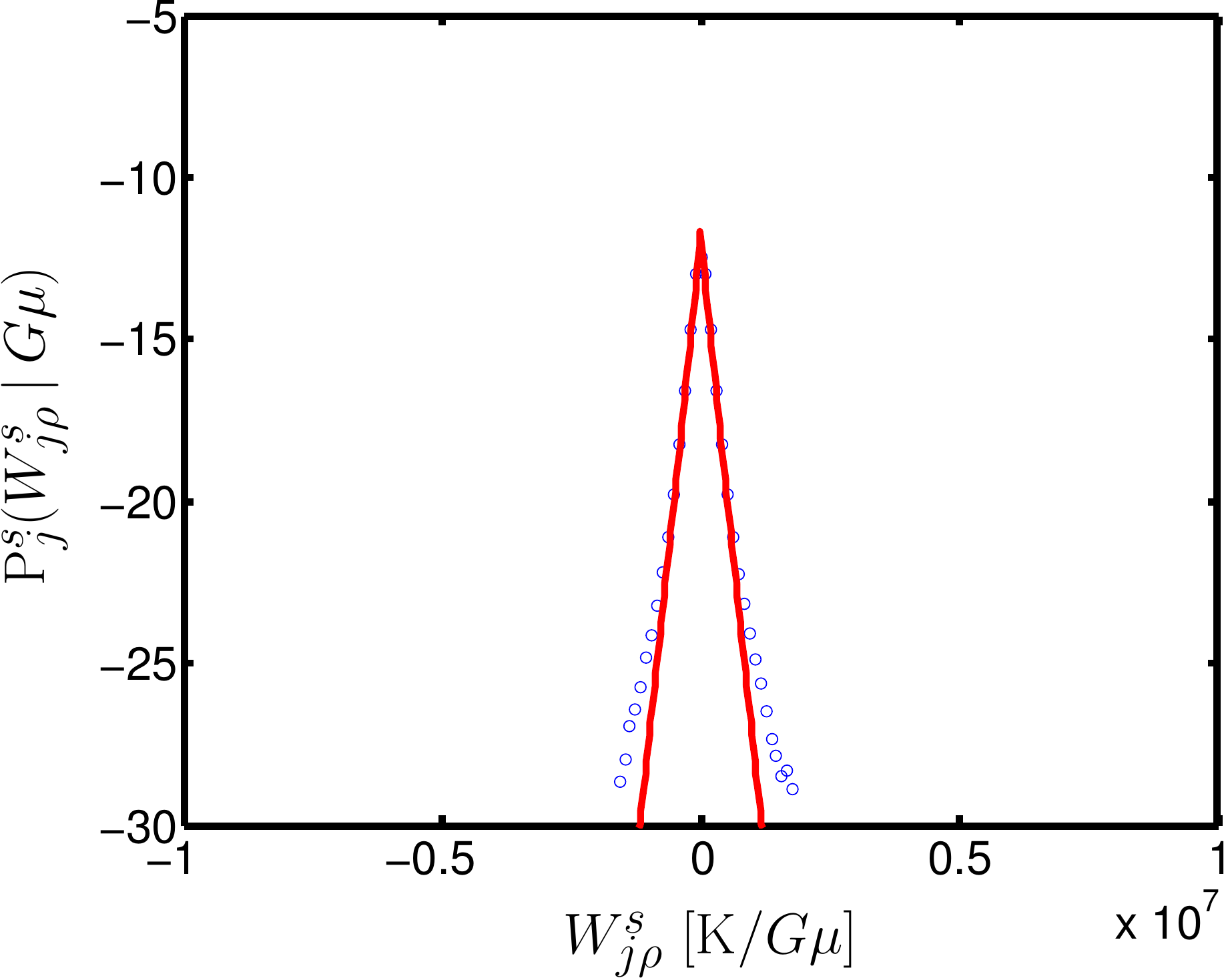}}\hfill
  \subfigure[$\wscale = 1$]{\includegraphics[width=.24\textwidth]{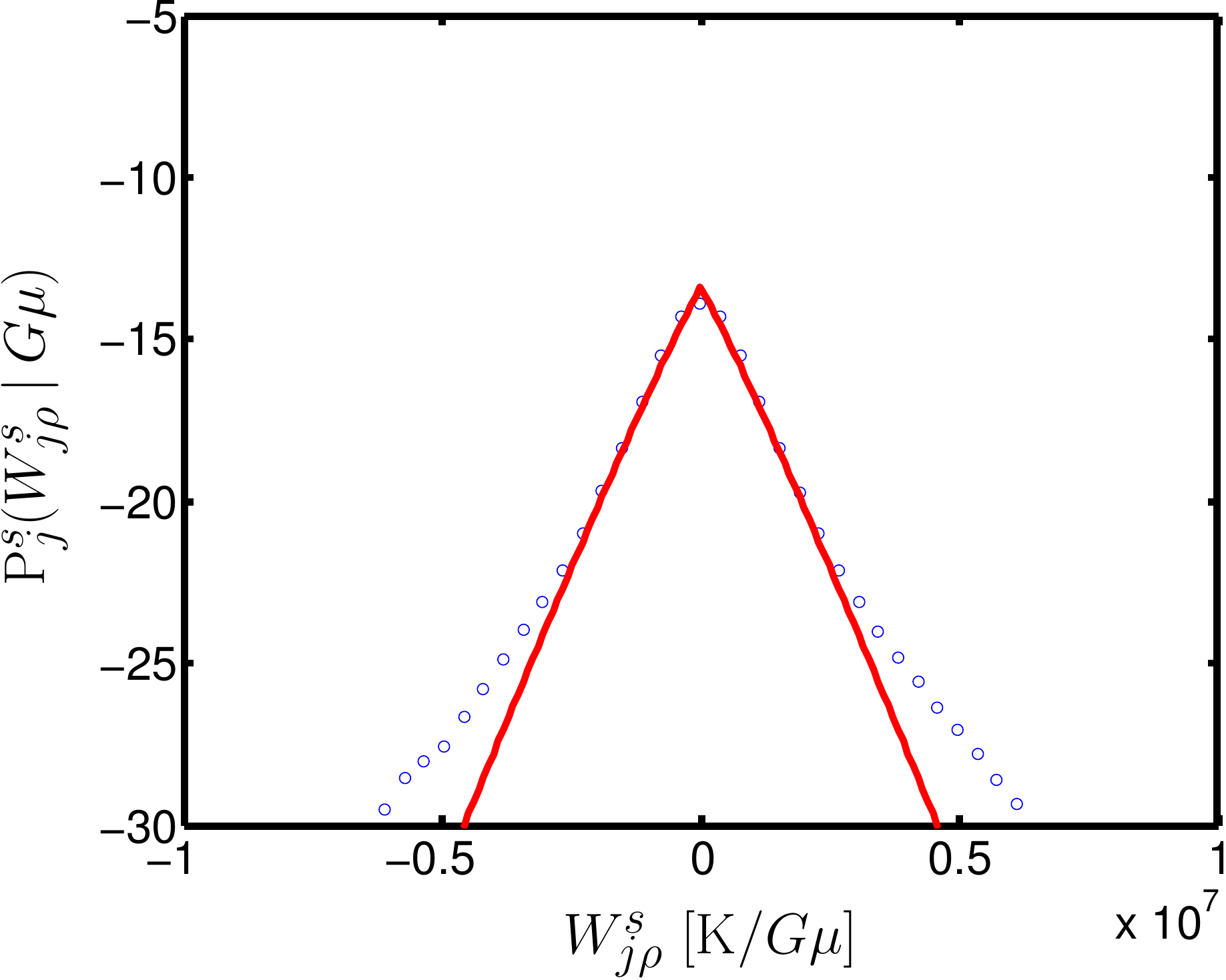}}\hfill
  \subfigure[$\wscale = 2$]{\includegraphics[width=.24\textwidth]{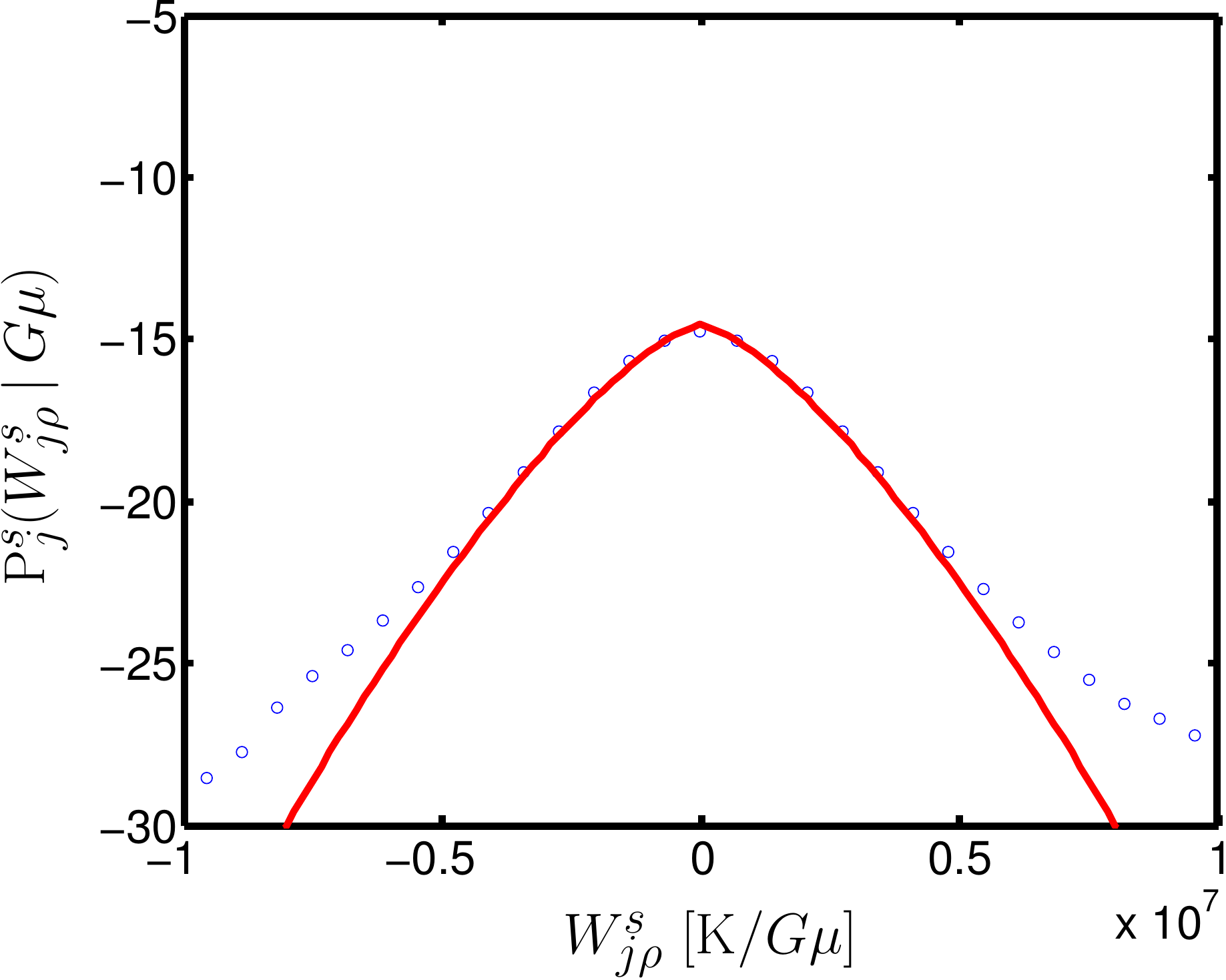}}\hfill
  \subfigure[$\wscale = 3$]{\includegraphics[width=.24\textwidth]{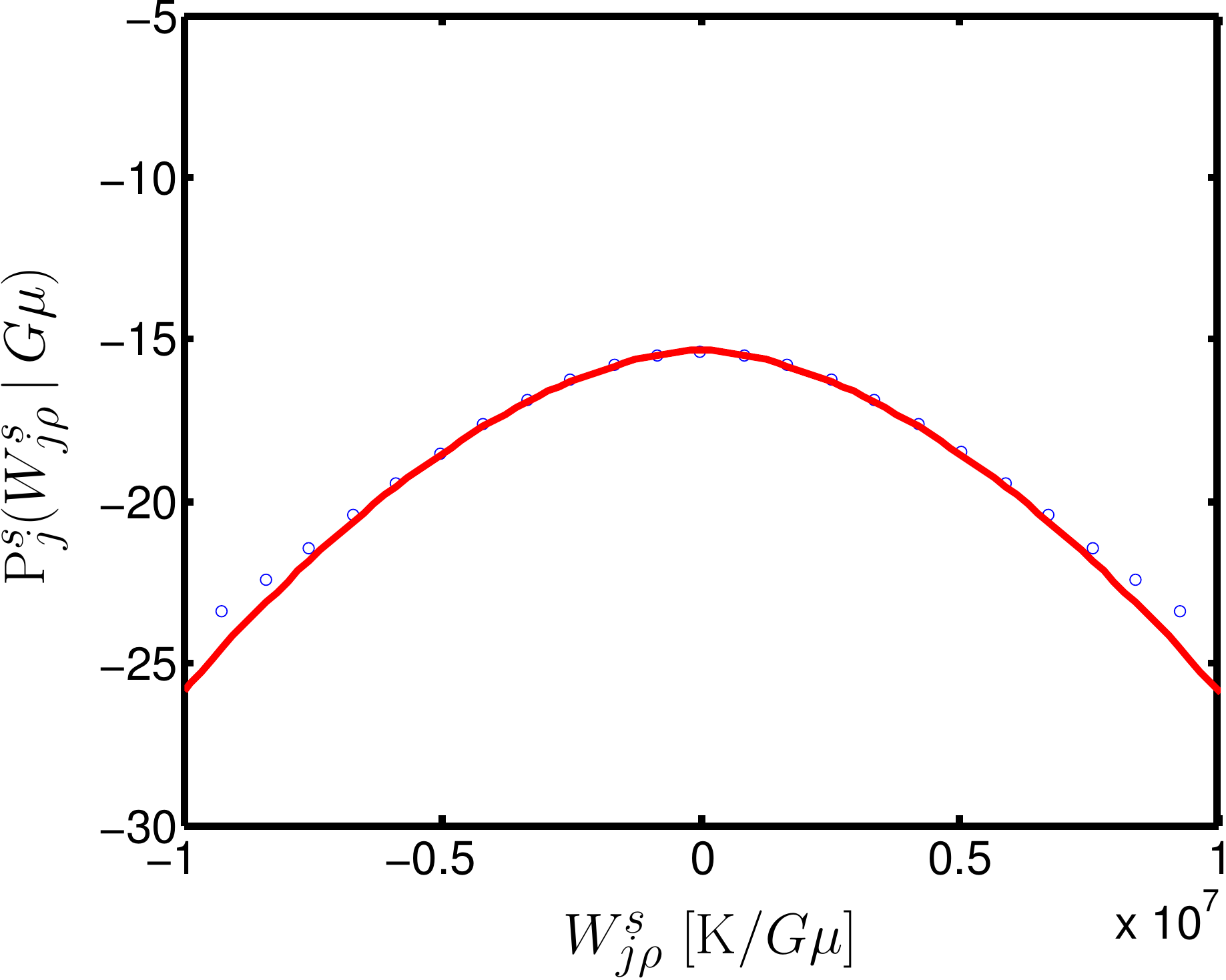}} \\
  \subfigure[$\wscale = 4$]{\includegraphics[width=.24\textwidth]{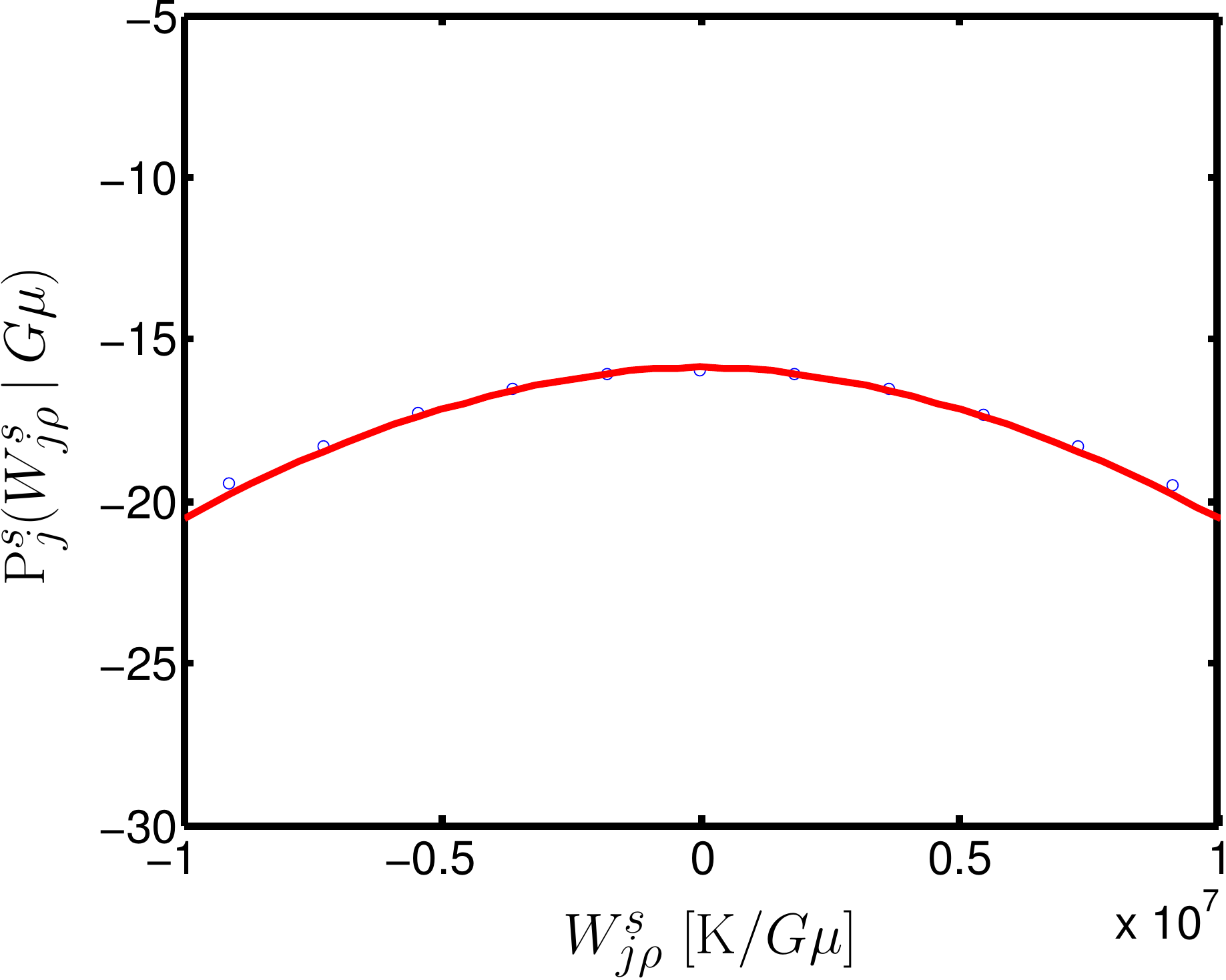}}\hfill
  \subfigure[$\wscale = 5$]{\includegraphics[width=.24\textwidth]{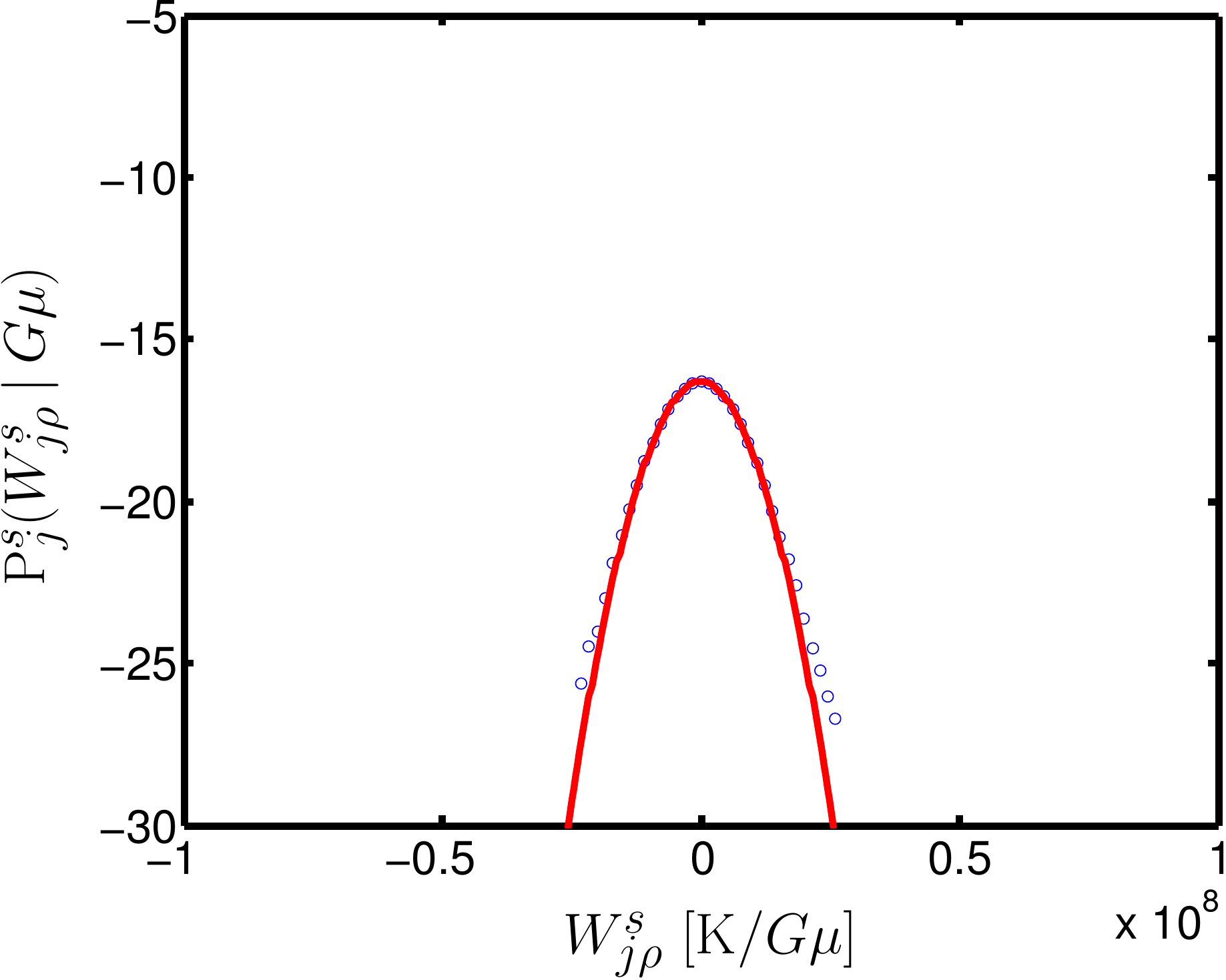}}\hfill
  \subfigure[$\wscale = 6$]{\includegraphics[width=.24\textwidth]{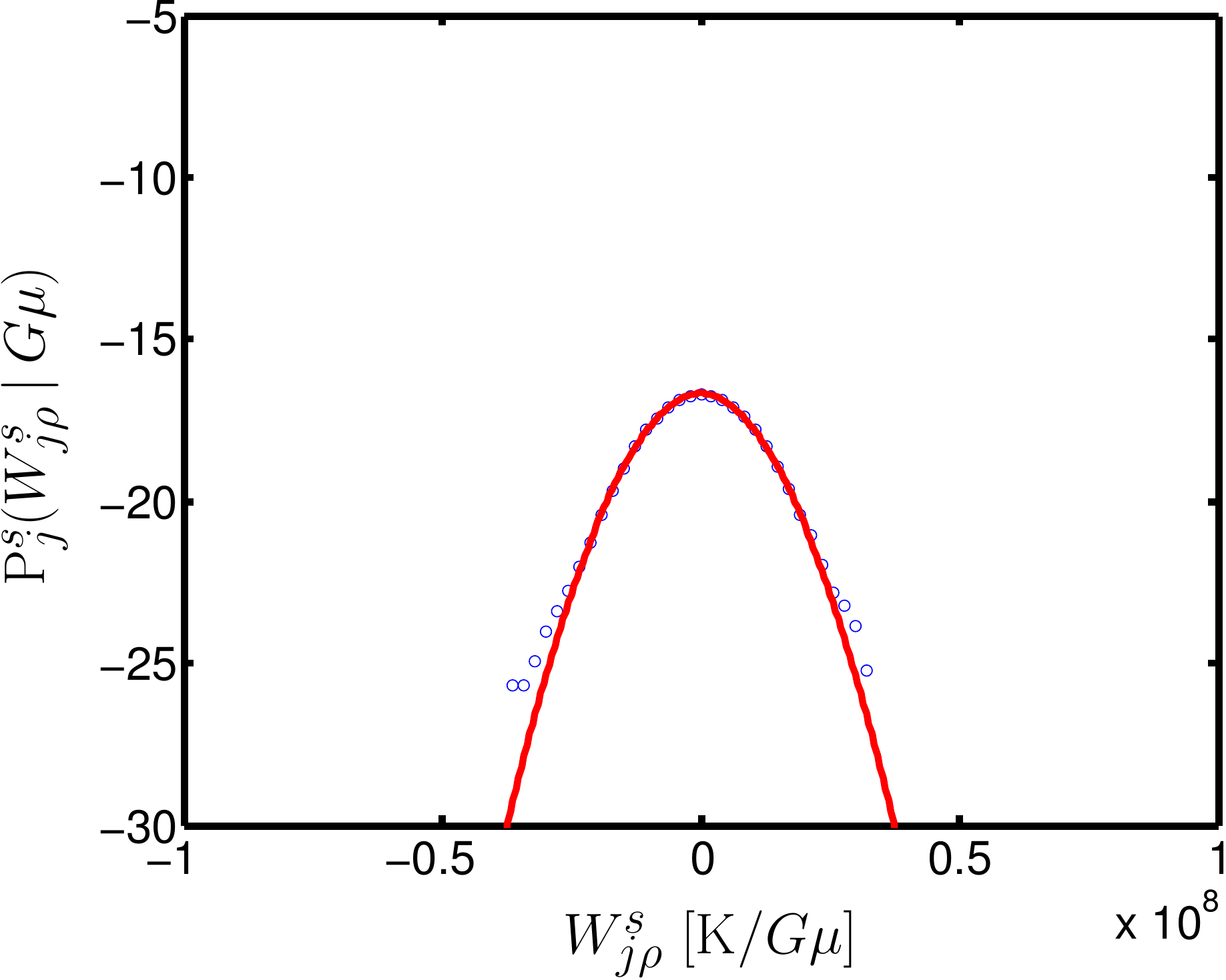}}\hfill
  \subfigure[$\wscale = 7$]{\includegraphics[width=.24\textwidth]{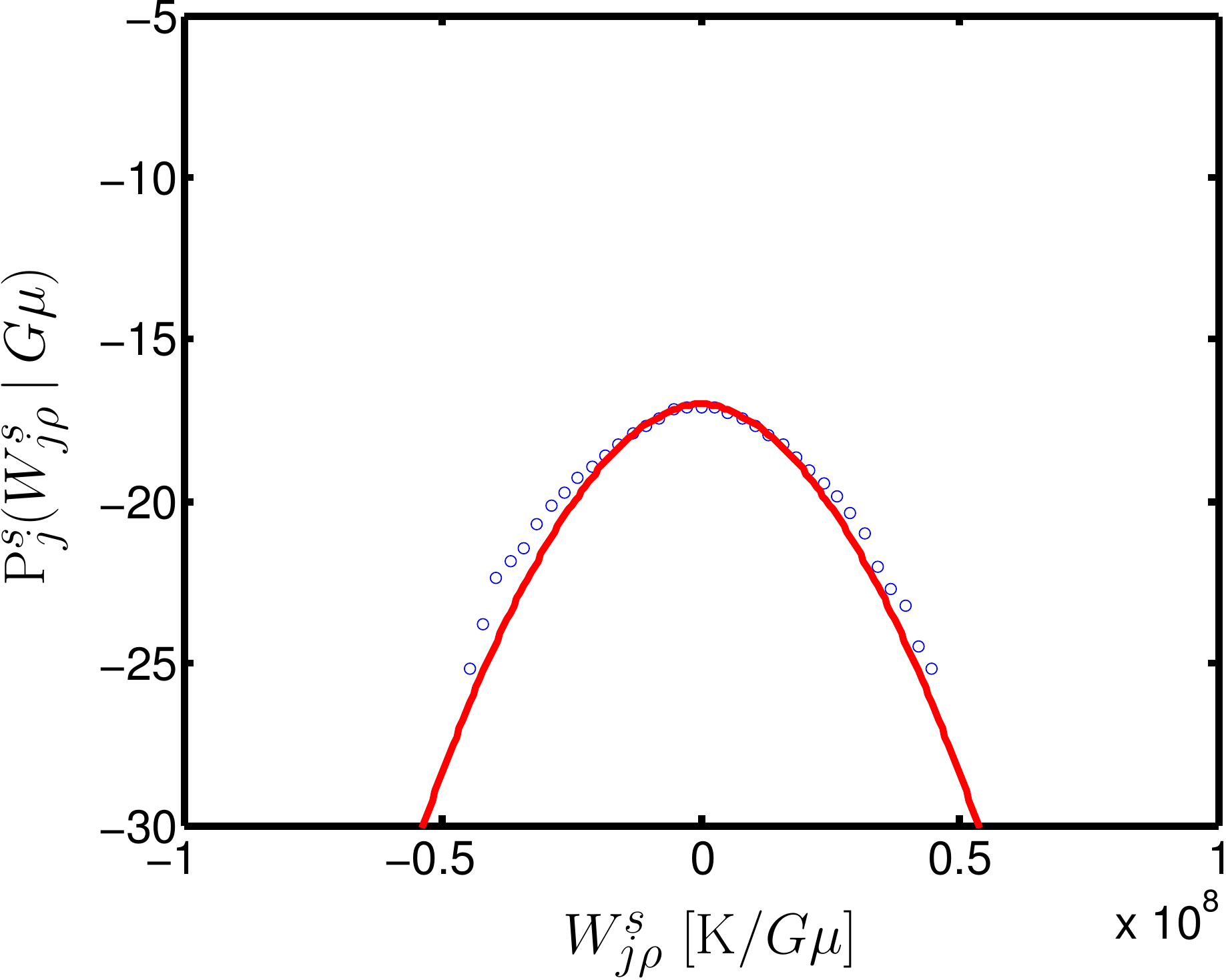}} \\
  \subfigure[$\wscale = 8$]{\includegraphics[width=.24\textwidth]{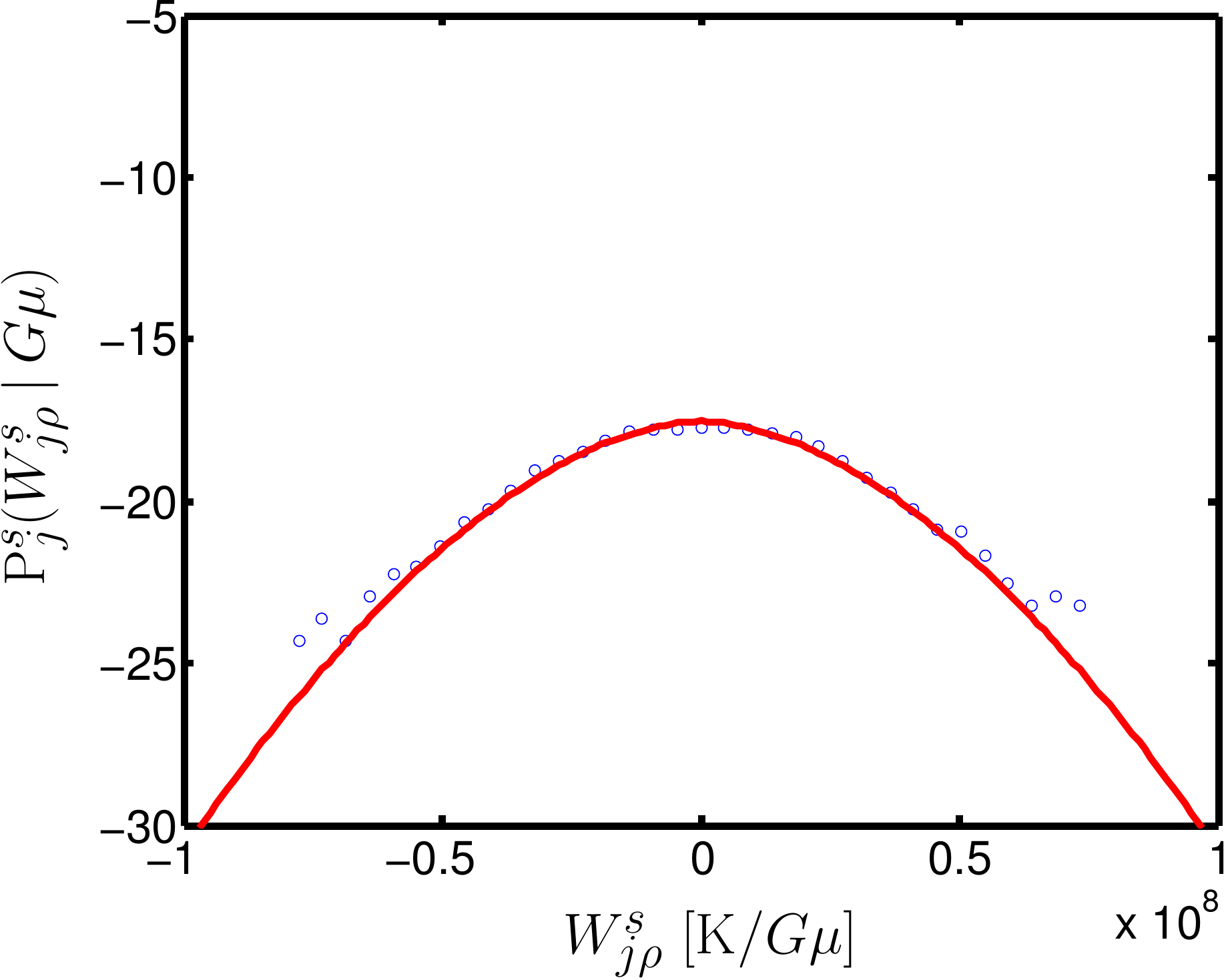}}\hfill
  \subfigure[$\wscale = 9$]{\includegraphics[width=.24\textwidth]{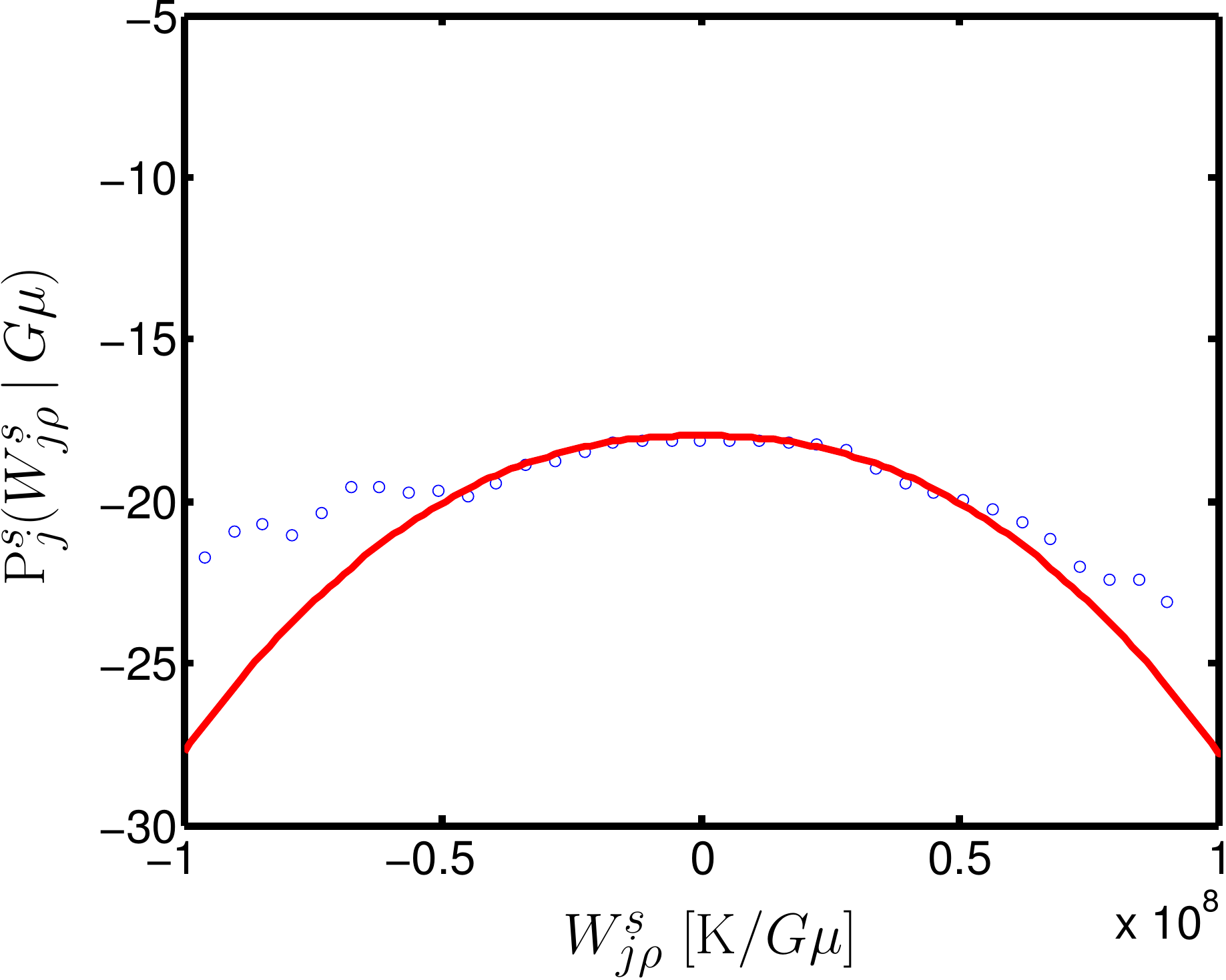}}\hfill
  \subfigure[$\wscale = 10$]{\includegraphics[width=.24\textwidth]{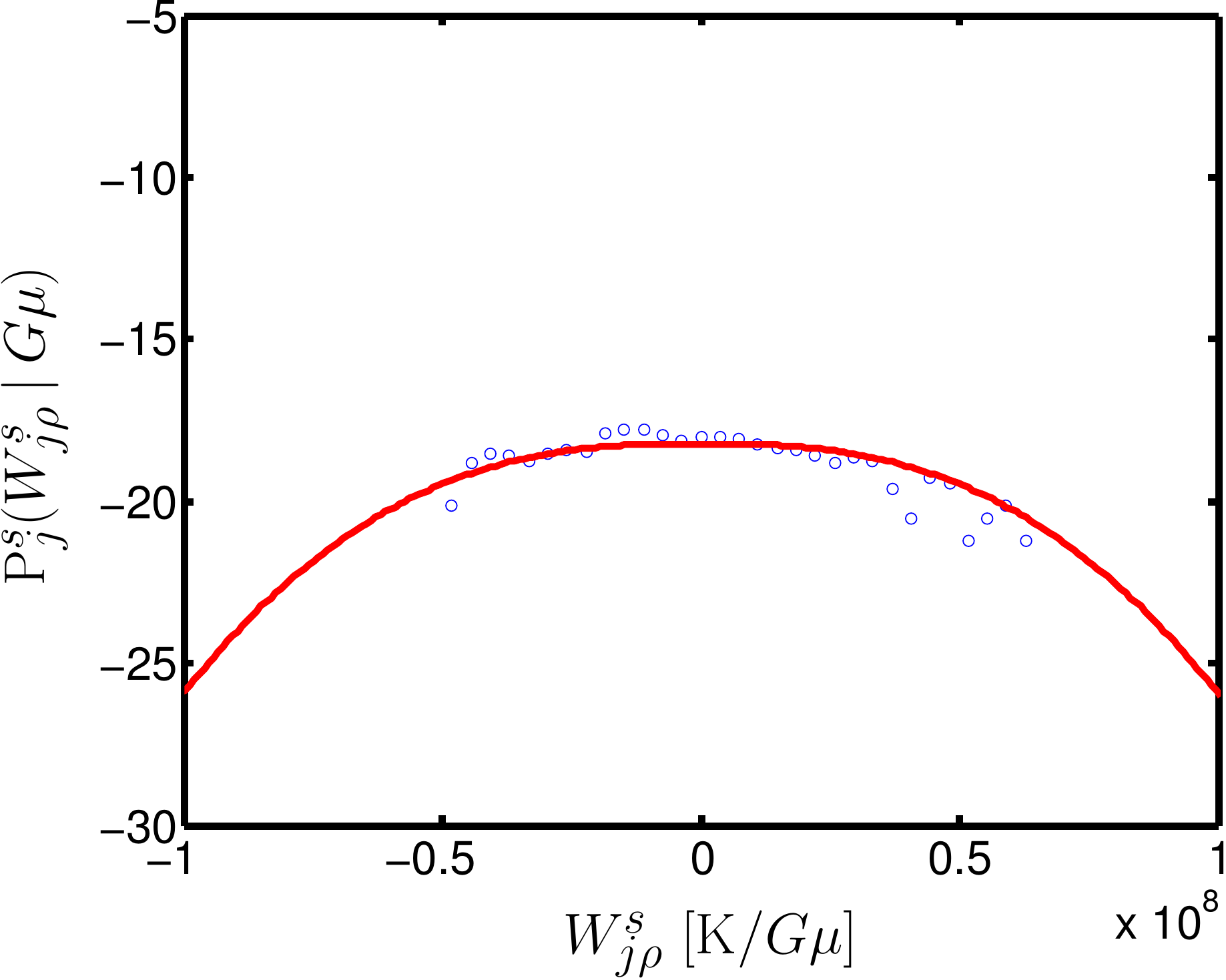}}\hfill
  \phantom{\includegraphics[width=.24\textwidth]{figures/psfrag/dist_j10_log_psfrag}}
  \caption{Same as \fig{\ref{fig:string_distributions_linear}} but plotted on a $\log_{10}$ scale.}
  \label{fig:string_distributions_log}
\end{figure*}

The distributions of the cosmic string maps are shown in \fig{\ref{fig:string_distributions_linear}} and \fig{\ref{fig:string_distributions_log}}, while the estimated GGD shape parameters are listed in \tbl{\ref{tbl:ggd_shape_parameters}}. The fitted GGD distribution of the training map matches the histogram of the testing map well for small scales (low \wscale), indicating that the learnt GGD accurately models the general statistical properties of cosmic string included CMB maps.  As the scale becomes larger (higher \wscale) the match becomes less accurate due to cosmic variance.
The distributions are also highly leptokurtic for small scales (low \wscale), \ie\ $\ggdshapew<2$, as apparent from the plots of the distributions (\fig{\ref{fig:string_distributions_linear}} and \fig{\ref{fig:string_distributions_log}}) and the fitted GGD shape parameters listed in \tbl{\ref{tbl:ggd_shape_parameters}}: the string map is indeed sparse in wavelet space, as expected.
As the scale increases the distribution becomes less leptokurtic, also as expected.
We therefore consider wavelet coefficients up to and including scale $\wscale=7$ only in the subsequent analysis, \ie\ we set $\wscalemax=7$.

While we focus on inference in the current article, as an aside we note that once we have learnt the statistical properties of string maps, we can use the learnt distribution to simulate realisations of string maps for very low computational cost.  However, in the current approach to training we do not learn the full covariance properties of the string components in wavelet space.  We leave the development of a computationally efficient approach to simulating high-resolution, full-sky cosmic string-induced CMB maps to future work.

\begin{table}
  \centering
  \caption{GGD shape parameter \ggdshapew\ fitted to the testing string map.  As expected the fitted GGD distributions are highly leptokurtic for small scales (low \wscale), with GGD shape parameter $\ggdshapew<2$, due to the sparse representation of the string-induced CMB component in wavelet space.}
  \label{tbl:ggd_shape_parameters}
  \begin{tabular}{cc}\toprule
       Wavelet scale $\wscale$ & GGD shape \ggdshapew \\ \midrule
       0   & 0.94  \\
       1   & 1.08  \\
       2   & 1.40  \\
       3   & 1.69  \\
       4   & 1.68  \\
       5   & 1.79  \\
       6   & 1.91  \\
       7   & 1.76  \\
       8   & 1.84  \\
       9   & 1.80  \\
      10   & 2.57  \\
  \bottomrule
  \end{tabular}
\end{table}

\subsection{String tension estimation}
\label{sec:inference:gmu_estimation}

In this section, we derive the posterior distribution for the string tension $\gmu$ under the string model $\modelstring$. By Bayes theorem the string tension posterior $\prob(\gmu \given \wcoeffdata)$ is related to the likelihood $\prob(\wcoeffdata \given \gmu)$ by
\begin{equation}
  \prob(\gmu \given \wcoeffdata) =
  \frac{\prob(\wcoeffdata \given \gmu) \: \prob(\gmu)}
  {\prob(\wcoeffdata)}
  \propto
  \prob(\wcoeffdata \given \gmu) \: \prob(\gmu)
  \spcend ,
\end{equation}
where $\prob(\gmu)$ is the prior distribution for the string tension.  For now we ignore the normalising denominator $\prob(\wcoeffdata)$ (the Bayesian evidence), which we return to in the following section.  Recall that $\wcoeffdata$ are the wavelet coefficients of the observed CMB data.

For each wavelet coefficient $\wcoeffdatap$ at scale $\wscale$ and position and orientation $\eul$ the likelihood can be calculated by
\begin{align}
  \prob(\wcoeffdatap \given \gmu)
  & =
  \prob(\wcoeffstringp + \wcoeffcmbnoisep \given \gmu) \\
  & =
  \int_{\reals}
  \dx \wcoeffstringp \:
  \prob^{\fcmbnoise}_\wscale(\wcoeffdatap - \wcoeffstringp) \:
  \prob^\fstring_\wscale(\wcoeffstringp \given \gmu)
  \spcend ,
  \label{eqn:lut_posterior}
\end{align}
where $\wcoeffcmbnoisep$ are the wavelet coefficients of the Gaussian component $\fcmbnoise$, which includes the inflationary CMB component and noise. The distributions comprising the integrand of \eqn{\ref{eqn:lut_posterior}} are precisely those described in \sectn{\ref{sec:inference:distributions}}, which we determine analytically or learn from a simulated string map.
To compute the overall likelihood of the data, for speed of processing we assume each wavelet coefficient is independent, in which case the overall likelihood reads:
\begin{equation}
  \prob(\wcoeffdata \given \gmu)
  =
  \prod_{\wscale,\wposn} \prob(\wcoeffdatap \given \gmu)
  \label{eqn:likelihood_full}
  \spcend .
\end{equation}
For numerical purposes we compute the log-likelihood, given by
\begin{equation}
  \ln \prob(\wcoeffdata \given \gmu)
  =
  \sum_{\wscale,\wposn} \ln \prob(\wcoeffdata_{\wscaleposn} \given \gmu)
  \label{eqn:likelihood_full_log}
  \spcend .
\end{equation}

The assumption of independence of wavelet coefficients is approximate. Nevertheless, the covariance of wavelet coefficients decays rapidly with spatial separation (relative to the spatial size of the wavelet considered) and is zero for non-adjacent scales (\ie\ for scales $\wscale$ and $\wscale\p$ such that $\vert \wscale - \wscale\p\vert \geq 2$).  We readdress the assumption of independence later and introduce measures to account for this approximation.

\begin{figure*}
  \centering
  \subfigure[$\wscale = 0$]{\includegraphics[width=.24\textwidth]{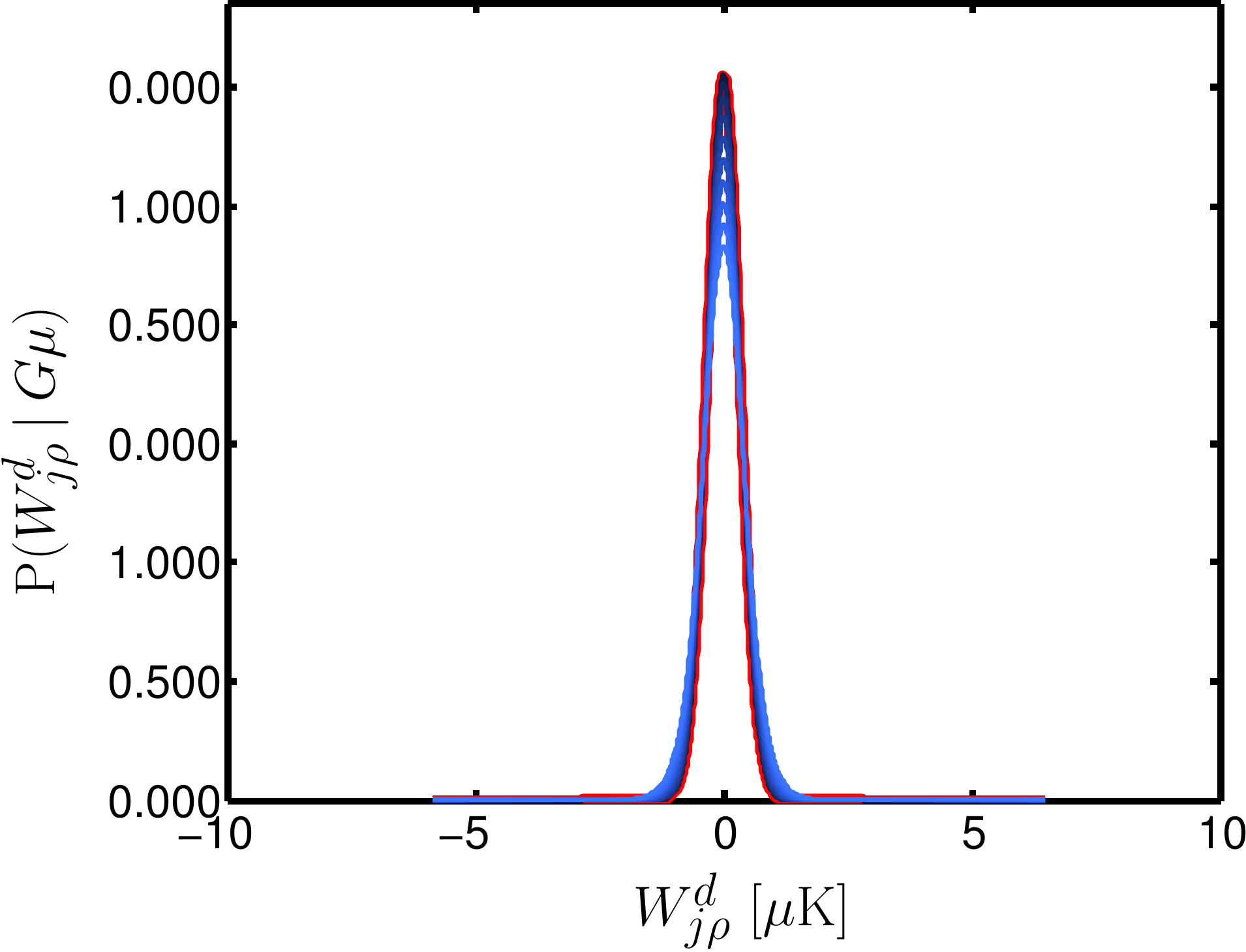}} \hfill
  \subfigure[$\wscale = 1$]{\includegraphics[width=.24\textwidth]{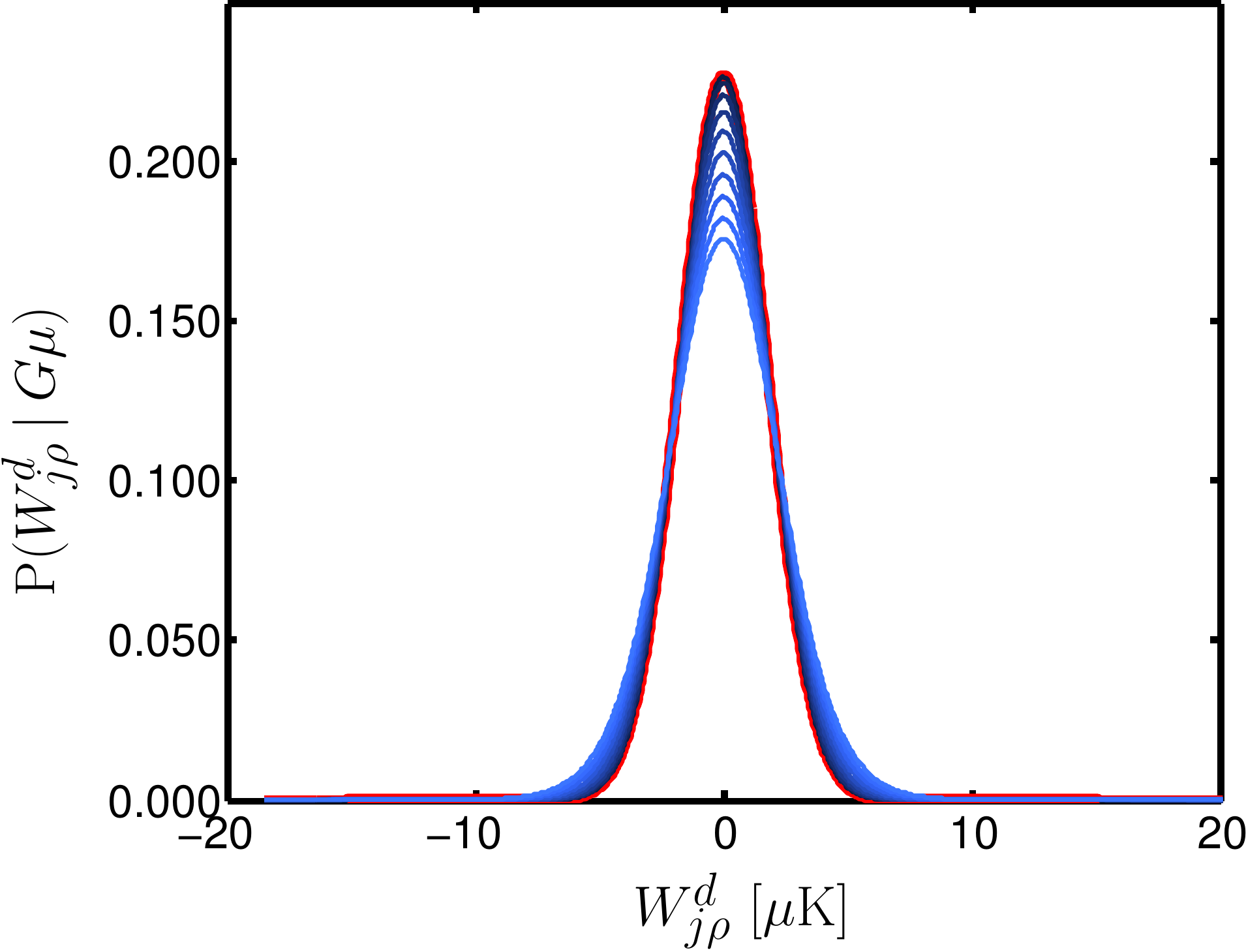}} \hfill
  \subfigure[$\wscale = 2$]{\includegraphics[width=.24\textwidth]{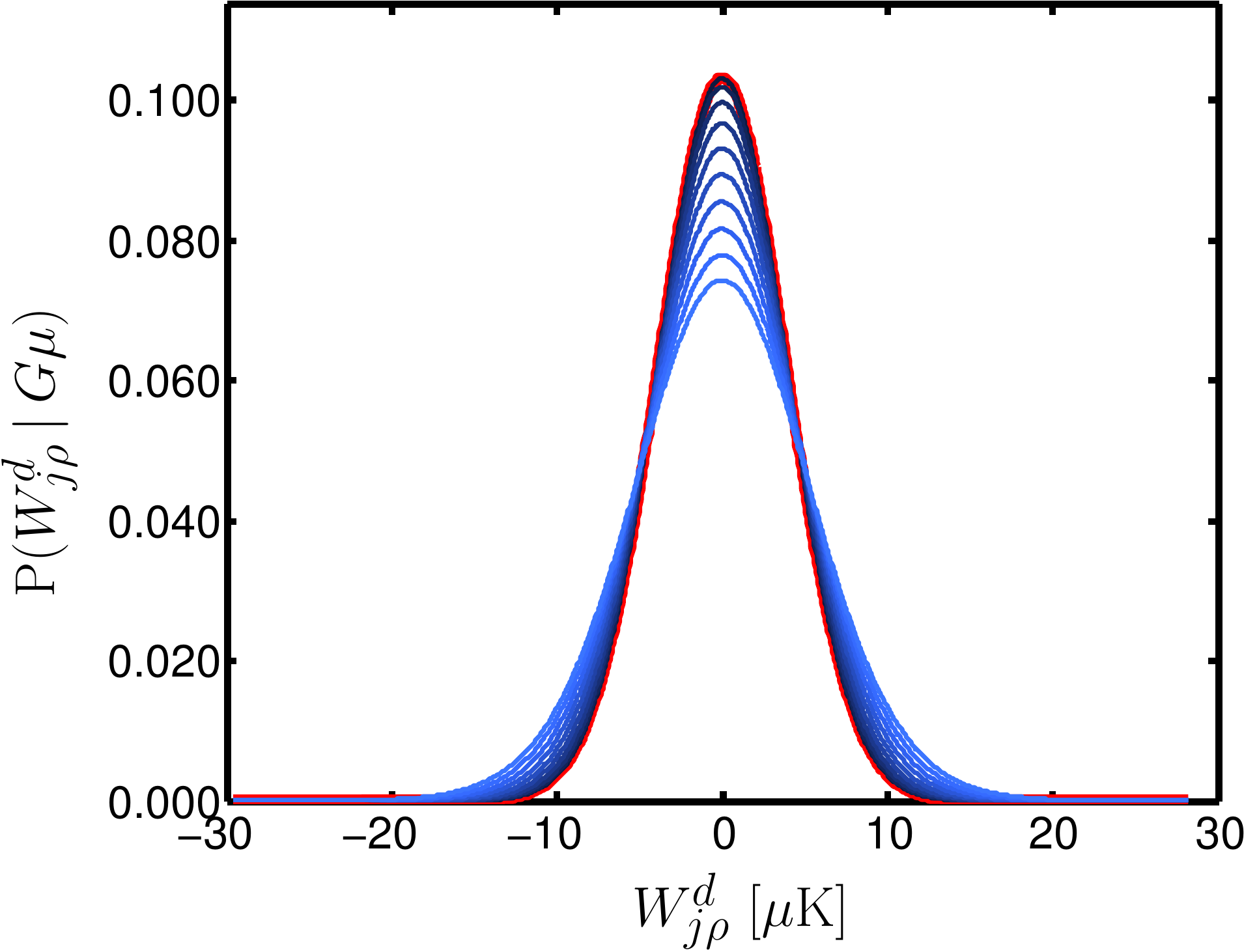}} \hfill
  \subfigure[$\wscale = 3$]{\includegraphics[width=.24\textwidth]{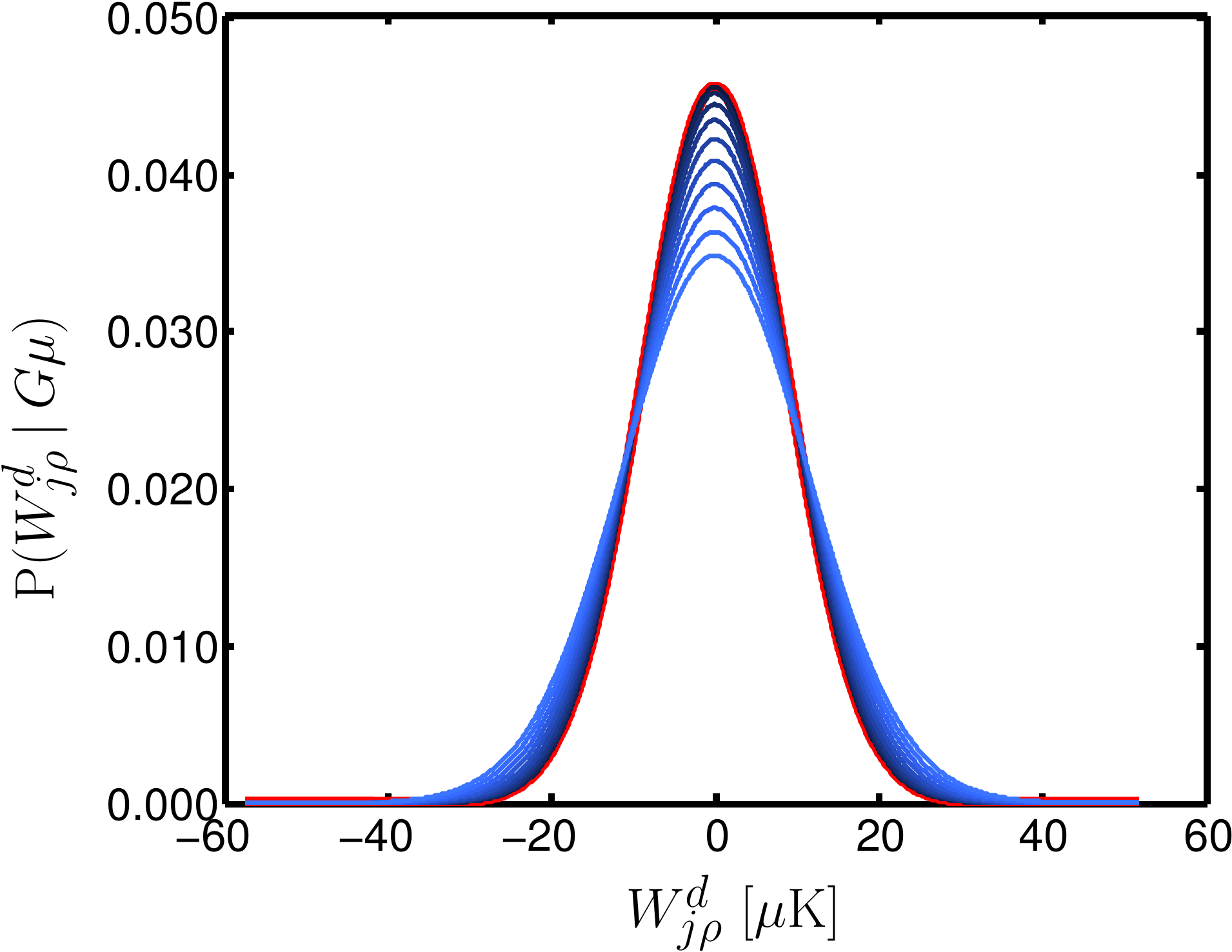}} \hfill
  \subfigure[$\wscale = 4$]{\includegraphics[width=.24\textwidth]{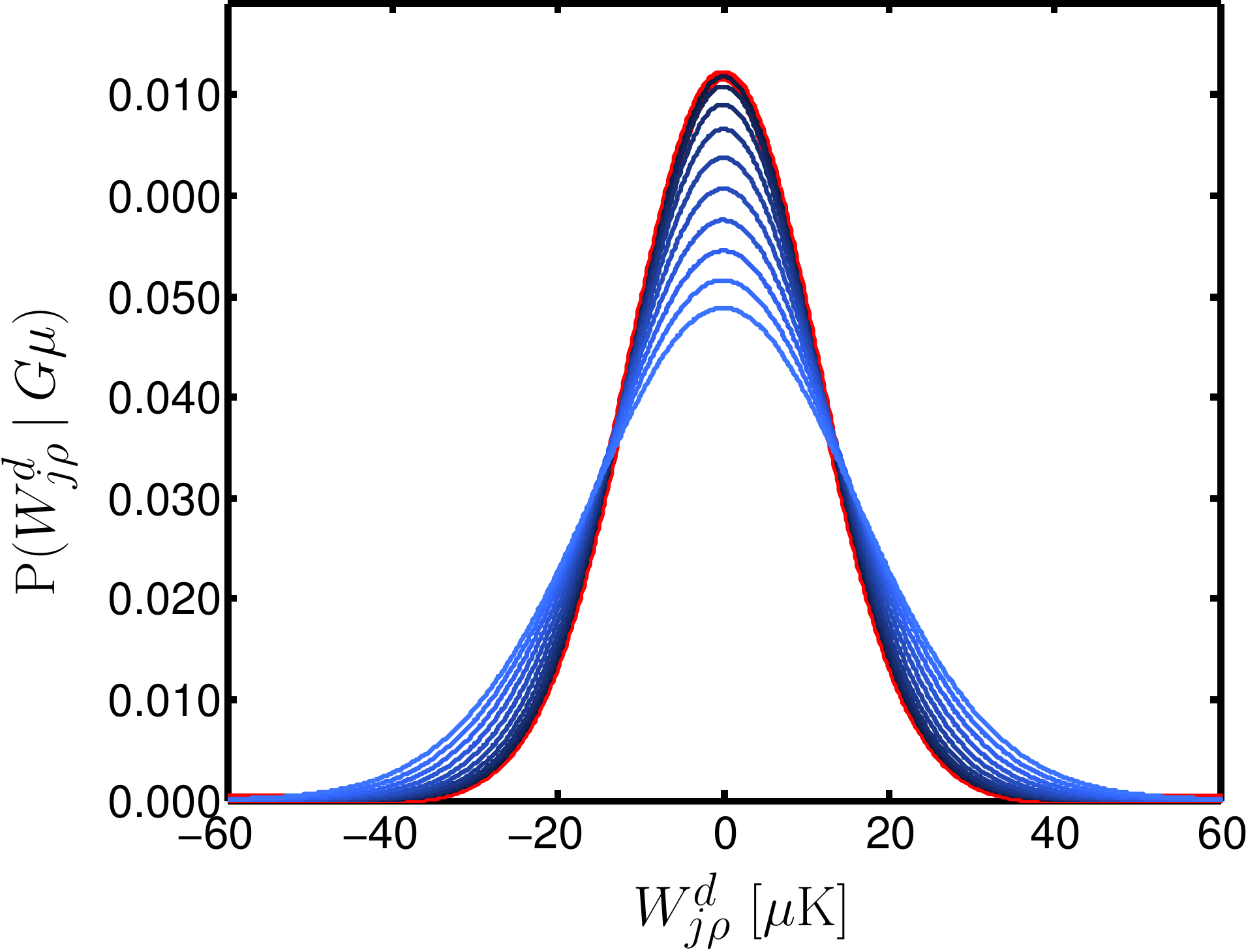}} \hfill
  \subfigure[$\wscale = 5$]{\includegraphics[width=.24\textwidth]{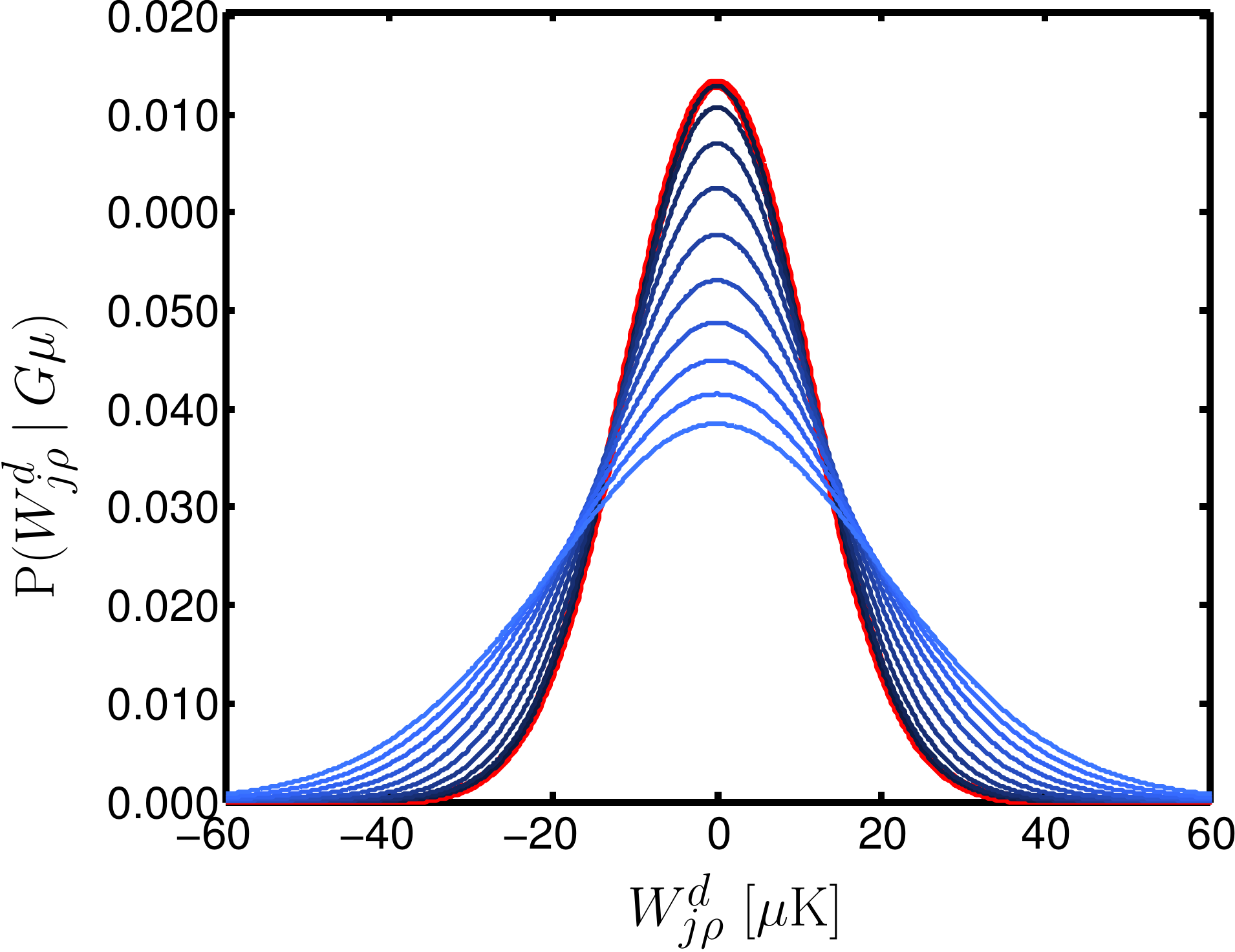}} \hfill
  \subfigure[$\wscale = 6$]{\includegraphics[width=.24\textwidth]{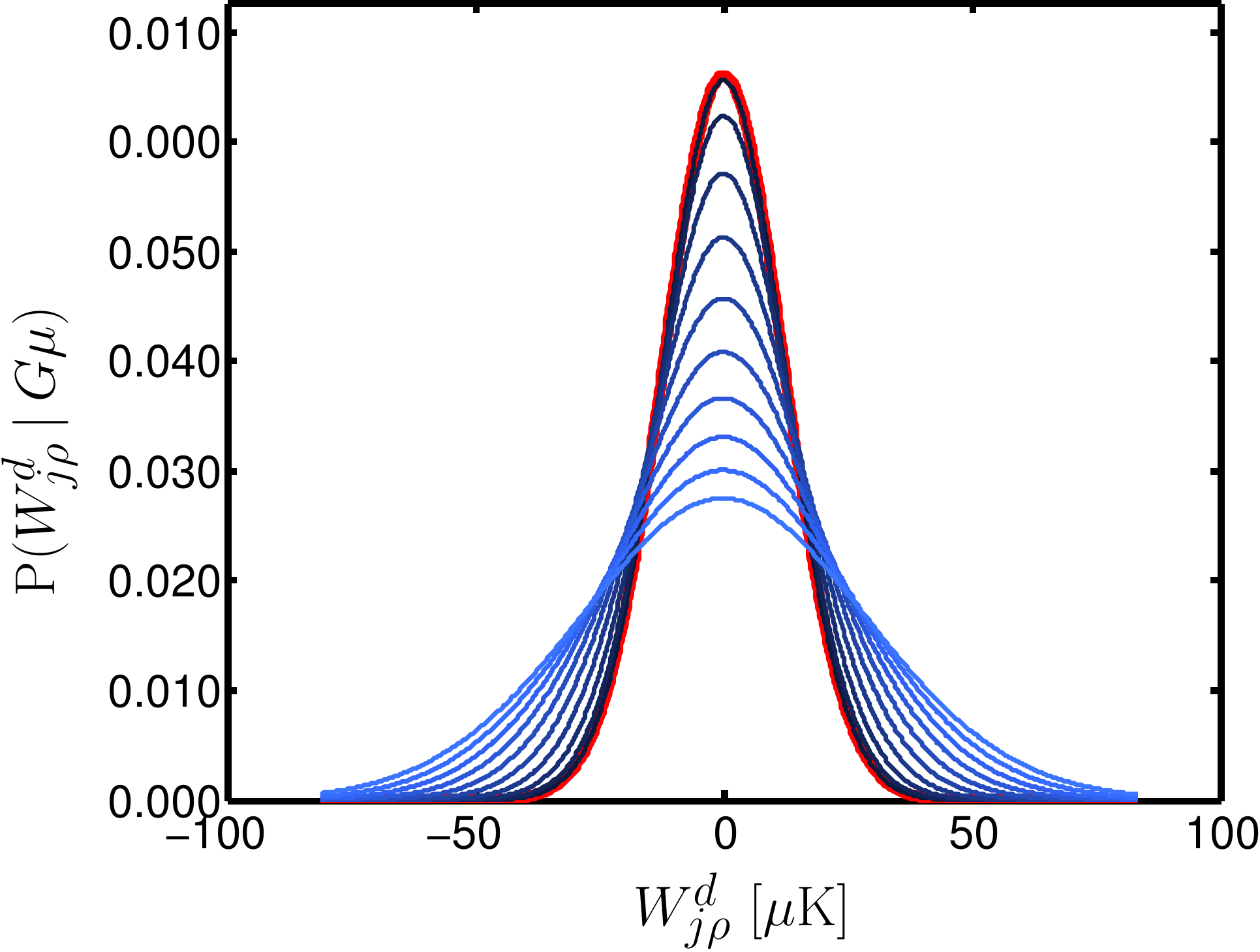}} \hfill
  \subfigure[$\wscale = 7$]{\includegraphics[width=.24\textwidth]{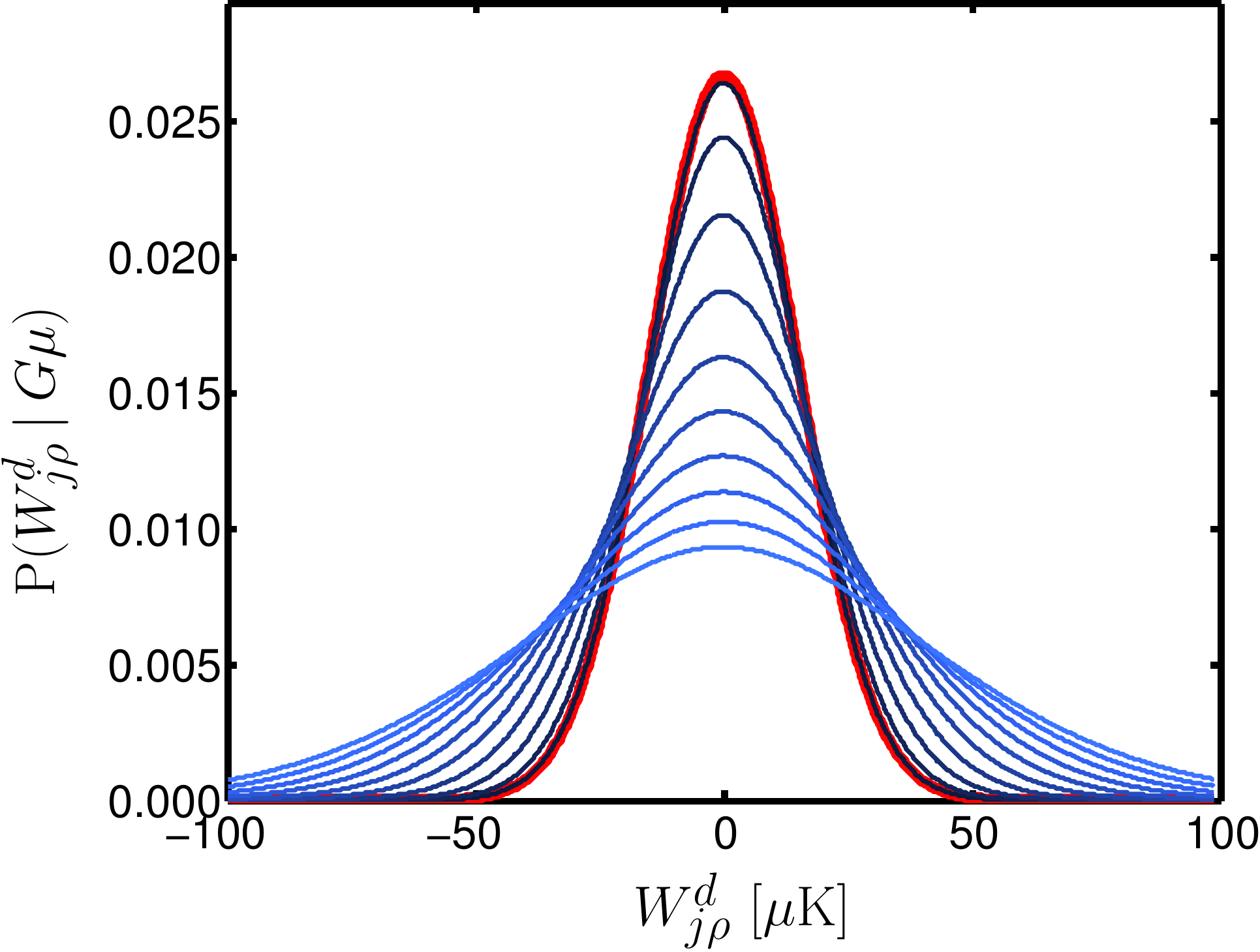}} \hfill
  \caption{Look-up-tables (LUTs) to precompute the mapping $\wcoeffdatap$ to $\prob(\wcoeffdata_{\wscaleposn} \given \gmu)$ of \eqn{\ref{eqn:lut_posterior}} for each $\wscale$.   Precomputing LUTs avoids recalculating integrals for identical (or similar) values of $\wcoeffdatap$ when computing the posterior distribution of the string tension.  The distributions $\prob(\wcoeffdata_{\wscaleposn} \given \gmu)$ are shown for different values of $\gmu$ by the light blue lines.  For comparison, the Gaussian CMB distribution $\prob_\wscale^\fcmb(\wcoeffcmbp)$ is shown by the heavy red line.  As $\gmu$ is reduced towards zero, the distributions $\prob(\wcoeffdata_{\wscaleposn} \given \gmu)$ approach $\prob_\wscale^\fcmb(\wcoeffcmbp)$ (the light blue curves darken and approach the heavy red curve). The LUTs are normalised probability distributions and integrate to unity. These plots are created using the testing string map and parameters $\elmax=2048$, $\nmax=4$, $\wscalemax=7$, and $\dilparam=2$.}
  \label{fig:lut_posterior}
\end{figure*}

In practice, to compute the posterior distribution it is necessary to first evaluate the likelihood for each individual wavelet coefficient by \eqn{\ref{eqn:lut_posterior}}, before combining these terms to compute the overall likelihood for the data by \eqn{\ref{eqn:likelihood_full}} or \eqn{\ref{eqn:likelihood_full_log}}.  In order to avoid recalculating integrals for identical (or similar) values of $\wcoeffdatap$, we precompute look-up-tables (LUTs) for \eqn{\ref{eqn:lut_posterior}}, storing the mapping from $\wcoeffdatap$ to $\prob(\wcoeffdata_{\wscaleposn} \given \gmu)$ for each $\wscale$. When evaluating the likelihood of a given dataset we linearly interpolate the wavelet coefficients onto the LUT grid. These LUTs are plotted in \fig{\ref{fig:lut_posterior}}.  Since the distributions $\prob^{\fcmbnoise}_\wscale$ and $\prob^{\fstring}_\wscale$ are properly normalised, the likelihood $\prob(\wcoeffdatap \given \gmu)$ for a given \gmu\ is also a normalised probability distribution and should integrate to unity.  To ensure the quadrature used to evaluate \eqn{\ref{eqn:lut_posterior}} is accurate, we check that the precomputed distributions $\prob(\wcoeffdata_{\wscaleposn} \given \gmu)$ integrate to unity (using the trapezium rule), which is indeed the case provided a sufficient number of samples is used to evaluate the integral.

\subsection{String model comparison}
\label{sec:inference:evidence}

To ascertain the overall evidence for cosmic strings we compare the Bayesian evidence of the string mode $\modelstring$, which includes string and inflationary induced CMB components, to the evidence of the standard inflationary model $\modelcmb$.  The Bayesian evidence of the string model is given by
\begin{equation}
  \evidence^{\rm \fstring}
  =
  \prob(\wcoeffdata \given \modelstring)
  =
  \int_{\reals}
  \dx(\gmu) \:
  \prob(\wcoeffdata \given \gmu, \modelstring) \: \prob(\gmu \given \modelstring)
  \spcend ,
\end{equation}
where now we make the dependence on the model explicit. The Bayesian evidence of the CMB model is given by
\begin{equation}
  \evidence^{\rm \fcmb}
  =
  \prob(\wcoeffdata \given \modelcmb)
  =
  \prod_{\wscale,\wposn} \prob^\fcmbnoise_\wscale(\wcoeffdatap)
  \spcend .
\end{equation}
For numerical purposes we compute the log-evidence, given by
\begin{equation}
  \ln \evidence^{\rm \fcmb}
  =
  \ln \prob(\wcoeffdata \given \modelcmb)
  =
  \sum_{\wscale,\wposn} \ln \prob^\fcmbnoise_\wscale(\wcoeffdatap)
  \spcend.
\end{equation}

In the absence of any prior information favouring either model, the ratio of the model posterior probabilities is given by the ratio of the Bayesian evidences:
\begin{equation}
  \frac{\prob(\modelstring \given \wcoeffdata)}{\prob(\modelcmb \given \wcoeffdata)}
  =
  \frac{\evidence^{\rm \fstring}}{\evidence^{\rm \fcmb}}
  \spcend.
\end{equation}
We compute the ratio of evidences to determine the model favoured by the data.  In practice we compute the difference in log-evidence (also called the Bayes factor):
\begin{equation}
\Delta \ln \evidence
= \ln(\evidence^{\rm \fstring} / \evidence^{\rm \fcmb})
= \ln \evidence^{\rm \fstring}  - \ln \evidence^{\rm \fcmb}
\spcend .
\end{equation}

The Jeffreys scale \citep{jeffreys:1961} is often used as a
rule-of-thumb when comparing models via their Bayes factor.  While we caution against using the Jeffreys scale as a strict test to classify models (since the boundaries of the scale are somewhat arbitrary), it can nevertheless be useful to gain some intuition for those not familiar with Bayesian model selection.  The
log-Bayes factor $\Delta \ln \evidence = \ln (\evidence^{(1)}/\evidence^{(2)})$ 
represents the degree by which model
$\fitmodelsel^{(1)}$ is favoured over model $\fitmodelsel^{(2)}$, assuming the 
models are equally likely {\em a priori}. On the Jeffreys scale log-Bayes
factors are given the following interpretation: $0 \leq
\Delta {\rm ln} E < 1$ is regarded as inconclusive; $1 \leq \Delta
{\rm ln} E < 2.5$ as significant; $2.5 \leq \Delta {\rm ln} E < 5$ as
strong; and $\Delta {\rm ln} E \geq 5 $ as conclusive (without loss of
generality we have assumed $E_1 \geq E_2$).  For reference, a
log-Bayes factor of 2.5 corresponds to odds of approximately 1 in 12,
while a factor of 5 corresponds to odds of approximately 1 in 150.

\section{Estimation of cosmic string maps}
\label{sec:denoising}

In addition to estimating the evidence for the cosmic string model and the posterior distribution of the string tension, we also recover a direct estimate of the string-induced CMB component itself.  To estimate the string contribution at the map level we develop a Bayesian estimation approach in wavelet space, generalising the technique described in \citet{hammond:2009} from a planar region to the spherical full-sky setting.  We first describe the string map estimation technique, before examining its properties as a Bayesian thresholding approach to denoise the inflationary CMB component from the observed data.

\subsection{String map estimation}
\label{sec:denoising:map_estimation}

Our inference of the wavelet coefficients of the underlying string map, and equivalently the string map itself, is encoded in the posterior probability distribution
$\prob(\wcoeffstringp \given \wcoeffdata)$.  Various estimators can be considered to recover the wavelet coefficients of string map from their posterior distribution.  We estimate the wavelet coefficients of the string map from the mean of the posterior distribution, which can be computed by
\begin{align}
  \wcoeffstringprecov
  & =
  \int_\reals \dx\wcoeffstringp \: \wcoeffstringp \:
  \prob(\wcoeffstringp \given \wcoeffdata) \\
  & =
  \int_\reals \dx\wcoeffstringp \: \wcoeffstringp
  \int_\reals \dx(\gmu) \:
  \prob(\wcoeffstringp \given \wcoeffdata, \gmu) \:
  \prob(\gmu \given \wcoeffdata) \\
  & =
  \int_\reals \dx(\gmu) \:
  \prob(\gmu \given \fdata) \: \wcoeffstringprecov(\gmu)
  \spcend,
  \label{eqn:string_estimate_gmu_all}
\end{align}
where
\begin{align}
  \wcoeffstringprecov(\gmu)
  & =
  \int_\reals \dx\wcoeffstringp \: \wcoeffstringp \:
  \prob(\wcoeffstringp \given \wcoeffdatap, \gmu) \\
  & =
  \frac{\int_\reals \dx\wcoeffstringp \: \wcoeffstringp \:
  \prob(\wcoeffdatap \given \wcoeffstringp, \gmu) \:
  \prob(\wcoeffstringp \given \gmu) }{\prob(\wcoeffdatap \given \gmu)}
  \label{eqn:string_estimate_gmu_given:1}
  \\
  & =
  \frac{\int_\reals \dx\wcoeffstringp \: \wcoeffstringp \:
  \prob^\fcmbnoise_\wscale(\wcoeffdatap - \wcoeffstringp \given \gmu) \:
  \prob^\fstring_\wscale(\wcoeffstringp \given \gmu)}{\prob(\wcoeffdatap \given \gmu)}
  \spcend .
  \label{eqn:string_estimate_gmu_given}
\end{align}
Note that we replace $\prob(\gmu \given \wcoeffdata)$ with $\prob(\gmu \given \fdata)$ in \eqn{\ref{eqn:string_estimate_gmu_all}} (since there is a one-to-one relationship between a map and its wavelet coefficients) and appeal to Bayes theorem in \eqn{\ref{eqn:string_estimate_gmu_given:1}}. To summarise, for each $\gmu$ we compute a denoised set of wavelet coefficients $\wcoeffstringprecov(\gmu)$ by \eqn{\ref{eqn:string_estimate_gmu_given}}.  We then combine these, taking the posterior distribution of the string tension $\prob(\gmu \given \fdata)$ into account, to compute the overall denoised set of wavelet coefficients $\wcoeffstringprecov$ by \eqn{\ref{eqn:string_estimate_gmu_all}}.  The denominator of \eqn{\ref{eqn:string_estimate_gmu_given}} is given by \eqn{\ref{eqn:lut_posterior}} for which LUTs have been precomputed already.  Similarly, LUTs for the numerator are precomputed for each $\wscale$.  Since we consider zero-mean signals, these LUTs should integrate to zero, which indeed they do provided a sufficient number of samples is used to evaluate the integrals.  As we assume independence of the wavelet coefficients, wavelet coefficients are denoised pointwise.

Once we have recovered the denoised wavelet coefficients, a string map can be recovered through an inverse wavelet transform:
\begin{equation}
  \fstringrecov = \mathcal{W}^{-1}\bigl(\wcoeffstringrecov\bigr)
  \spcend .
\end{equation}
Alternatively, string maps could also be estimated for each $\gmu$ through an inverse wavelet transform:
\begin{equation}
  \fstringrecov(\gmu) = \mathcal{W}^{-1}\bigl(\wcoeffstringrecov(\gmu)\bigr)
  \spcend.
\end{equation}
Since the wavelet transform is linear, the overall string map could then be recovered by
\begin{equation}
  \fstringrecov
  =
  \int_\reals \dx(\gmu) \:
  \prob(\gmu \given \fdata) \: \fstringrecov(\gmu)
  \spcend.
\end{equation}

While we use the string tension posterior distribution $\prob(\gmu \given \fdata)$ estimated in wavelet space by the approach outlined in \sectn{\ref{sec:inference:gmu_estimation}}, one is free to substitute a posterior distribution estimated by alternative methods.  The resulting recovered string maps could be considered as a pre-processed input to other map-based methods for estimating the string tension from the non-Gaussian structure of the string-induced CMB component, such as edge detection \citep[\eg][]{lo:2005,amsel:2008,stewart:2009,danos:2010}.  The enhanced string component and reduced background is likely to boost the effectiveness of subsequent string tension estimation.

An estimate of the variance of wavelet coefficients of the string component could also be performed in order to provide a measure of the accuracy of the recovered string-induced component.  For this to be most useful it would be necessary to express the variance in map space, which could be computed by an inverse wavelet transform with $\wav_\wscaleposn$ and $\wavs_{\sa\p}$ substituted by $\vert \wav_\wscaleposn \vert^2$ and $\vert \wavs_{\sa\p} \vert^2$, respectively (for a related discussion see \citealt{rogers:s2let_ilc_temp}).  We leave this to future work.

\subsection{Bayesian thresholding}

\begin{figure*}
  \centering
  \subfigure[$\wscale = 0$]{\includegraphics[width=.24\textwidth]{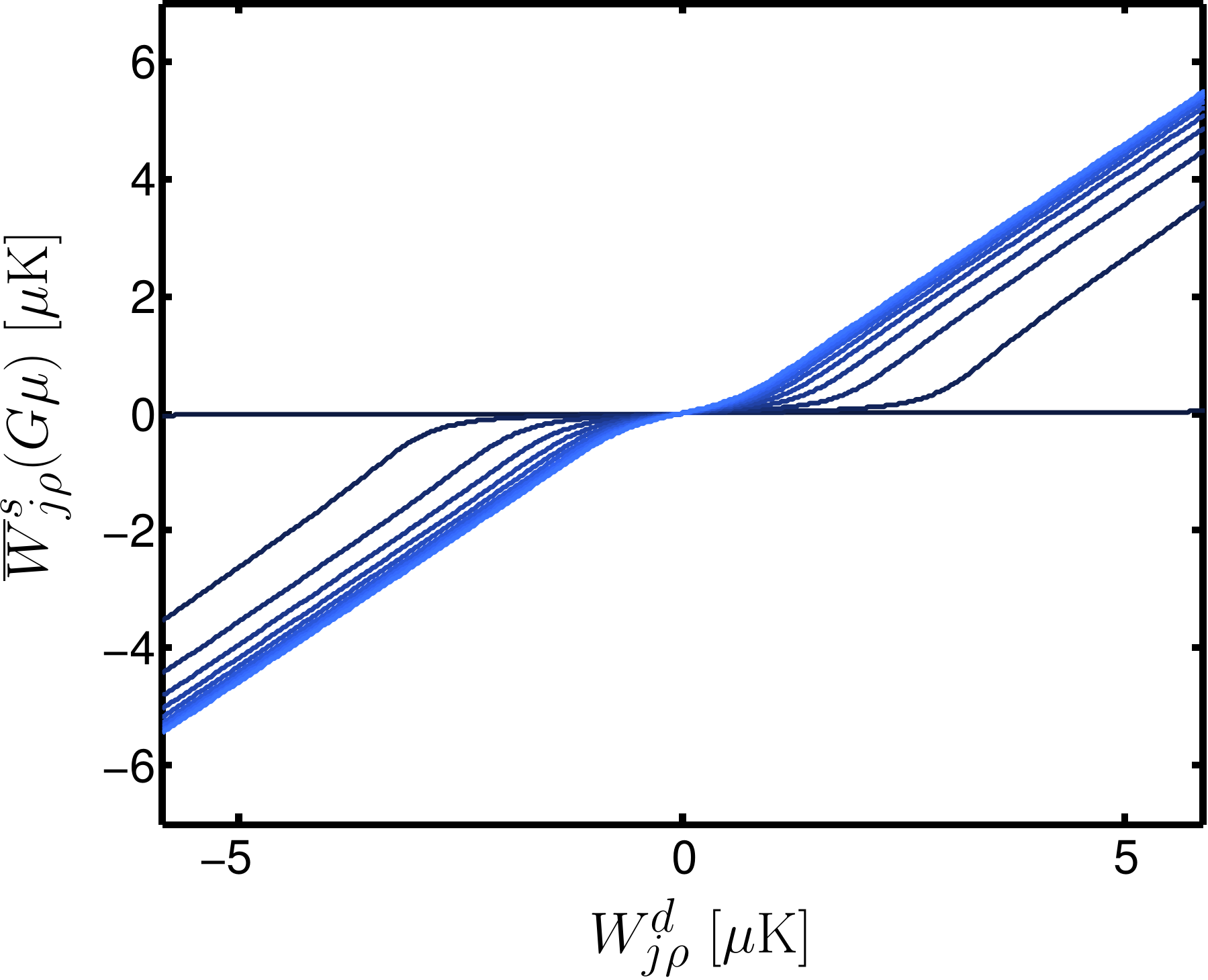}} \hfill
  \subfigure[$\wscale = 1$]{\includegraphics[width=.24\textwidth]{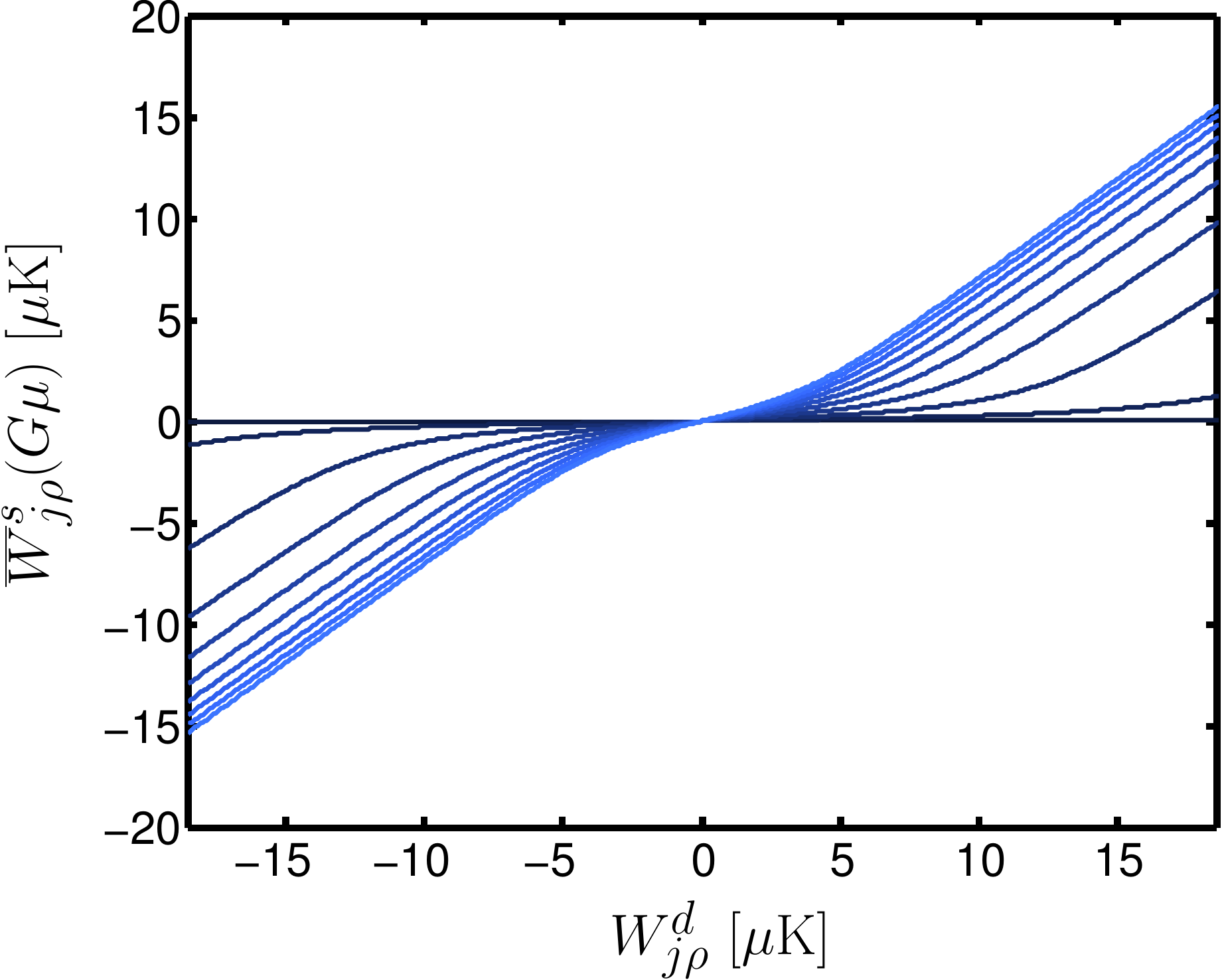}} \hfill
  \subfigure[$\wscale = 2$]{\includegraphics[width=.24\textwidth]{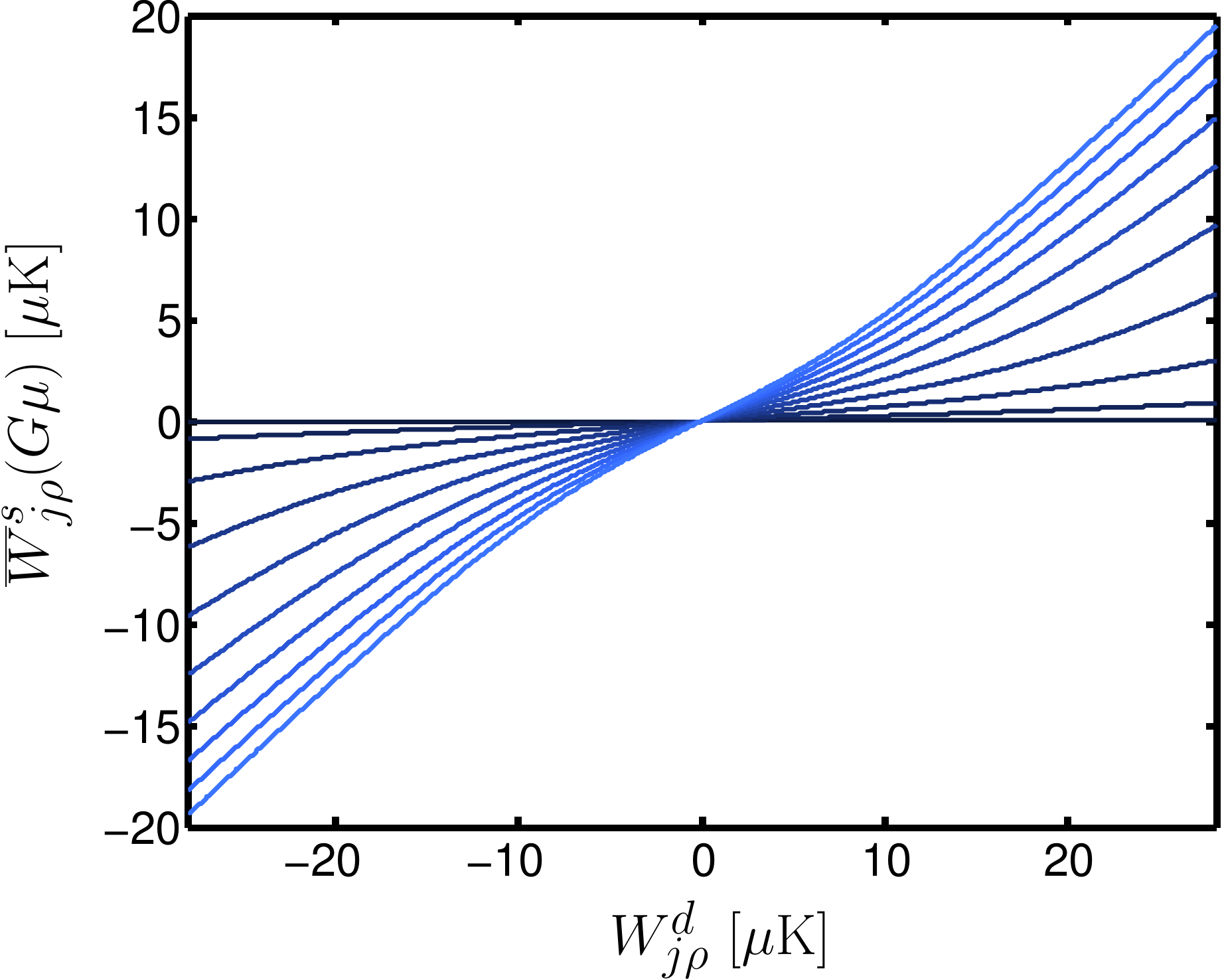}} \hfill
  \subfigure[$\wscale = 3$]{\includegraphics[width=.24\textwidth]{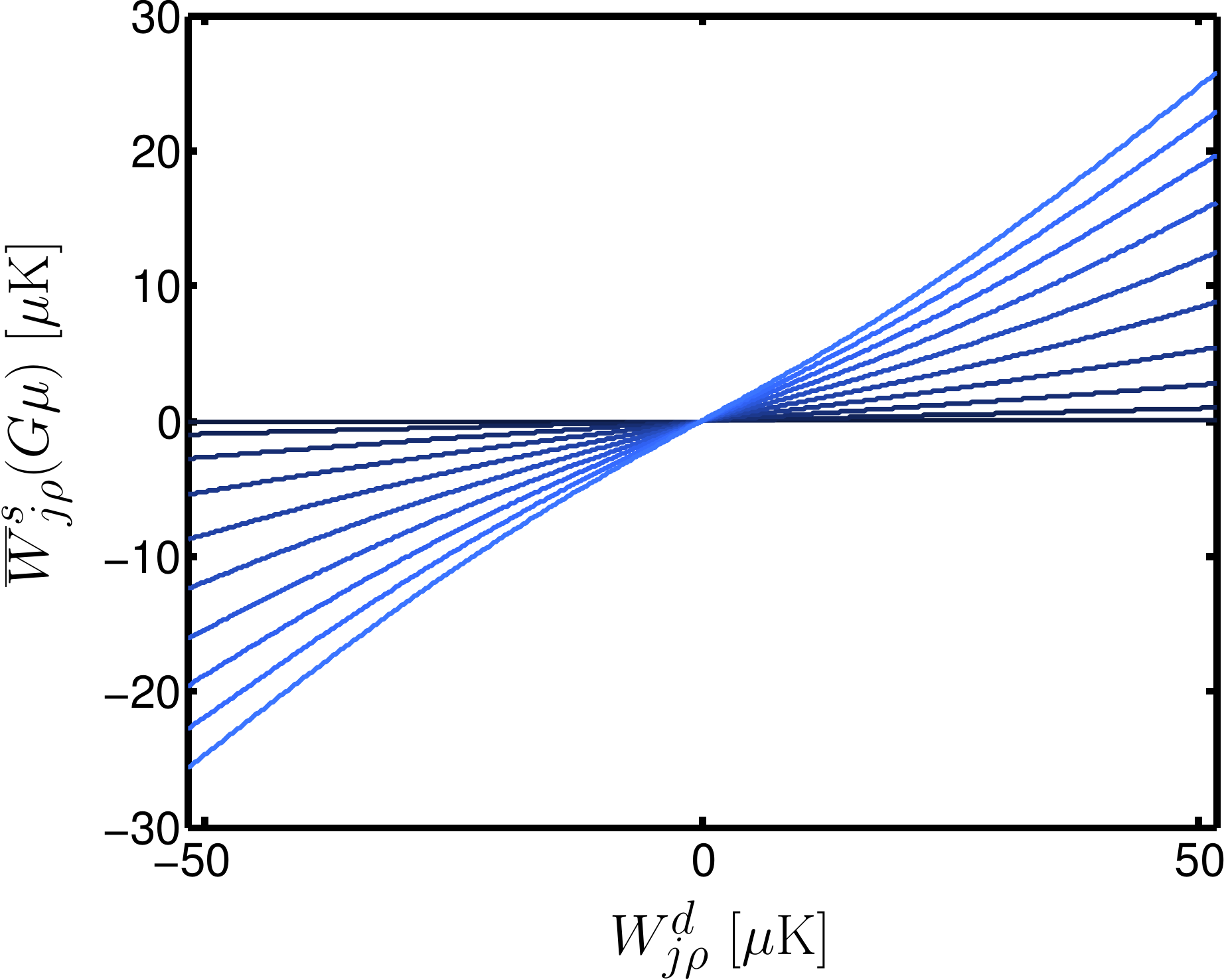}} \hfill
  \subfigure[$\wscale = 4$]{\includegraphics[width=.24\textwidth]{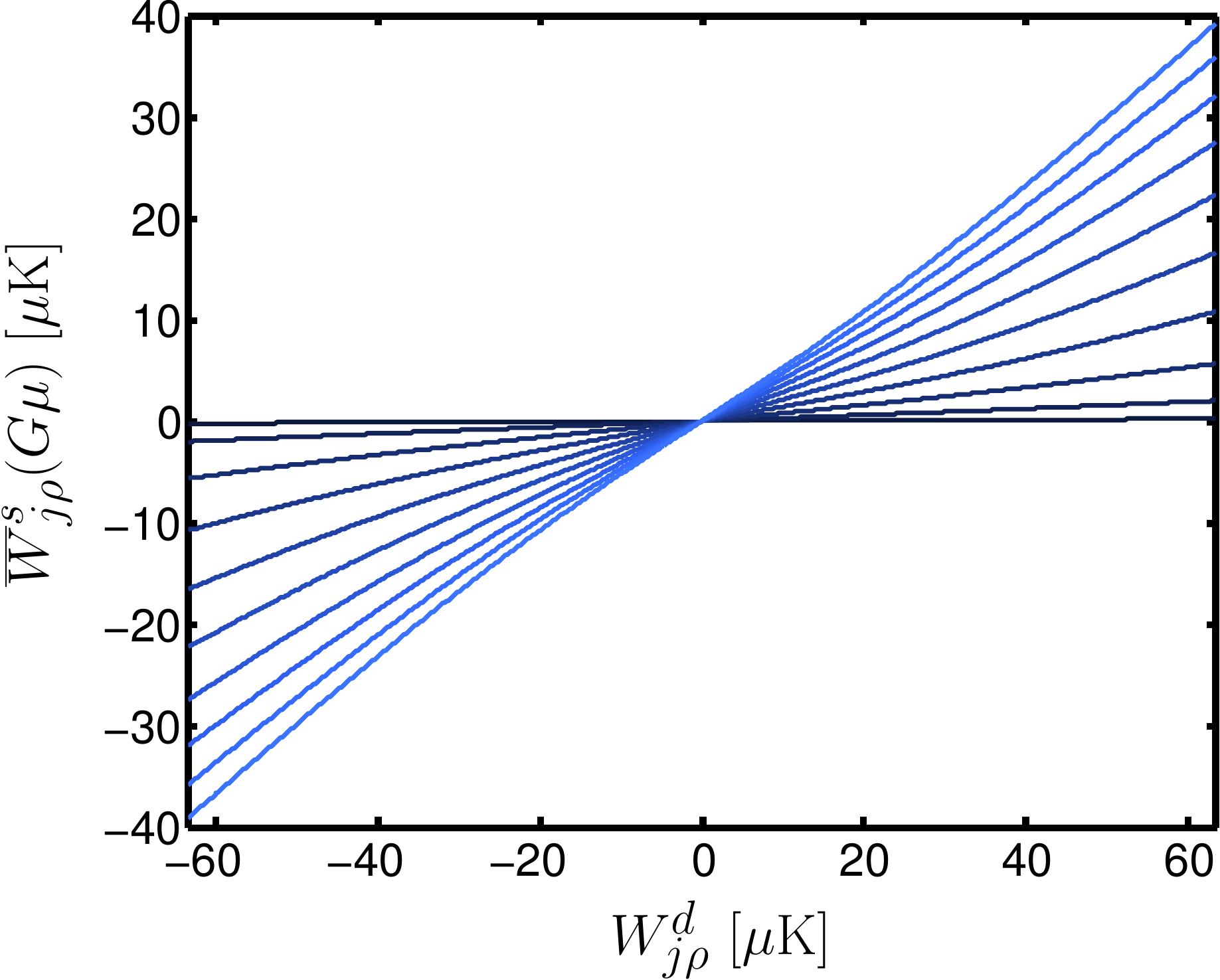}} \hfill
  \subfigure[$\wscale = 5$]{\includegraphics[width=.24\textwidth]{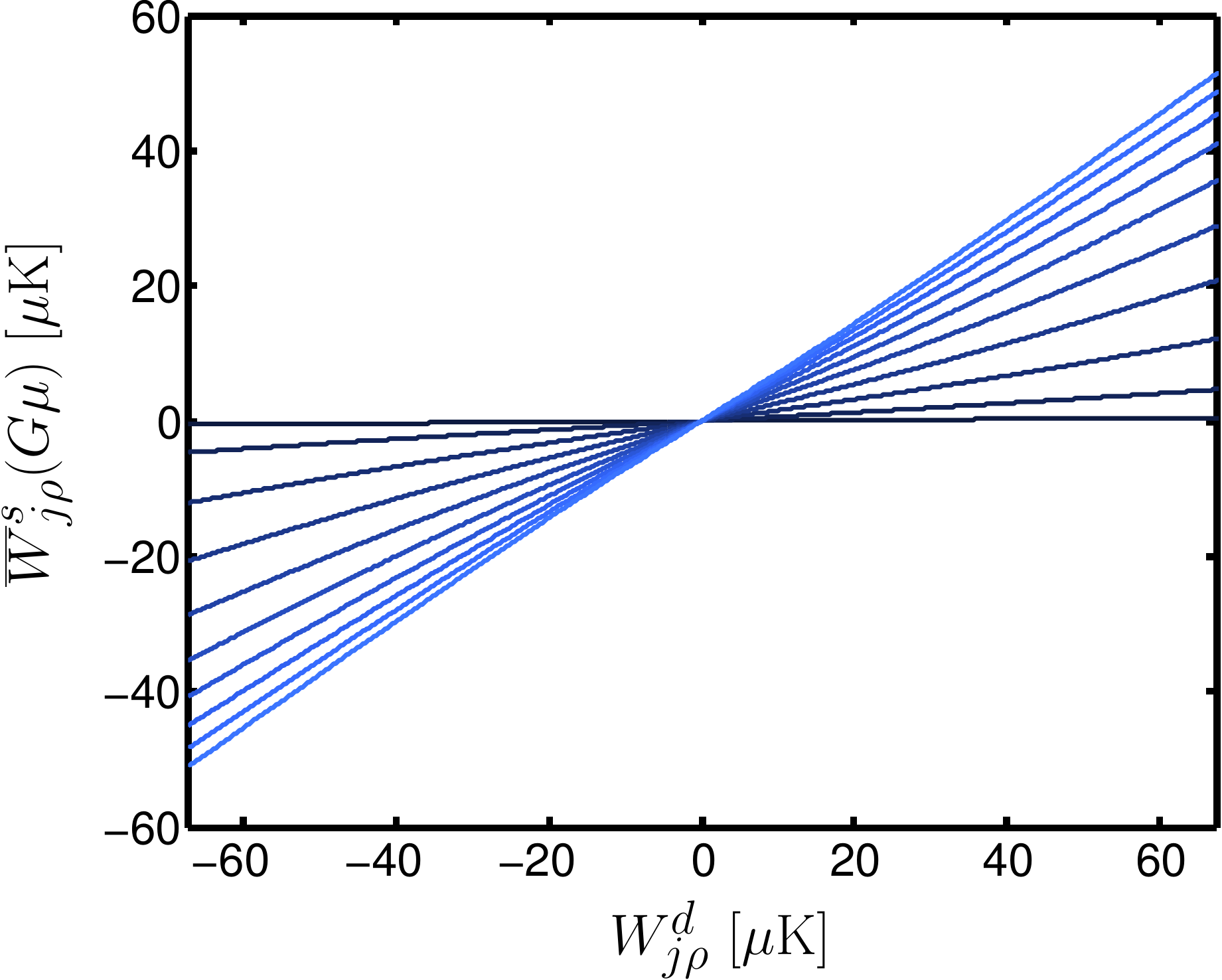}} \hfill
  \subfigure[$\wscale = 6$]{\includegraphics[width=.24\textwidth]{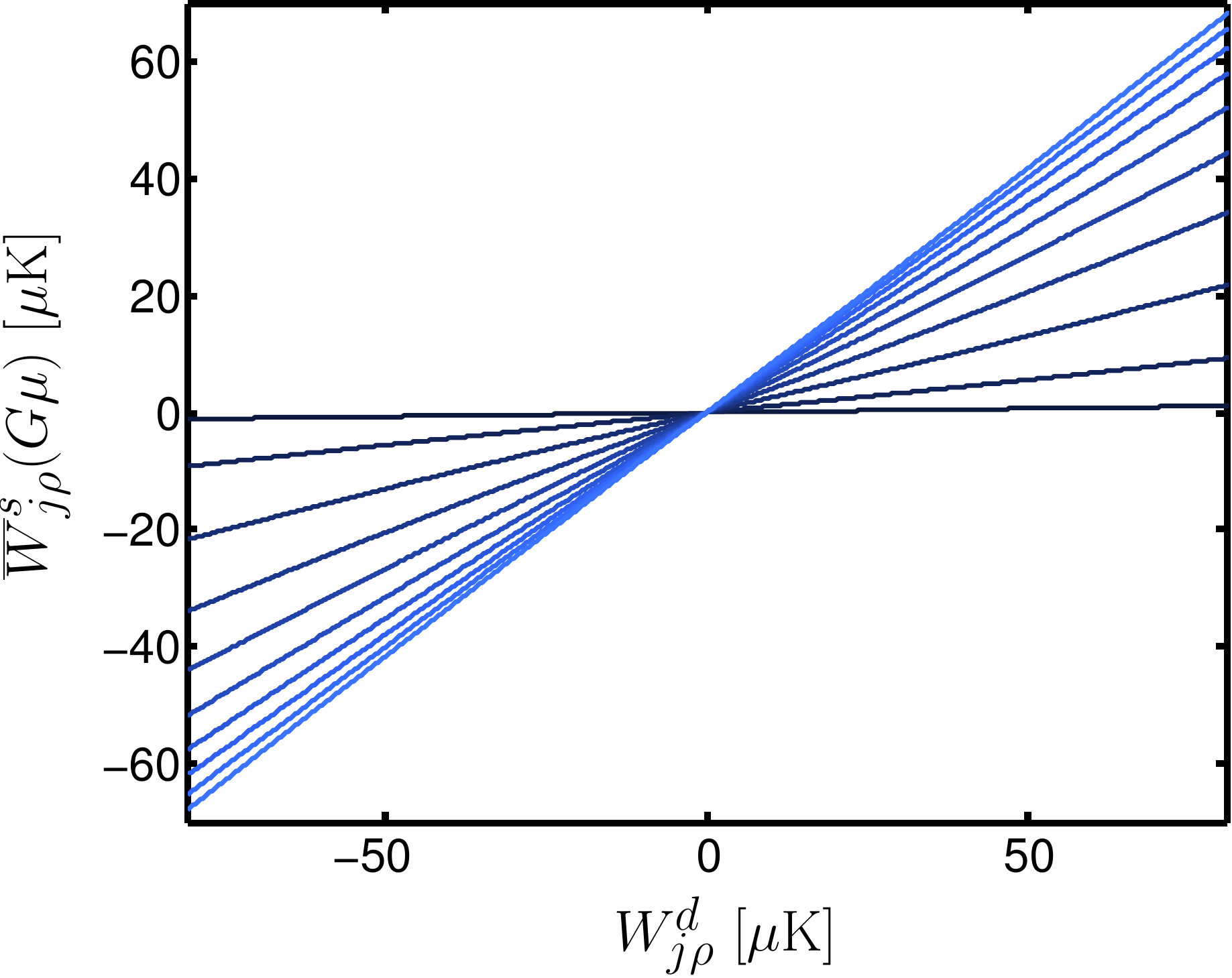}} \hfill
  \subfigure[$\wscale = 7$]{\includegraphics[width=.24\textwidth]{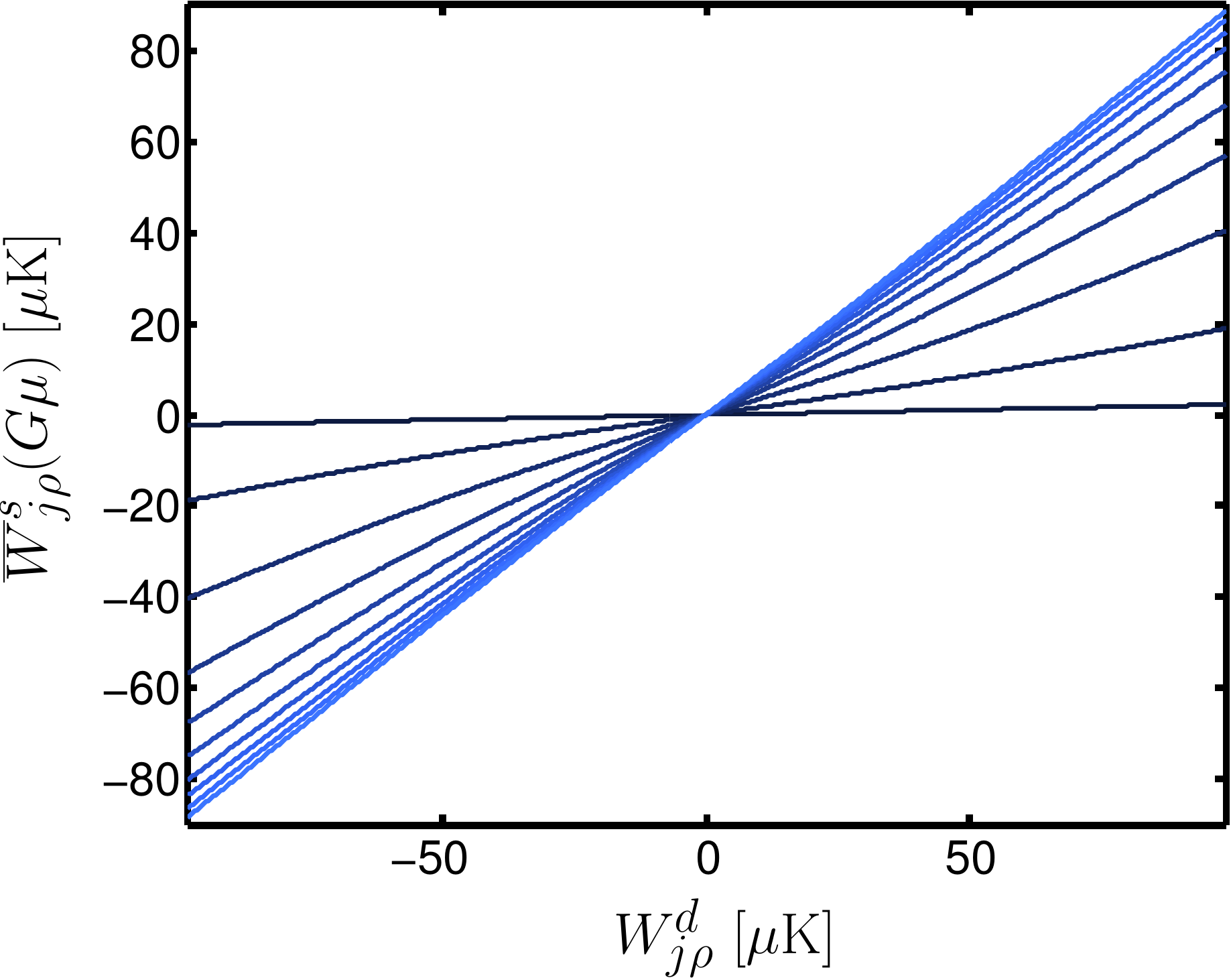}} \hfill
  \caption{Bayesian thresholding functions for each wavelet scale \wscale. Each curve in a given panel shows a different value of \gmu, with \gmu\ approaching zero as the shade of the curve darkens.  As \gmu\ is reduced the amplitude of the string contribution is reduced relative to the inflationary component and the thresholding curves approach zero, as expected.
  As the wavelet scale \wscale\ increases, larger scale features are probed by the wavelet and the thresholding functions become more linear since  the statistical distributions of the string and inflationary CMB components become more similar (see text for further discussion).}
  \label{fig:bayesian_soft_thresholding}
\end{figure*}

The hybrid wavelet-Bayesian string map estimation technique outlined in \sectn{\ref{sec:denoising:map_estimation}} can be viewed as a Bayesian thresholding approach to denoise the observed data.  The estimation of the wavelet coefficients of the string signal by \eqn{\ref{eqn:string_estimate_gmu_given}} can be viewed as a mapping from the wavelet coefficients of the data $\wcoeffdatap$ to the estimated string signal $\wcoeffstringprecov(\gmu)$ for a given \gmu.  One then marginalises over the prior distribution for \gmu\ by \eqn{\ref{eqn:string_estimate_gmu_all}}.
The thresholding mapping functions defined by \eqn{\ref{eqn:string_estimate_gmu_given}} are plotted in \fig{\ref{fig:bayesian_soft_thresholding}}.

As the wavelet scale \wscale\ increases larger scale features are probed, for which the string distribution in wavelet space becomes less leptokurtic (as shown in \tbl{\ref{tbl:ggd_shape_parameters}}), \ie\ more Gaussian.  Consequently, the thresholding functions become more linear as it becomes more difficult to distinguish the string and inflationary CMB distributions.  For the small scales, corresponding to low \wscale, the thresholding functions are less linear, with the energy in large coefficients more likely to be retained since these are more likely due to the string component, while the energy of small coefficients is more likely to be curtailed.

For each wavelet scale \wscale, curves are plotted for different values of \gmu, with \gmu\ approaching zero as the shade of the curve darkens.  As \gmu\ is reduced the amplitude of the string component is reduced relative to the inflationary component and the thresholding curves approach zero, as expected.

\section{Simulations and results}
\label{sec:results}

In this section, we demonstrate the application of our wavelet-Bayesian framework for cosmic string inference to simulated \planck\ observations.  We do not optimise the parameters of the analysis and consider the standard dyadic wavelet scaling (\ie\ $\dilparam=2$).  Alternative wavelet scalings, like that considered in \citet{rogers:s2let_ilc_temp} and \citet{rogers:s2let_ilc_pol}, are likely to improve performance.  The application to \planck\ data and the optimisation of the parameters of the method is left to future work.  We first describe the CMB simulations performed, before presenting results from applying the framework outlined previously to these simulations for differing values of \gmu.  We show results estimating the posterior distribution of the string tension, comparing the string model \modelstring\ with the standard inflationary model \modelcmb, and recovering maps of the string-induced CMB component.

\subsection{Simulations}
\label{sec:results:simulations}

Our simulations model idealised observations of combined string-induced and inflationary CMB skies by the \planck\ satellite's 143 GHz detectors, making heavy use of the \stwocode\ code. As high-resolution, full-sky string simulations are computationally challenging to produce (see \sectn{\ref{sec:inference:distributions:strings}}), we base all string simulations on the single testing string simulation (\fig{\ref{fig:input_string_sims:testing}}), smoothing with a $7.3$ arcmin Gaussian beam and rescaling by the appropriate $T_0 \gmu$ (assuming the mean CMB temperature of $T_0 = 2.725$ K; \citealt{mather:1999}).  We do not touch the training string simulation (\fig{\ref{fig:input_string_sims:training}}) since this was used to fit the GGD distributions modelling string-induced CMB components. We model the CMB and noise as pure Gaussian random fields and hence draw realisations directly from their combined power spectrum, using a band-limit of $\elmax = 2500$. We calculate the CMB power spectrum using \cambcode\footnote{\url{http://camb.info}} \citep{lewis:2000}, assuming the best-fit cosmology from \planck's analysis of a compilation of CMB, lensing, baryon acoustic oscillation, supernova and expansion datasets \citep{planck2014-a15}. To create the final power spectrum, we multiply the CMB power spectrum by the instrumental beam and \healpix\ window function, and then add white noise at 4.3 $\mu$K per beam-sized pixel (the final sensitivity of \planck's 143 GHz channel).  The resulting power spectra are shown in \fig{\ref{fig:input_power_spectra}}.  All maps are simulated at \healpix\ resolution $\nside=2048$.

Examples of simulated CMB maps, with and without a string-induced component, are plotted in \fig{\ref{fig:cmb_string_sims}} for a string tension of $\gmu=5 \times 10^{-7}$.  It is not possible to determine the presence cosmic strings by eye.

\setlength{\plotwidth}{1.0\columnwidth}

\begin{figure}
  \centering
  \subfigure[Inflationary contribution only]{\includegraphics[width=\plotwidth, trim=4mm 8mm 4mm 8mm, clip=true]{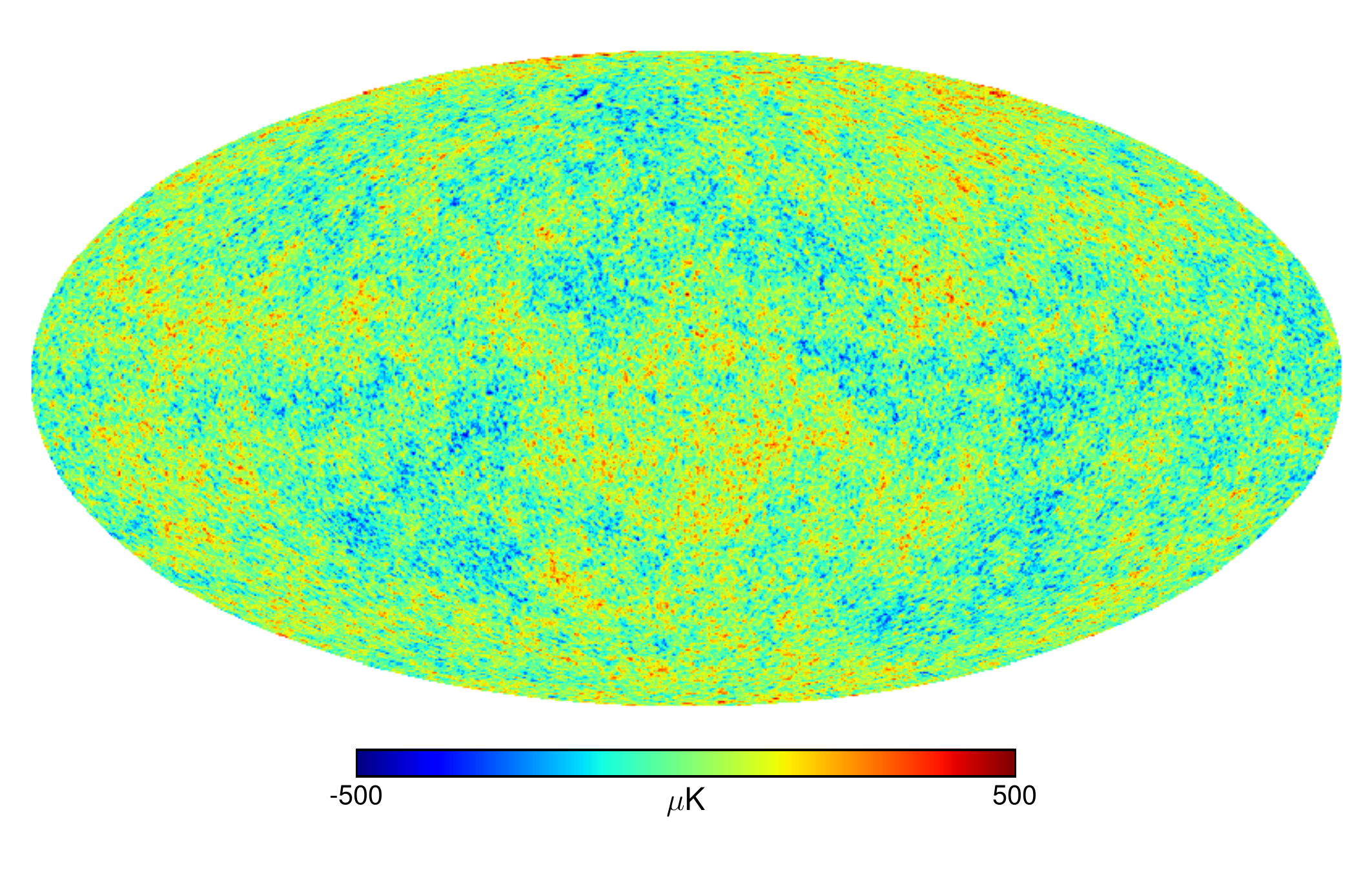}}
  \subfigure[Inflationary and string-induced contributions]{\includegraphics[width=\plotwidth, trim=4mm 8mm 4mm 8mm, clip=true]{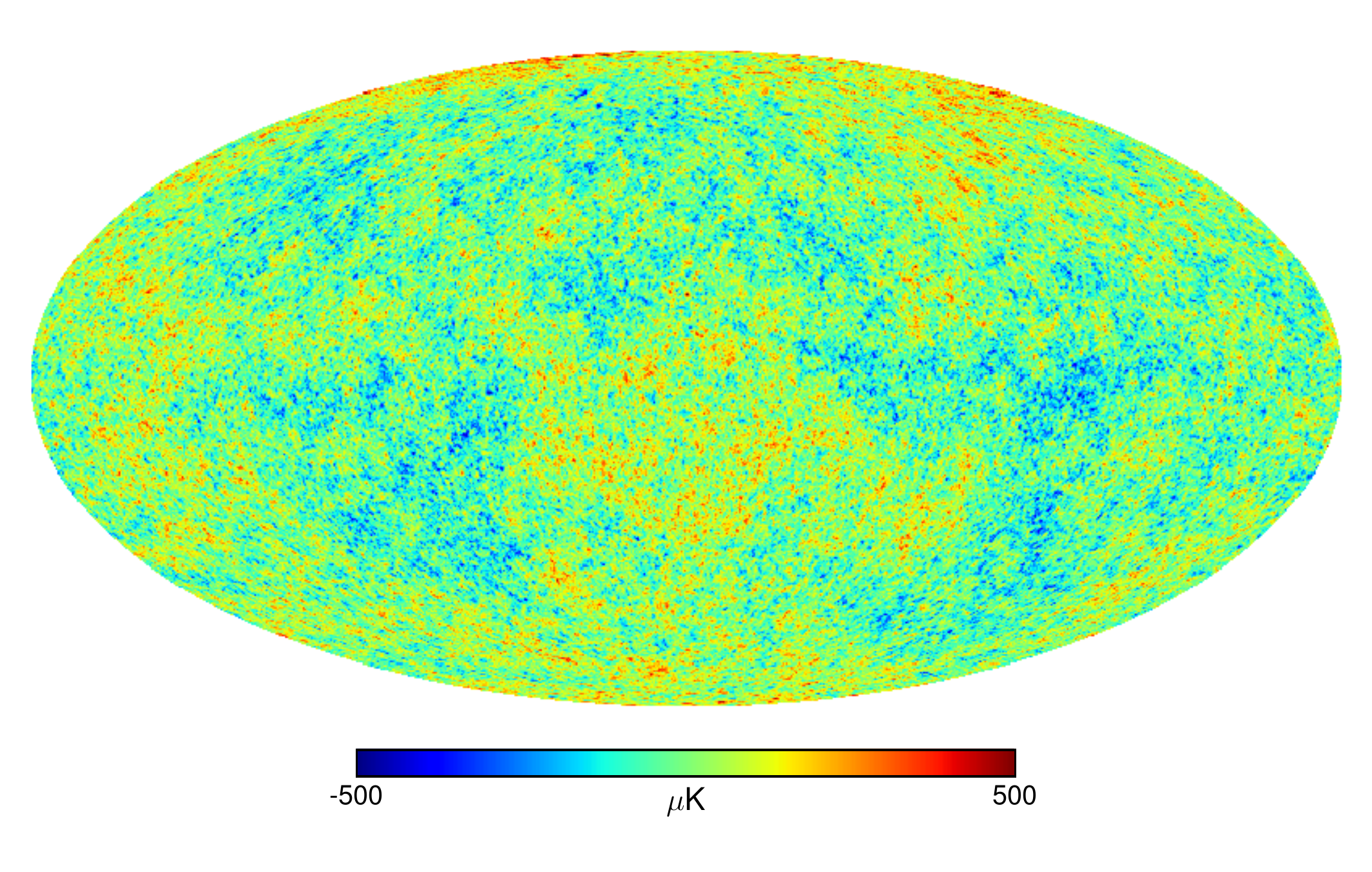}}
  \caption{Simulated CMB maps with and without a string-induced component.  Panel~(a) includes an inflationary CMB contribution only, whereas panel~(b) includes an inflationary contribution and the string-induced contribution shown in \fig{\ref{fig:input_string_sims:testing}}, scaled to $\gmu=5\times 10^{-7}$. It is not possible to determined the presence of the string-induced component by eye.}
  \label{fig:cmb_string_sims}
\end{figure}

\subsection{String tension estimation}
\label{sec:results:tension_estimation}

We perform the analysis outlined in \sectn{\ref{sec:inference:gmu_estimation}} on a number of simulations with embedded string contributions of varying string tension \gmu\ to estimate the posterior distribution of the string tension.

In the framework presented in \sectn{\ref{sec:inference:gmu_estimation}} we assume wavelet coefficients are independent. In practice, wavelet coefficients are not independent, but the covariance of wavelet coefficients does decay rapidly with spatial separation (relative to the spatial size of the wavelet considered) and is identically zero for non-adjacent scales (\ie\ for scales $\wscale$ and $\wscale\p$ such that $\vert \wscale - \wscale\p\vert \geq 2$).  To better account for the covariance of wavelet coefficients, we fold into the analysis only those wavelet coefficients that are essentially uncorrelated.  To achieve this, we compute a correlation length for each wavelet scale, which we define by the fifth zero crossing of the theoretical wavelet covariance when assuming an inflationary CMB power spectrum (for the derivation of the theoretical wavelet covariance see \citealt{mcewen:s2let_localisation}).  We then downsample wavelet coefficients to the resolution defined by the correlation length and use the resulting downsampled maps of wavelet coefficients, for non-adjacent wavelet scales \wscale\ only, when computing the full log-posterior by \eqn{\ref{eqn:likelihood_full_log}}.

Here and subsequently we consider a dyadic wavelet scaling with $\dilparam=2$, as discussed previously.  We consider a maximum wavelet scale of $\wscalemax=7$, as also discussed previously, since for these wavelet scales the GGD modelling the string component is highly leptokurtic (see \sectn{\ref{sec:inference:distributions:strings}}).  For the string tension \gmu, we assume a uniform prior over the domain \mbox{$(1\times10^{-10},\  4\times10^{-6})$}, sampled with 200 uniformly spaced gridpoints. When constructing the LUTs, we evaluate tables sampled over a domain of 1000 uniformly spaced gridpoints for the wavelet coefficients of the data and use 9000 uniformly spaced gridpoints for the string wavelet coefficients when computing integrals (by the trapezium rule).  The limits of the coefficient ranges are specified by the minimum and maximum values of the wavelet coefficients of the data.  As discussed, we perform a number of tests to ensure the LUTs are evaluated accurately.

Since the string tension is a scaling parameter, an uninformative (Jeffreys) prior for the string tension would be a log-uniform prior.  However, for this first work we instead choose to adopt a uniform prior so that the string tension posterior and likelihoods correspond, which can be useful for gaining further intuition regarding the effectiveness of the method.  By using a uniform prior the impact of alternative priors can be approximately inferred by a kind of ``posterior-by-eye'' approach.  In future, for applications to data, a log-uniform prior or indeed other priors can be considered.

\begin{figure*}
  \centering
  \subfigure[$G\mu = 3 \times 10^{-7}$]{\includegraphics[width=.3\textwidth]{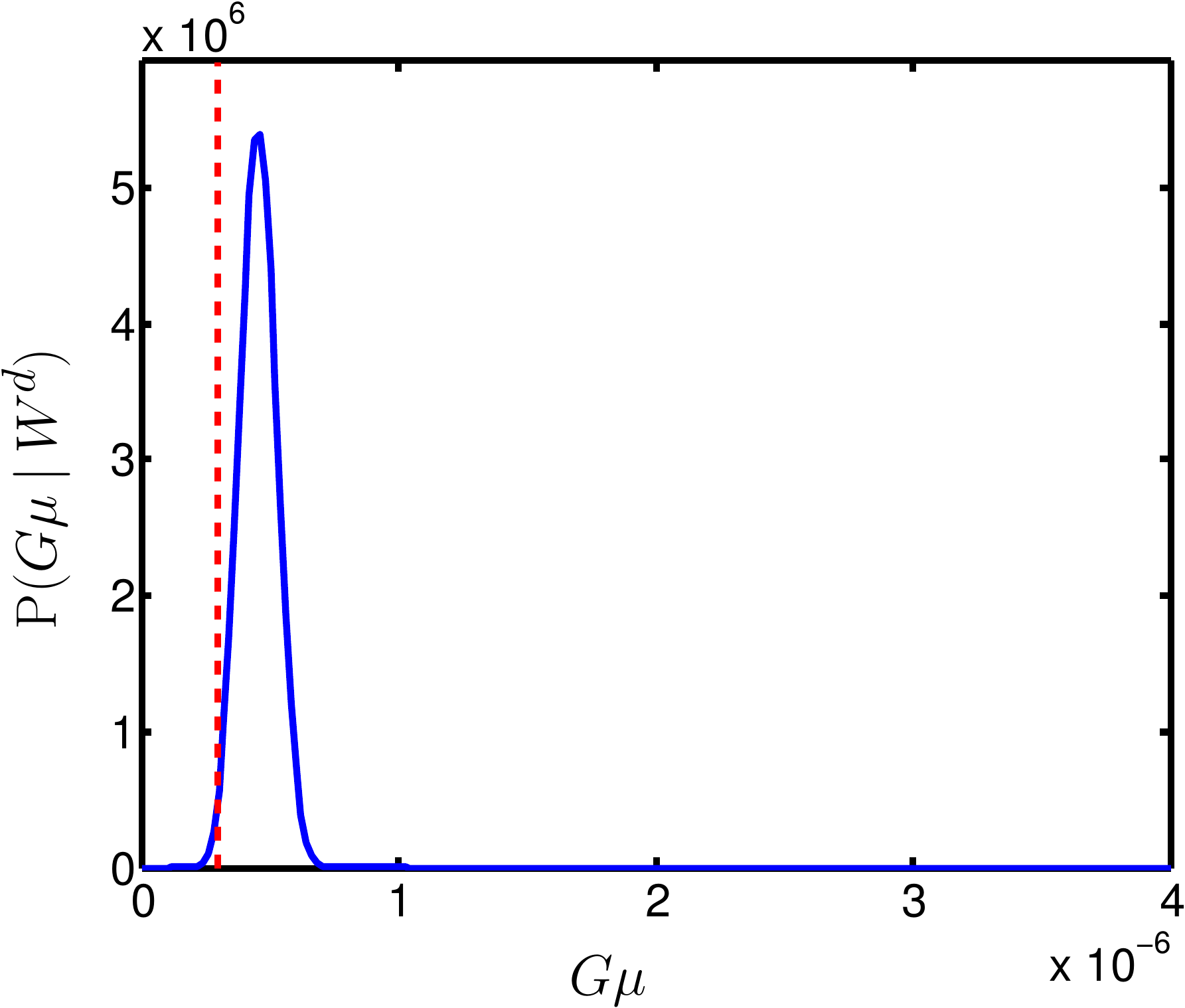}} \hfill
  \subfigure[$G\mu = 5 \times 10^{-7}$]{\includegraphics[width=.3\textwidth]{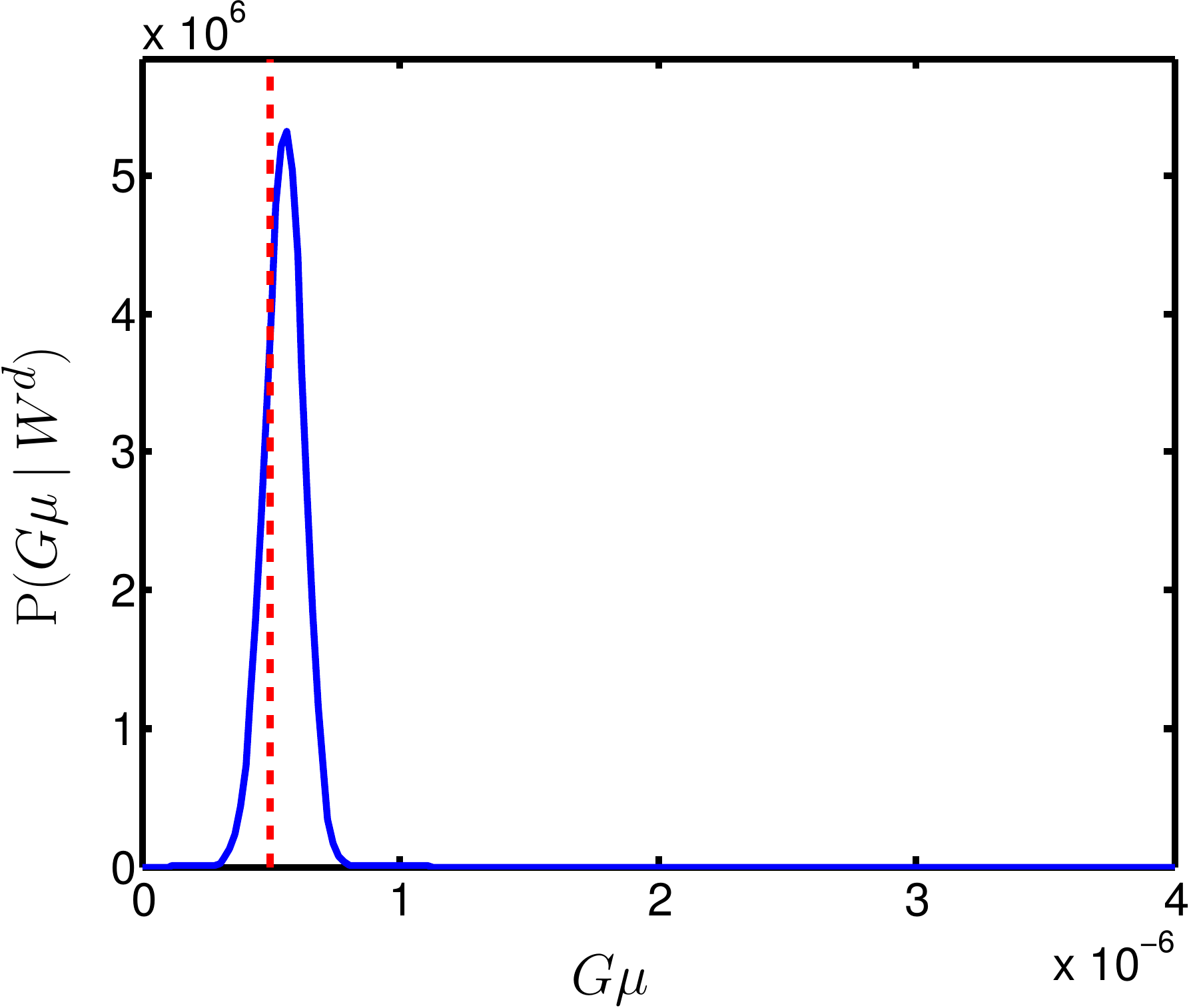}} \hfill
  \subfigure[$G\mu = 7 \times 10^{-7}$]{\includegraphics[width=.3\textwidth]{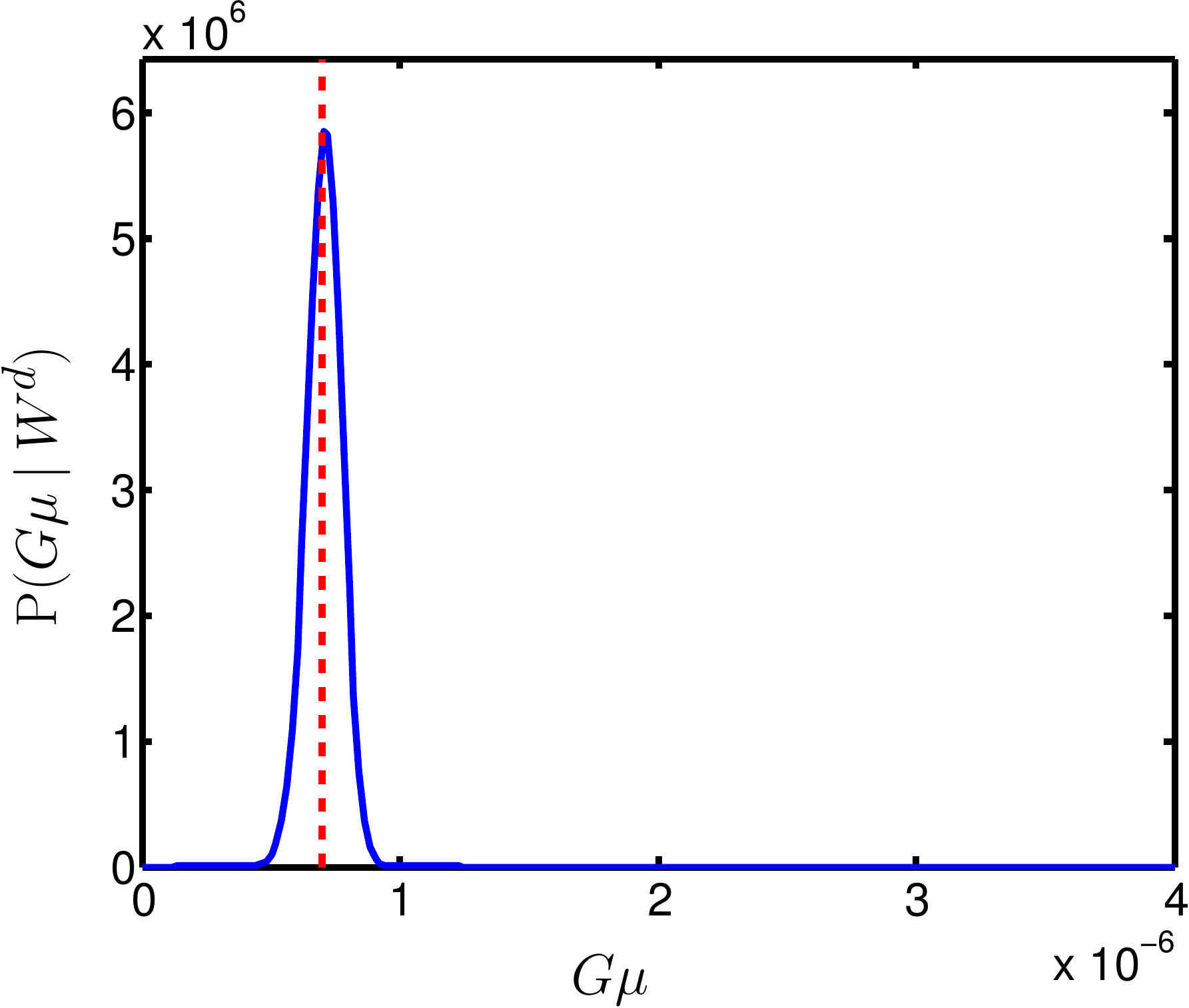}} \hfill
  \subfigure[$G\mu = 1 \times 10^{-6}$]{\includegraphics[width=.3\textwidth]{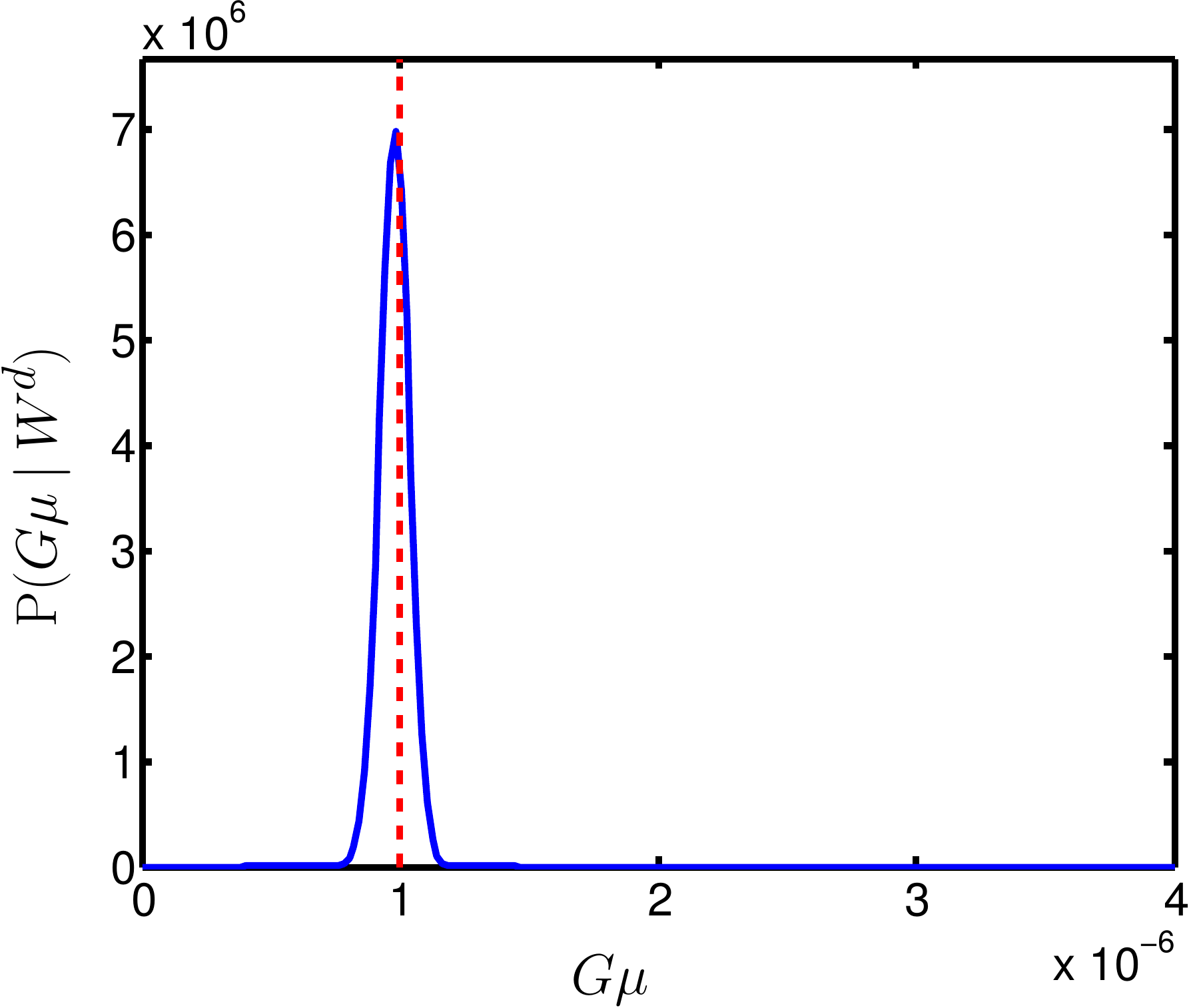}} \hfill
  \subfigure[$G\mu = 2 \times 10^{-6}$]{\includegraphics[width=.3\textwidth]{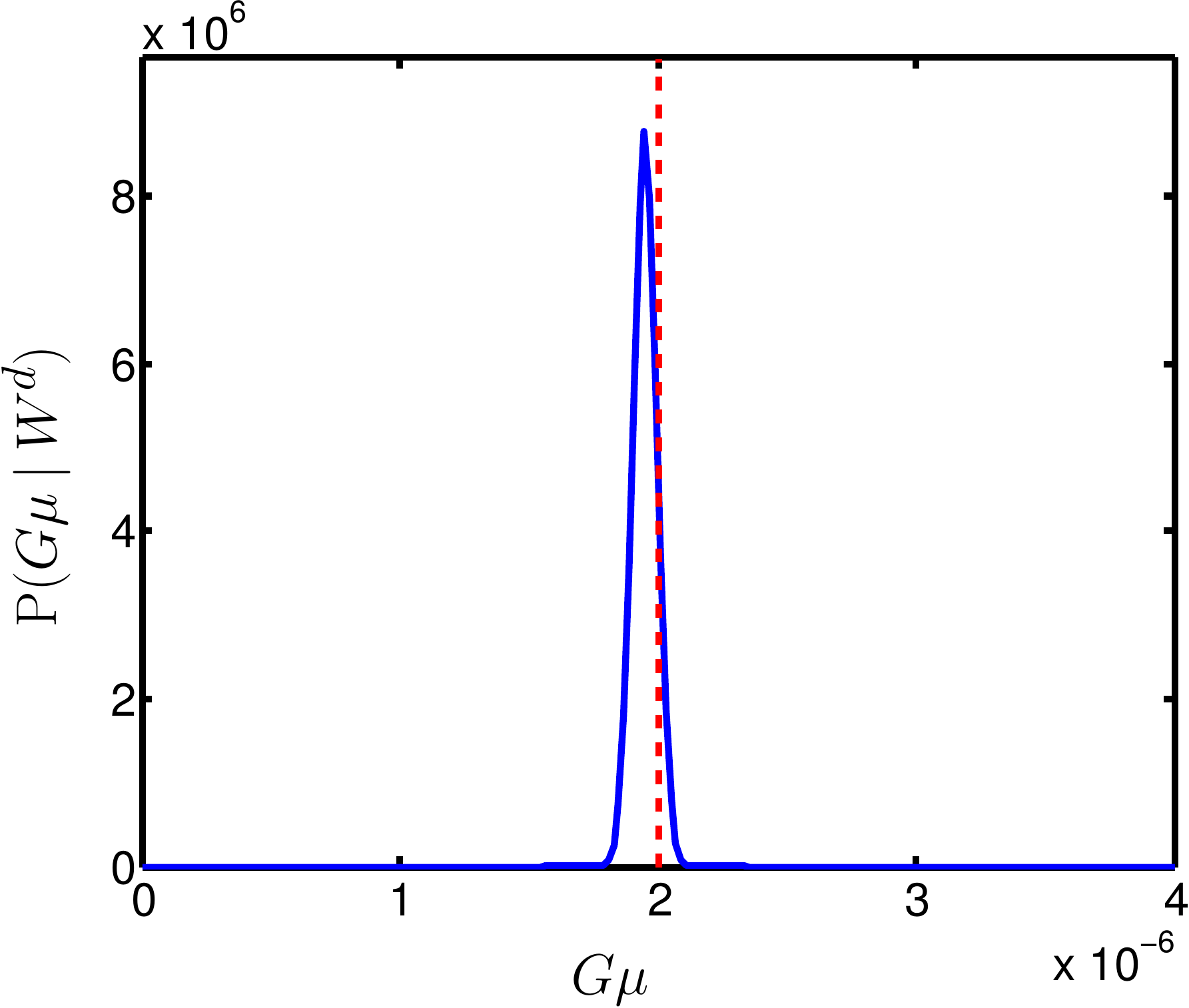}} \hfill
  \subfigure[$G\mu = 3 \times 10^{-6}$]{\includegraphics[width=.3\textwidth]{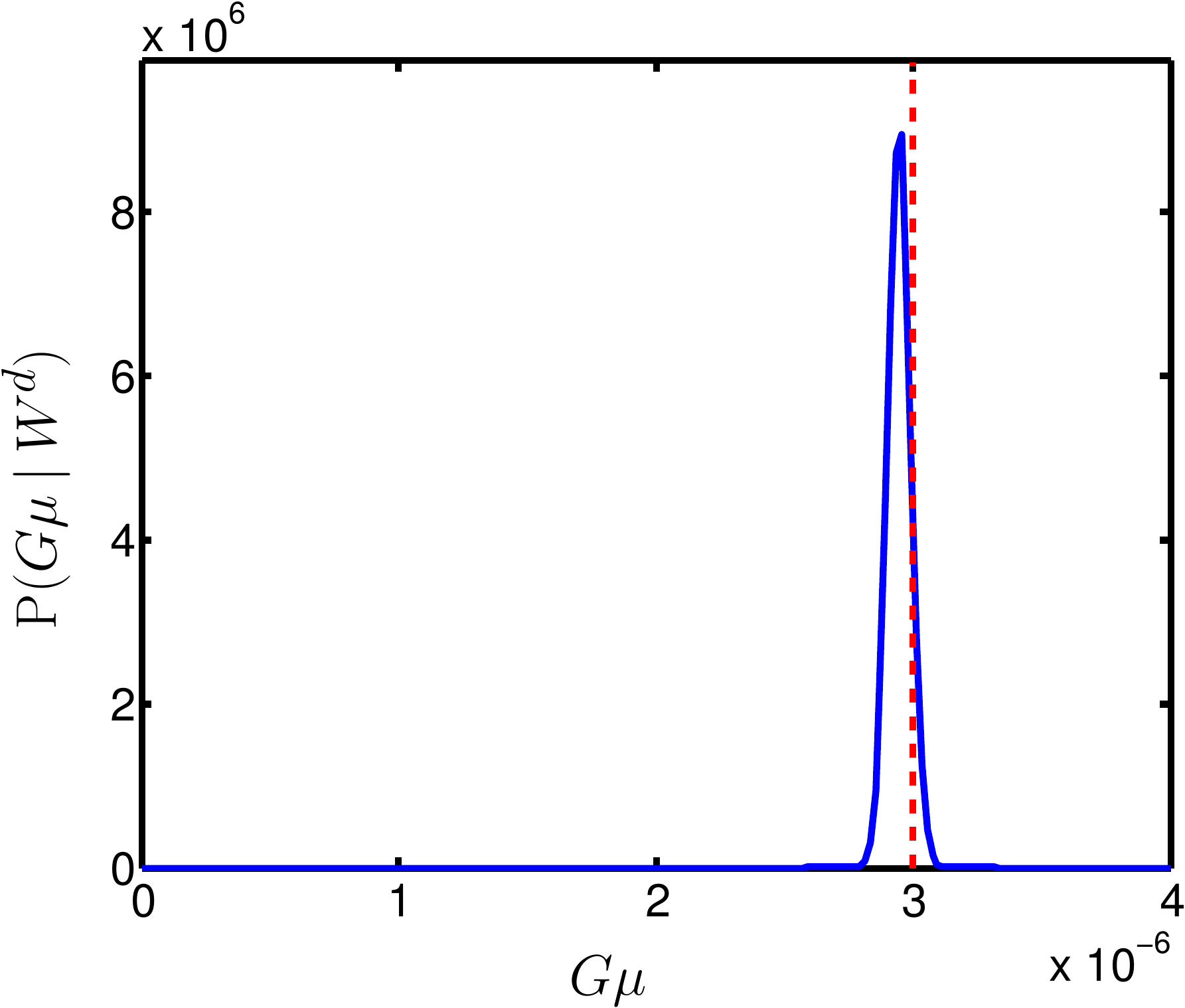}} \hfill
  \caption{Posterior distributions of the string tension \gmu\ recovered from simulations.  Each panel shows the recovered posterior distribution for a different ground truth value of \gmu\ as a solid blue curve; the ground truth value of \gmu\ is indicated by a vertical red dashed line.  The ground truth value of \gmu\ is estimated accurately above $\sim 5 \times 10^{-7}$. For lower string tensions the posterior distribution is biased high, illustrating the sensitivity of the (unoptimised) method.}
  \label{fig:posteriors}
\end{figure*}

The estimated posterior distributions are shown in \fig{\ref{fig:posteriors}} for a representative subset of the ground truth string tension values used in generating the simulated data.  The full set of string tension values considered is shown in the first column of \tbl{\ref{tbl:gmu_estimates}}.  An estimate of \gmu\ and the corresponding error, for each simulation, is recovered from the mean and the standard deviation of the posterior distribution and also shown \tbl{\ref{tbl:gmu_estimates}}.
The ground truth string tension values used to embed the string-induced CMB component in the simulated data are recovered accurately above $\sim 5 \times 10^{-7}$.  Below this approximate transition value the recovered estimates are biased high, likely due to unmodelled residual correlations, indicating the limit of the sensitivity of this unoptimised method.

\begin{table}
  \centering
  \caption{String tension values considered in simulations, with recovered estimates and corresponding Bayesian evidence ratio (positive evidence favours the string model \modelstring).  The ground truth string tension is recovered accurately above $\sim 5 \times 10^{-7}$ but is biased high below this transition region.  The Bayesian evidence ratio favours the string model also above $\sim 5 \times 10^{-7}$  but favours the standard inflationary model below this transition, illustrating the sensitivity of the (unoptimised) method.}
  \label{tbl:gmu_estimates}
  \begin{tabular}{ccc}\toprule
  \gmu\ truth / $10^{-7}$ & \gmu\ estimate / $10^{-7}$ & Evidence ratio [$\log_e$]\\ \midrule
    $30.0$ & $29.58 \pm 0.45$ & $2\:020$ \\
    $20.0$ & $19.60 \pm 0.47$ & $563$ \\
    $10.0$ & $9.90 \pm 0.58$ & $51.4$ \\
    $9.00$ & $8.97 \pm 0.61$ & $34.6$ \\
    $8.00$ & $8.06 \pm 0.65$ & $21.9$ \\
    $7.00$ & $7.18 \pm 0.69$ & $12.5$ \\
    $6.00$ & $6.36 \pm 0.73$ & $5.88$\\
    $5.00$ & $5.63 \pm 0.75$ & $1.19$\\
    $4.00$ & $5.06 \pm 0.75$ & $-1.86$\\
    $3.00$ & $4.66 \pm 0.73$ & $-3.87$\\
    \bottomrule
  \end{tabular}
\end{table}

\subsection{String model comparison}

For the same set of simulations we compute the Bayesian evidence ratio of the string model \modelstring\ and standard inflationary model \modelcmb, performing the calculation outlined in \sectn{\ref{sec:inference:evidence}}.
Again, we fold into the analysis only those wavelet coefficients that are essentially uncorrelated, following the approach outlined in \sectn{\ref{sec:results:tension_estimation}}.

The computed evidence ratios are shown in \tbl{\ref{tbl:gmu_estimates}}, where a positive value favours the string model \modelstring\ over the standard inflationary model \modelcmb.  For values of the string tension \gmu\ greater than $\sim 5 \times 10^{-7}$ the string model is preferred.  Interestingly, this is the same approximate transition value of \gmu\ as found for the estimation of the string tension in \sectn{\ref{sec:results:tension_estimation}}, further confirming the sensitivity of the (unoptimised) method for statistical inference.

\subsection{String map recovery}

While we have examined the effectiveness of our wavelet-Bayesian method for statistical inference in the previous subsections, a significant advantage of our approach is the ability to also recover estimates of any embedded string-induced CMB component at the map level.  For the same set of simulations we recover estimated string maps following the calculations outlined in \sectn{\ref{sec:denoising}}.

Maps of the recovered string-induced CMB component are illustrated in \fig{\ref{fig:denoised_maps}}.  String maps are recovered well for large values of \gmu.  As \gmu\ is reduced, the fidelity of the recovered maps is reduced as small scale features are washed out.

\begin{figure}
  \centering
  \subfigure[Root-mean-squared (RMS) error \label{sec:denoised_errors:rms}]{\includegraphics[width=.7\columnwidth]{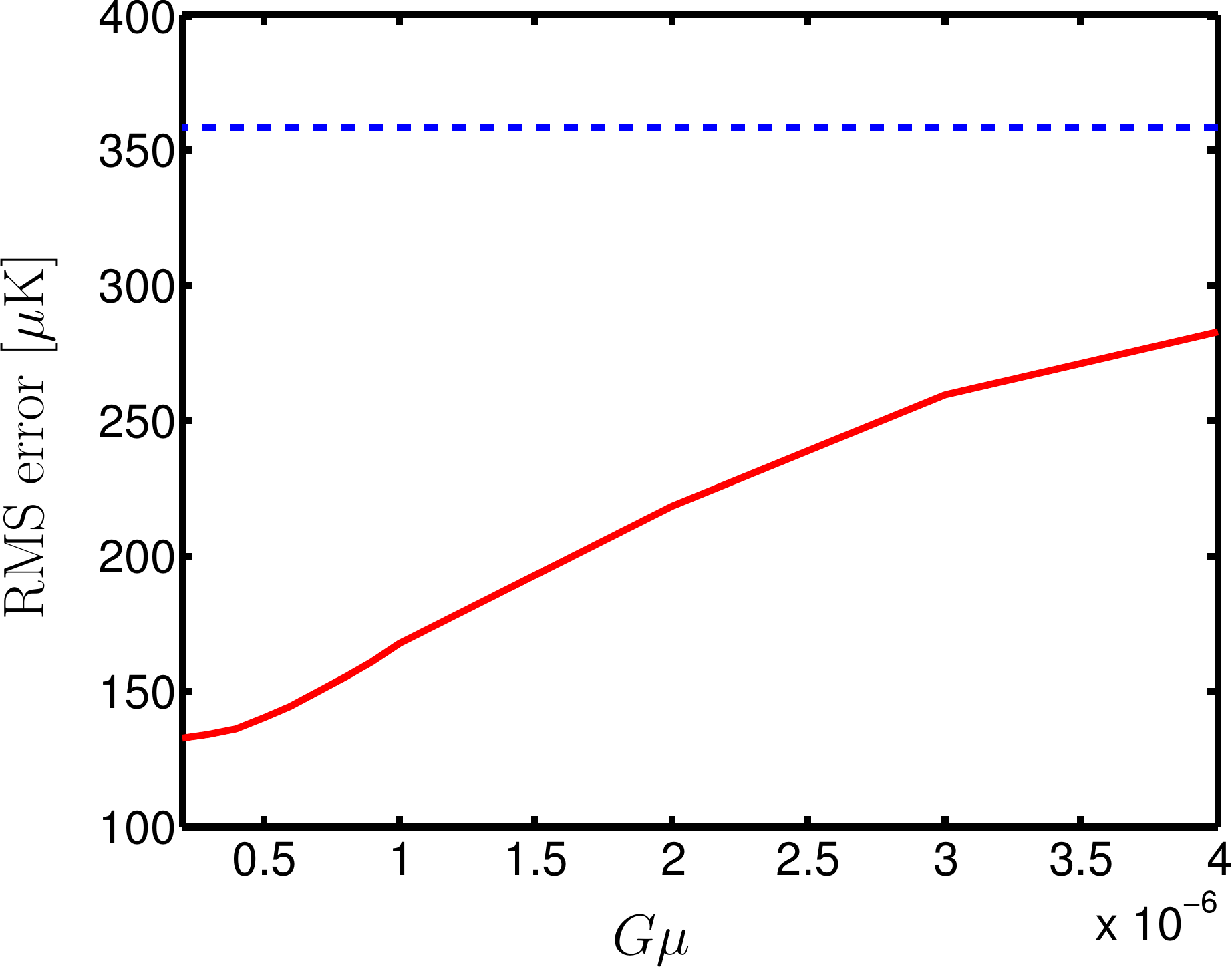}}
  \subfigure[Signal-to-noise-ratio (SNR) \label{sec:denoised_errors:snr}]{\includegraphics[width=.7\columnwidth]{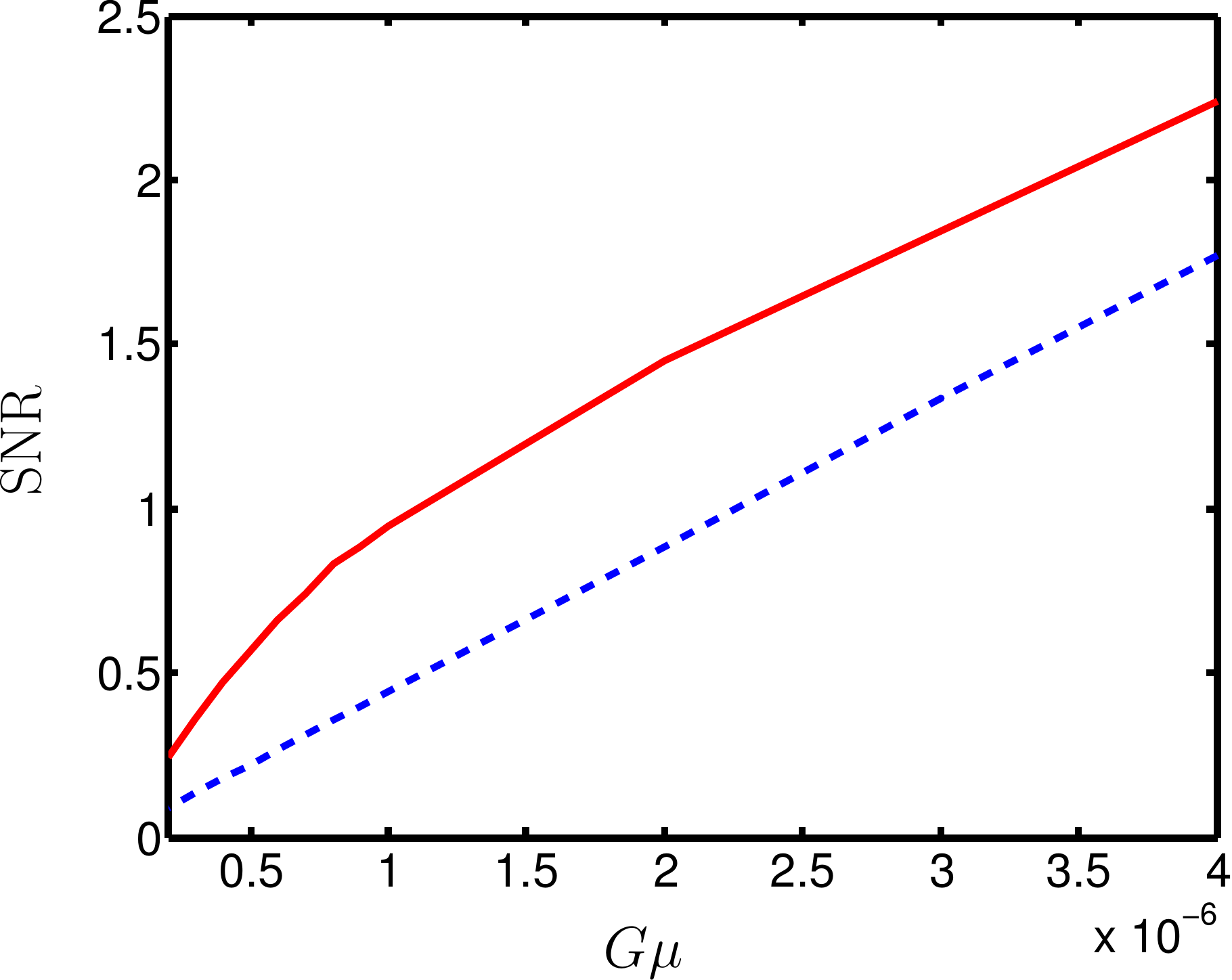}}
  \caption{Error metrics quantifying the difference between the recovered string-induced CMB component and the ground truth map (solid red curve).  For comparison, differences without estimating the string-induced component are also shown (dashed blue curve).}
  \label{sec:denoised_errors}
\end{figure}

To assess the performance of the recovery of string maps quantitatively we plot in \fig{\ref{sec:denoised_errors}} the root-mean-squared (RMS) error and the signal-to-noise ratio (SNR) to quantify the error between the recovered string map and the ground truth string map.  The SNR is defined as the ratio of the RMS value of the ground truth string map to the RMS error.   We compute these error metrics for the simulations performed with varying values of the string tension $\gmu$.  For comparison, results are also shown when not directly estimating the string-induced component.  In this case, we simply consider the residuals between the observed data and the ground truth string maps. The RMS error is then simply given by the RMS of the inflationary CMB component and noise, hence the constant dashed blue curve in \fig{\ref{sec:denoised_errors:rms}}.  As \gmu\ is reduced, though it is difficult to recover small scale string features (as shown in \fig{\ref{fig:denoised_maps}}) the RMS error of the estimated string components is nevertheless reduced: by greater than 200 $\mu$K for the lowest values of \gmu\ considered.
From the SNR plotted in \fig{\ref{sec:denoised_errors:snr}} it is clear that the relative improvement provided by the string estimation approach is reasonably constant over much of the domain considered.

Recall that, as discussed in \sectn{\ref{sec:denoising:map_estimation}}, in order to recover the string-induced CMB component a posterior distribution for the string tension is required.  Here, we adopt the posterior distributions recovered in \sectn{\ref{sec:results:tension_estimation}} and shown in \fig{\ref{fig:posteriors}}; however, if an alternative method provides a better estimate of the posterior distribution then the alternative posterior distribution can be substituted.

\setlength{\plotwidth}{0.78\columnwidth}

\begin{figure*}
  \centering
  \subfigure[$\gmu = 3 \times 10^{-6}$]{\includegraphics[width=\plotwidth, trim=4mm 8mm 4mm 8mm, clip=true]{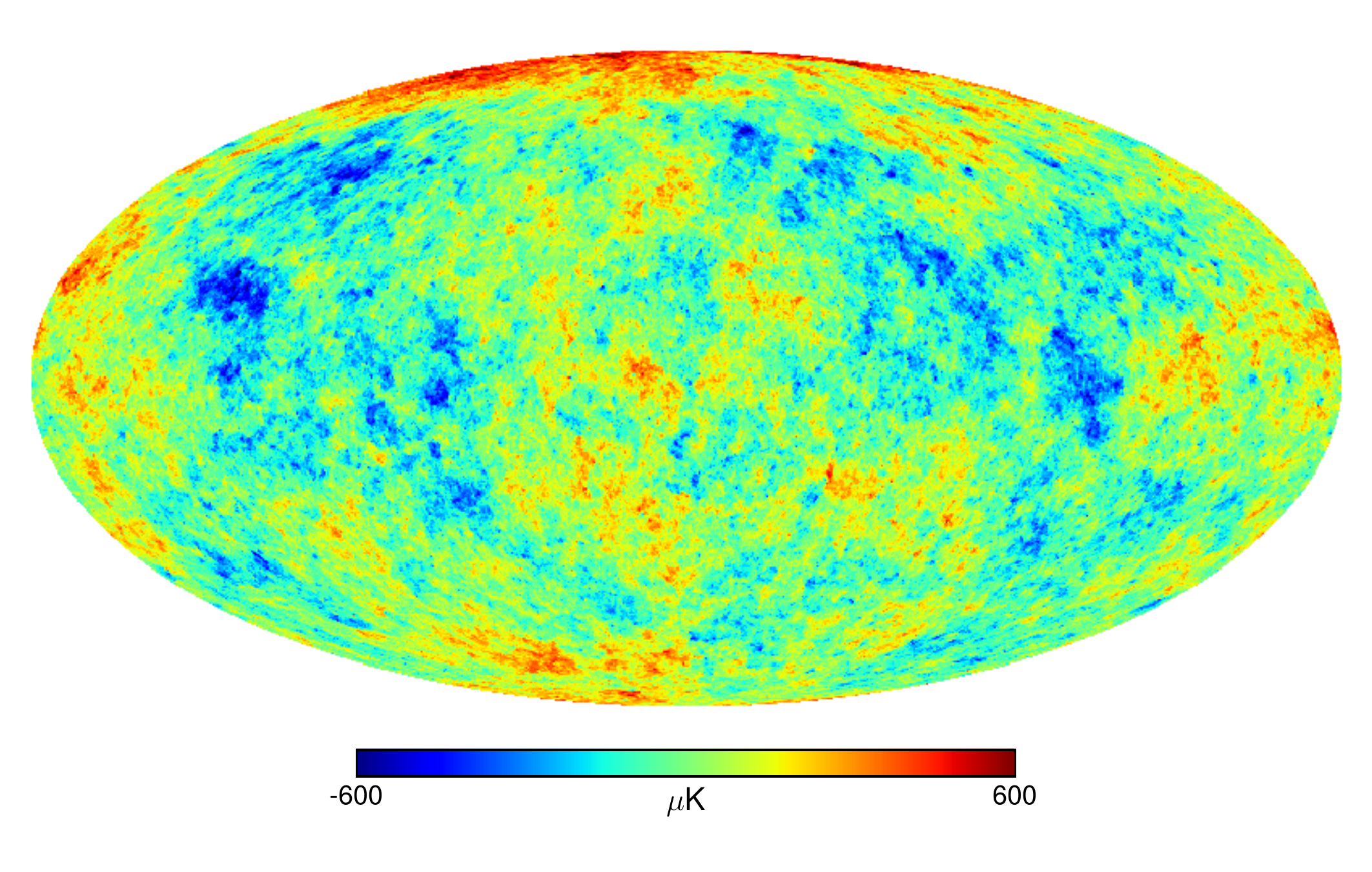}
  \includegraphics[width=\plotwidth, trim=4mm 8mm 4mm 8mm, clip=true]{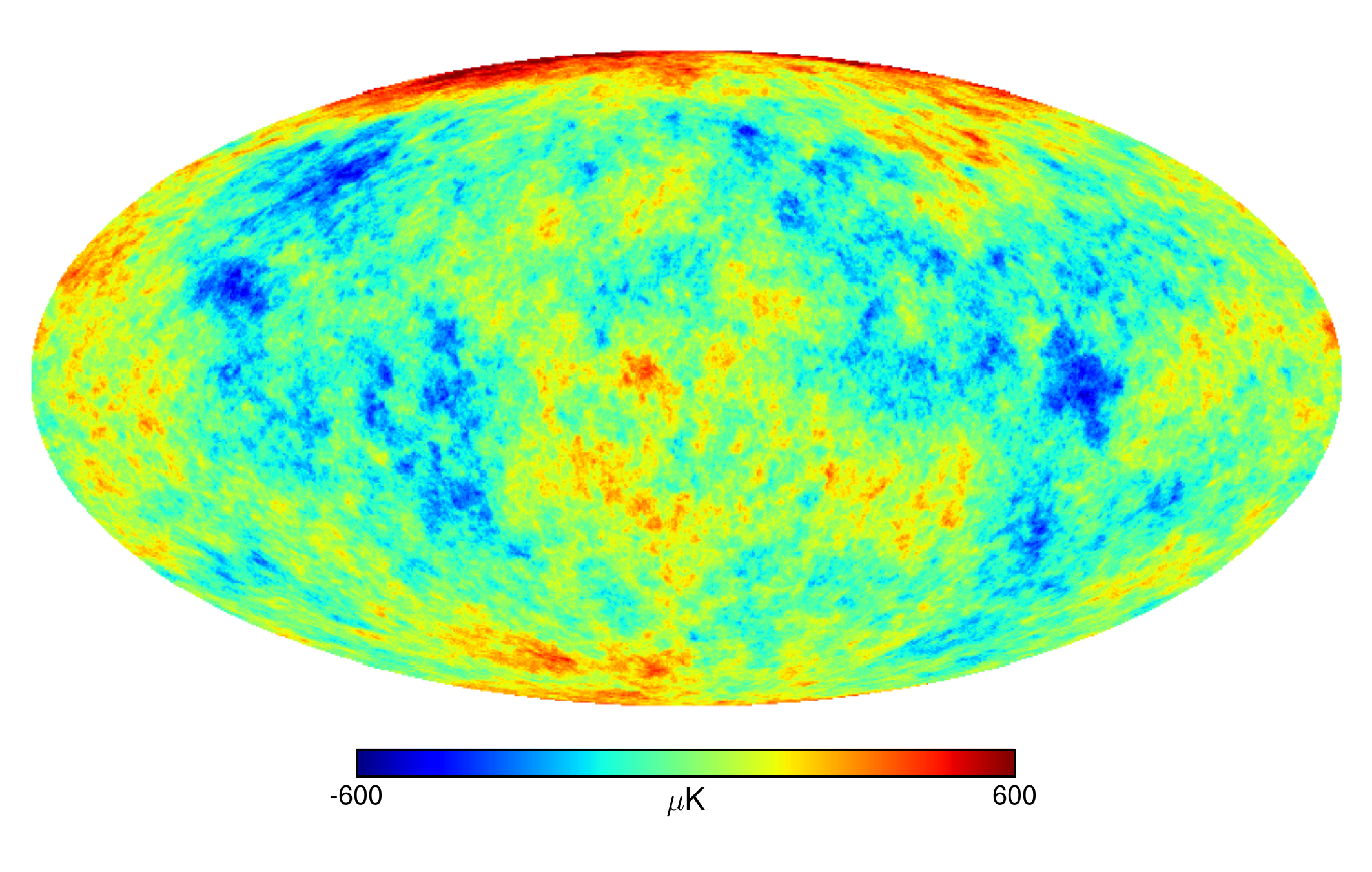}}
  \subfigure[$\gmu = 2 \times 10^{-6}$]{\includegraphics[width=\plotwidth, trim=4mm 8mm 4mm 8mm, clip=true]{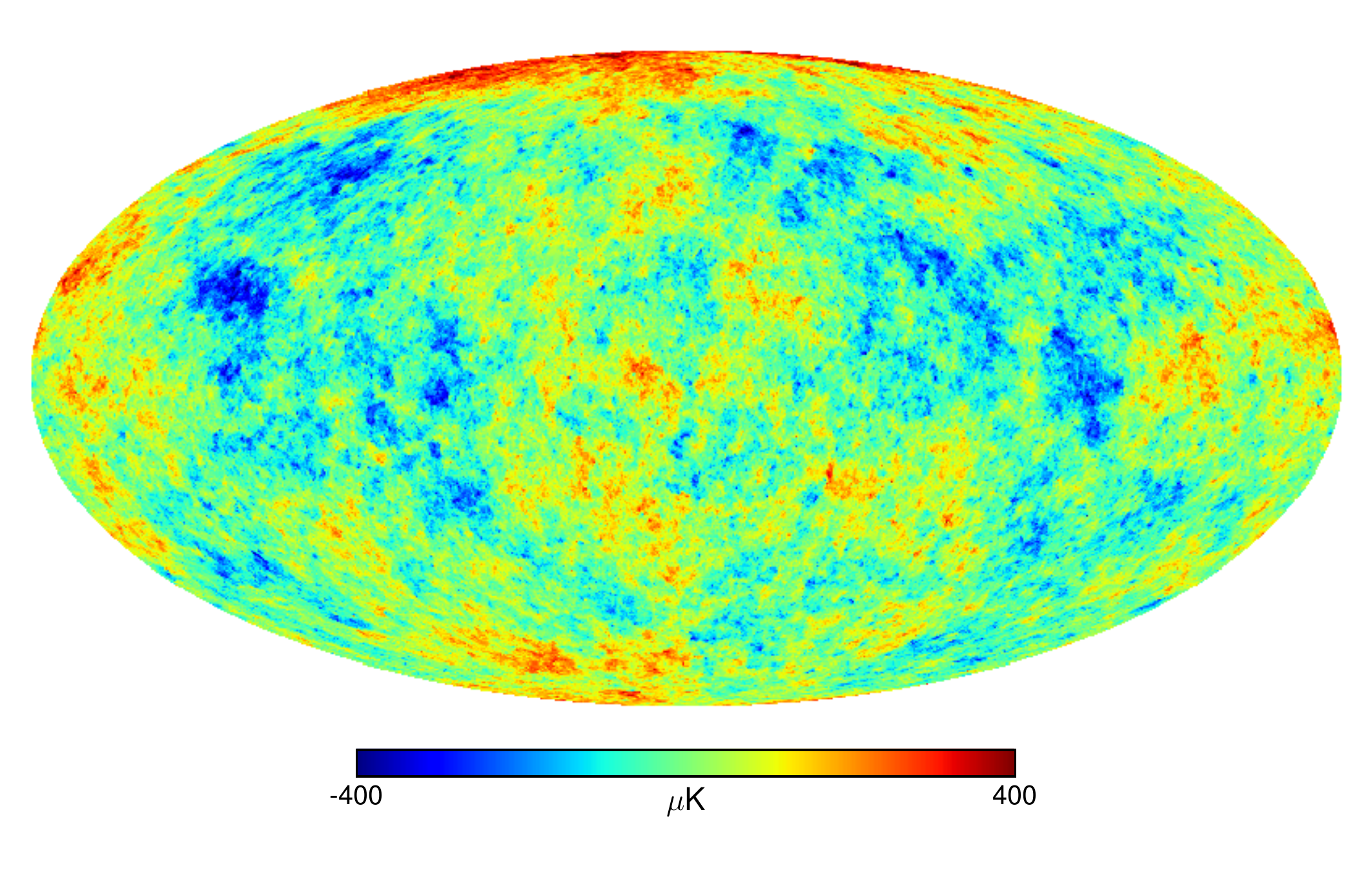}
  \includegraphics[width=\plotwidth, trim=4mm 8mm 4mm 8mm, clip=true]{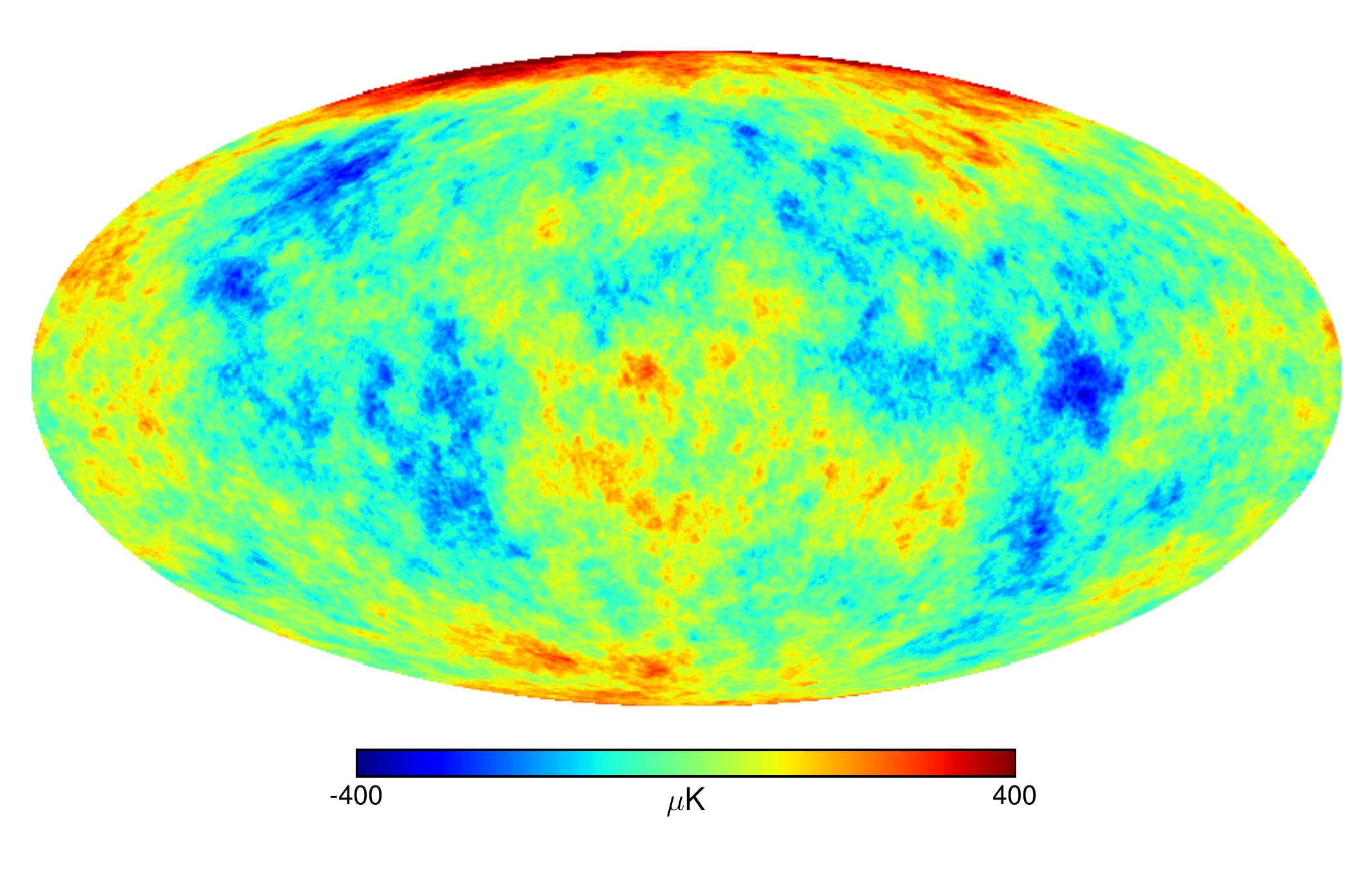}}
  \subfigure[$\gmu = 1 \times 10^{-6}$]{\includegraphics[width=\plotwidth, trim=4mm 8mm 4mm 8mm, clip=true]{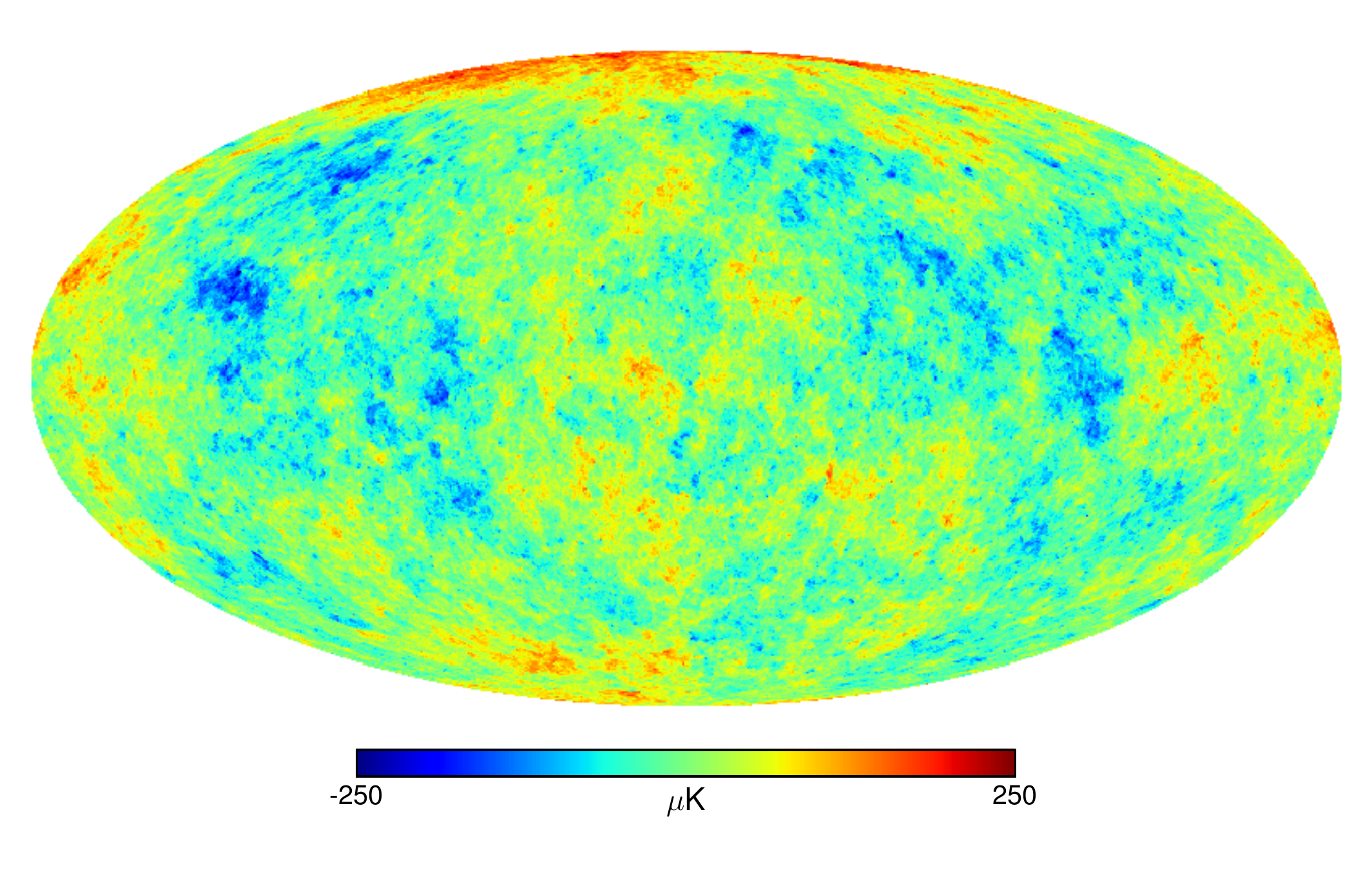}
  \includegraphics[width=\plotwidth, trim=4mm 8mm 4mm 8mm, clip=true]{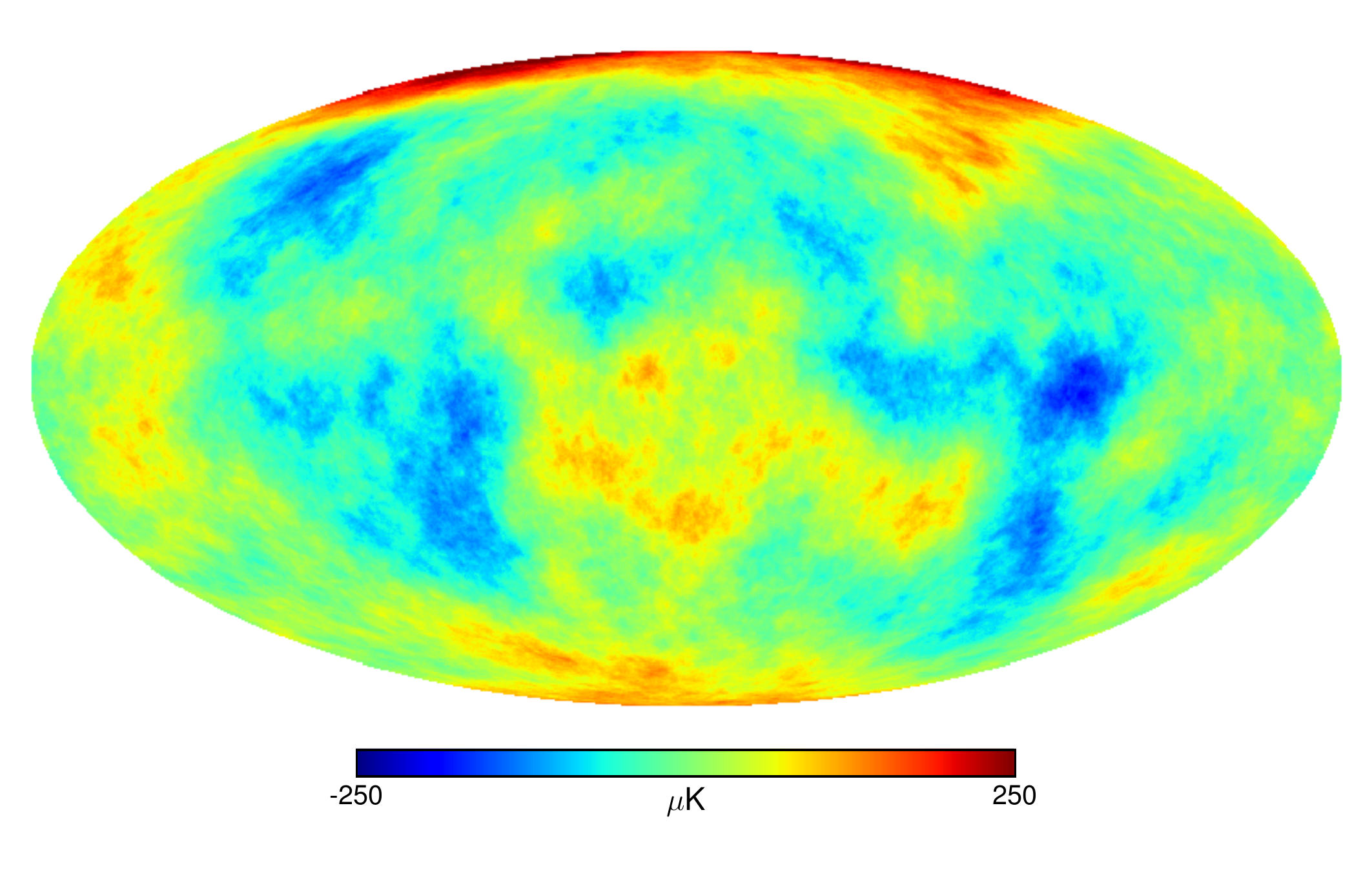}}
  \subfigure[$\gmu = 5 \times 10^{-7}$]{\includegraphics[width=\plotwidth, trim=4mm 8mm 4mm 8mm, clip=true]{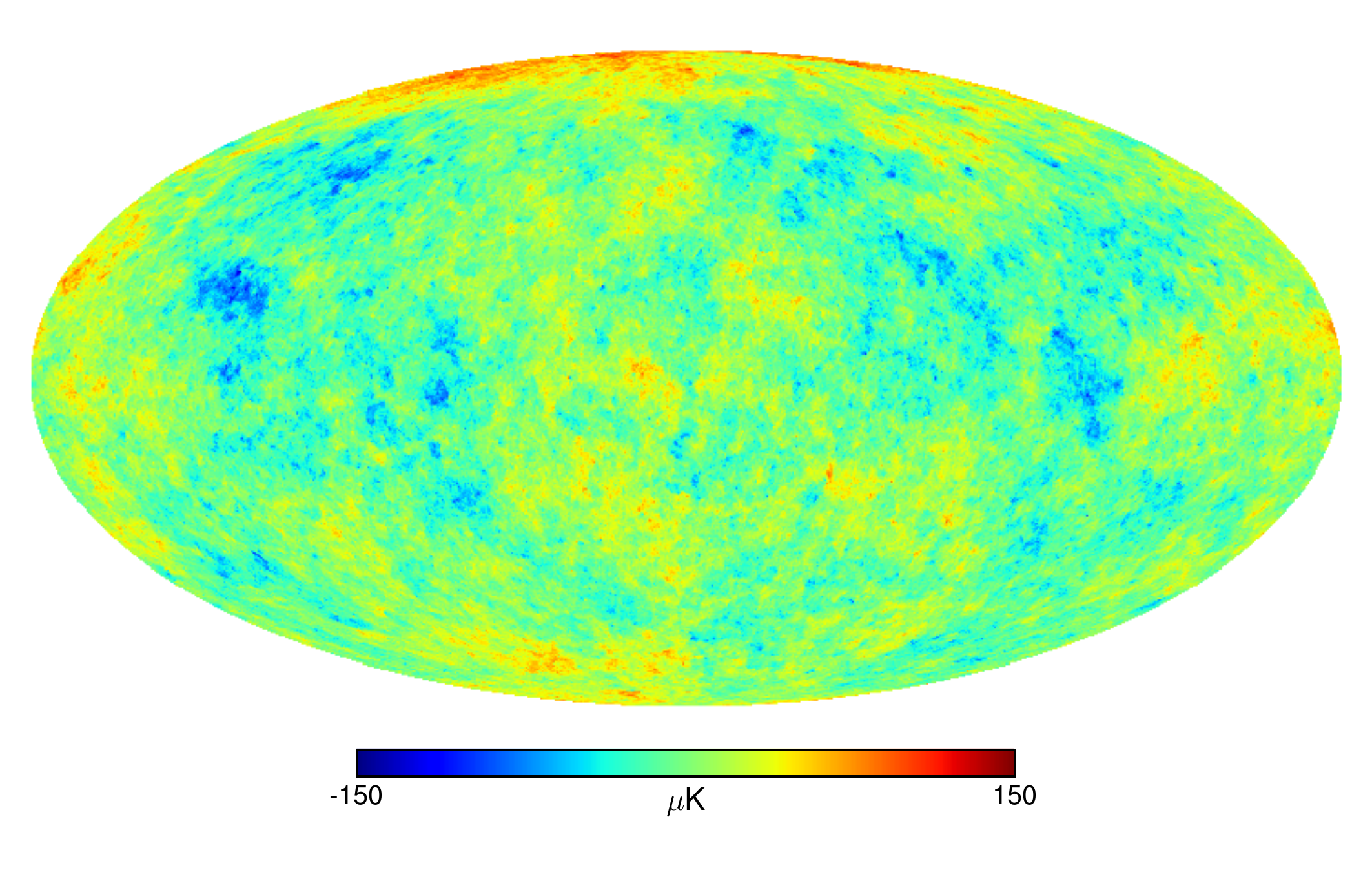}
  \includegraphics[width=\plotwidth, trim=4mm 8mm 4mm 8mm, clip=true]{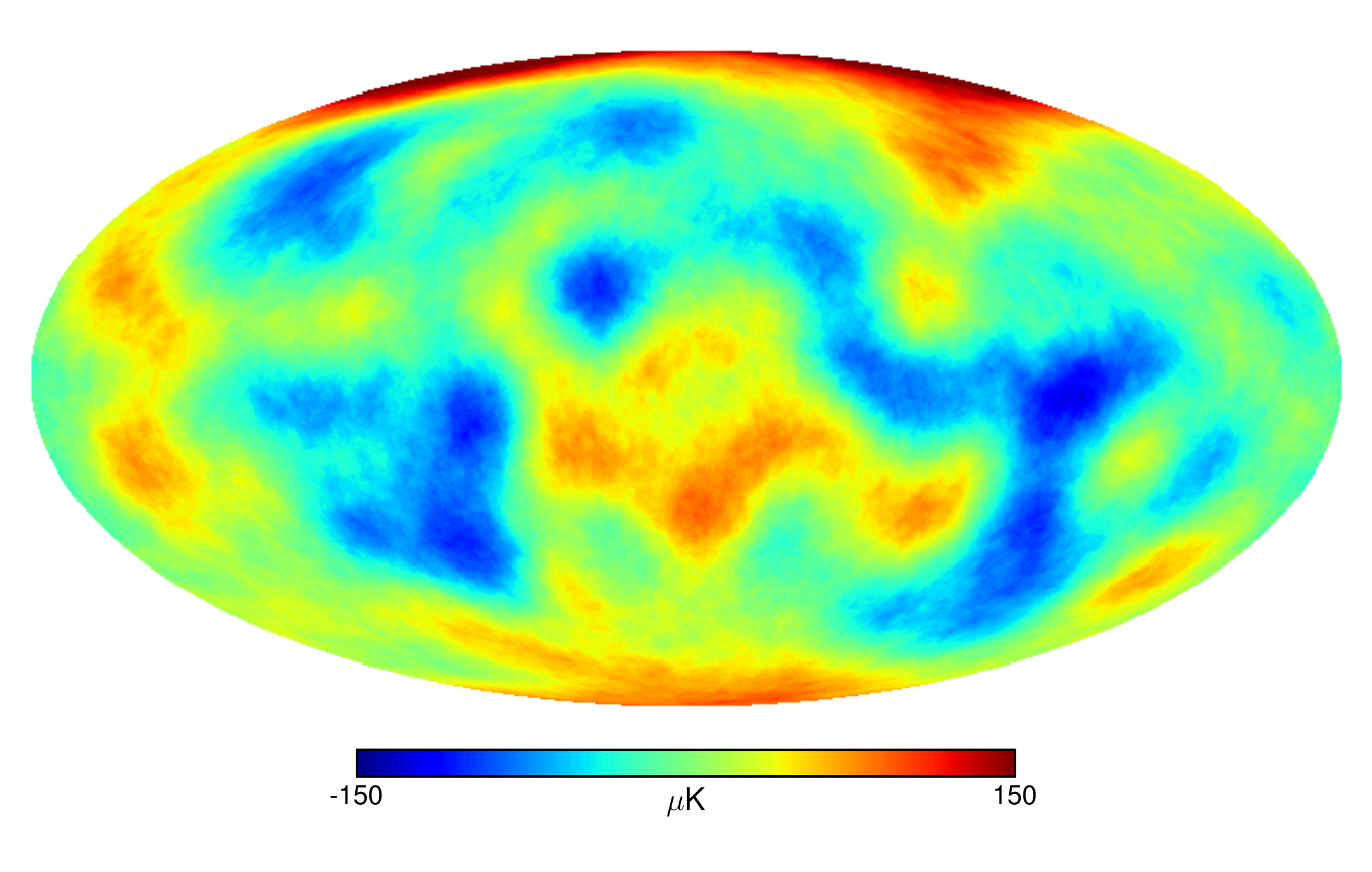}}
  \subfigure[$\gmu = 3 \times 10^{-7}$]{\includegraphics[width=\plotwidth, trim=4mm 8mm 4mm 8mm, clip=true]{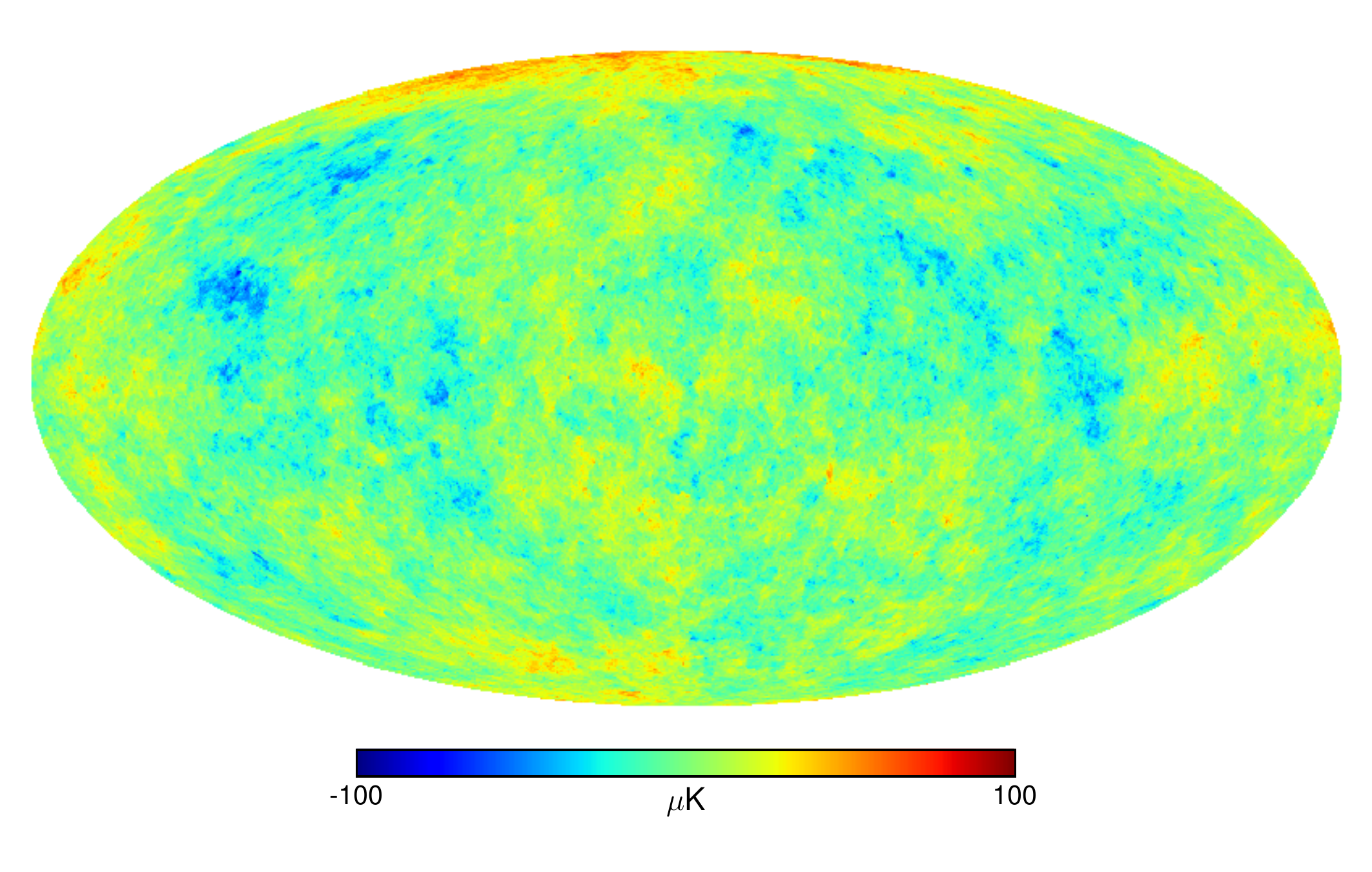}
  \includegraphics[width=\plotwidth, trim=4mm 8mm 4mm 8mm, clip=true]{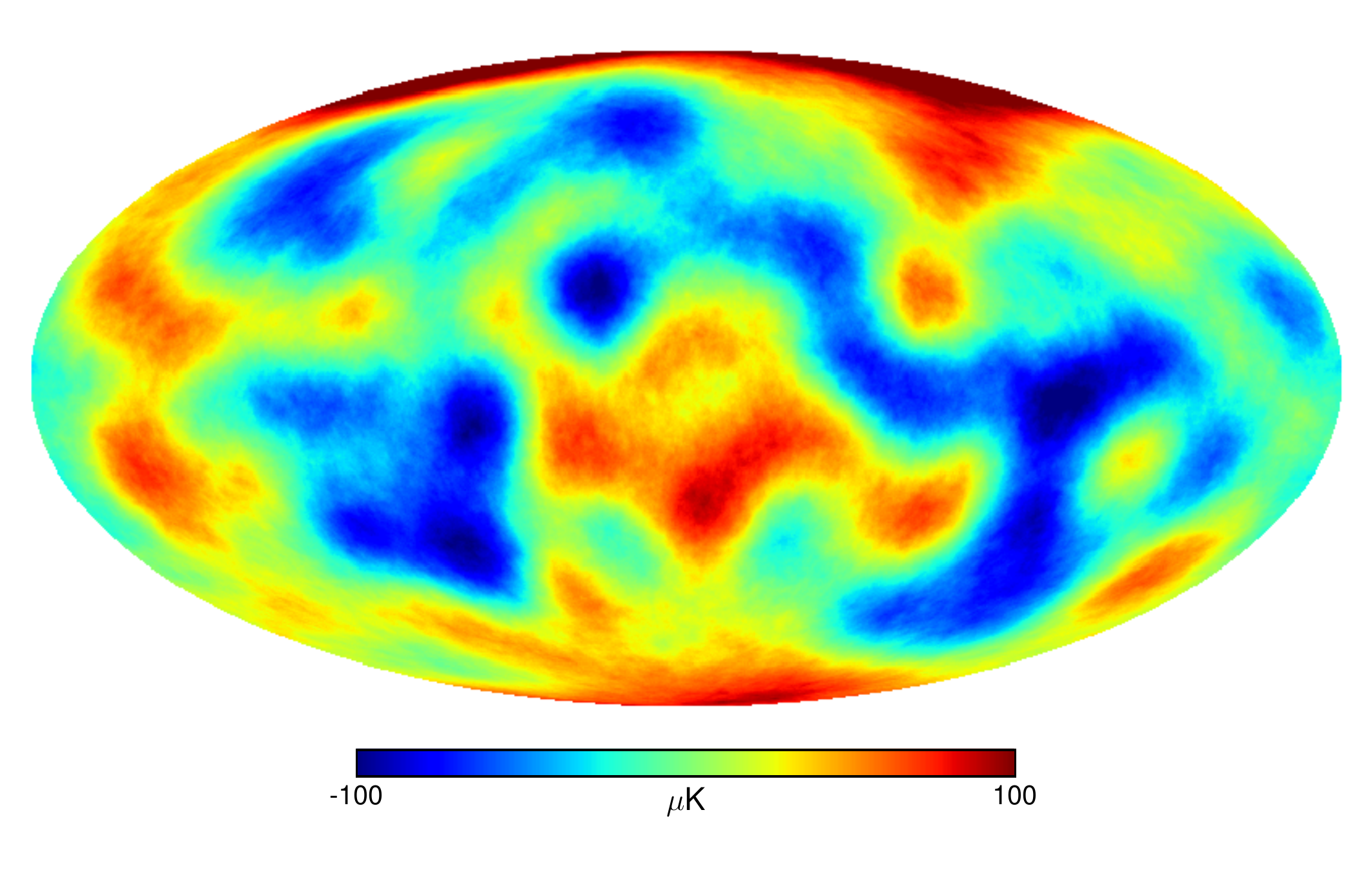}}
  \caption{Recovered string-induced CMB maps for various ground truth values of the string tension \gmu.  Ground truth maps are shown on the left and recovered maps on the right.  String-induced CMB contributions are recovered well for large values of \gmu.  As \gmu\ is reduced, the fidelity of the recovered maps is reduced as small scale features are washed out.}
  \label{fig:denoised_maps}
\end{figure*}

\section{Conclusions}
\label{sec:conclusions}

Cosmic strings are a well-motivated extension to the standard cosmological model and could induce a subdominant component in the anisotropies of the CMB.  Detecting such a component would provide a direct probe of corresponding symmetry breaking phase transitions in the early Universe at very high energy scales.  However, due to the weak nature of any string component its detection presents a significant observational challenge.

We present a hybrid wavelet-Bayesian framework for cosmic string inference, constructing a Bayesian analysis in wavelet space where the string-induced CMB component has very different statistical properties to the inflationary component. We learn and exploit the complex non-Gaussian structure of string-induced CMB contributions, rather than considering (insufficient) summary statistics like many alternative methods (\eg\ the kurtosis), for which the origin of any non-Gaussian component cannot be rigorously determined.  Our approach allows the full posterior distribution of the string tension to be estimated, from which a best estimate of the string tension and an associated error can be computed.  The Bayesian evidence ratio comparing the string model, including an inflationary component and a subdominant string-induced component, and the standard inflationary model can also be computed.  Moreover, it is also possible to recover an estimate of the string-induced component in the CMB at the map level.

We demonstrate the application of our wavelet-Bayesian framework and evaluate its performance using idealised simulations of CMB observations made by the \planck\ satellite, where a string component is embedded for a range of values of the string tension \gmu.  For values of the string tension \gmu\ above $\sim 5 \times 10^{-7}$, we recover accurate estimates of its posterior distribution, which can be used to provide accurate point estimates of the string tension and associated error.  The Bayesian evidence values computed also correctly favour the string model for values of the string tension \gmu\ above $\sim 5 \times 10^{-7}$, further highlighting the sensitivity of the method.  The performance of our approach compares favourably with current constraints obtained using the same string simulations (that obtain the constraint $\gmu < 7.8 \times 10^{-7}$; \citealt{planck2013-p20}) and, moreover, is based on a principled statistical framework. A more robust and principled analysis is inevitably more conservative than less well-motivated alternatives but, nevertheless, we find our method generally compares favourable with other map-based techniques.  While we consider slightly idealised \planck\ simulations we have not yet optimised the parameters of the analysis (alternative wavelet scalings, for example, are likely to improve performance; \cf\ \citealt{rogers:s2let_ilc_temp,rogers:s2let_ilc_pol}).

We find that the embedded string maps are recovered accurately for large values of the string tension \gmu.  As \gmu\ is reduced, small-scale features in the recovered string maps are washed out, but the RMS error of the recovered maps is nevertheless reduced considerably.  While maps of the string-induced CMB component are of interest in their own right, they can also be used as pre-processed inputs for alternative techniques to estimate the string tension from the non-Gaussian structure of the string-induced CMB component, such as computing the gradient.  We leave post-processing of the recovered string maps for further work.

This is one of many areas to be considered in future work.  First, more realistic \planck\ simulations will be considered, along with a mask to remove foreground emission (masking can be integrated in the wavelet analysis in a straightforward manner, following the approach of, \eg, \citealt{mcewen:2005:ng,mcewen:2006:isw,leistedt:ebsep}).  Second, the parameters of the analysis will be optimised for \planck\ observations (\cf\ \citealt{rogers:s2let_ilc_temp,rogers:s2let_ilc_pol}).  Third, the steerability of  scale-discretised wavelets will be exploited to provide more accurate inference when computing the posterior distribution of the string tension and the Bayesian evidence (\cf\ \citealt{planck2013-p20}).  Fourth, an estimate of the standard deviation of the recovered string map will be developed to characterise its accuracy, as outlined in \sectn{\ref{sec:denoising:map_estimation}}.  Fifth, techniques will be developed to better model the full covariance structure of signals in wavelet space.

Our framework will in future be applied to observational data from \planck\ and other CMB experiments to provide constraints on the string tension that are based on a principled statistical analysis of the non-Gaussian structure of string-induced CMB contributions.  While we focus in the current article on cosmic strings, the framework can also be adapted to other settings, such as other components embedded in the CMB.

\section*{Acknowledgements}

This work is supported by the UK Engineering and Physical Sciences Research
Council (EPSRC) by grant EP/M011852/1.  SMF and HVP were partially supported by the European Research Council under the European Community's Seventh Framework Programme (FP7/2007-2013) ERC GA n.~306478-CosmicDawn. SMF was additionally supported by the UK Science and Technology Facilities Council (STFC). We acknowledge use of the following
public software packages: \healpix\ \citep{gorski:2005}; \cambcode\
\citep{lewis:2000}; \stwodwcode\ \citep{wiaux:2007:sdw}; \stwocode\
\citep{mcewen:2006:fcswt,mcewen:2006:filters}.

\bibliographystyle{mymnras_eprint}
\bibliography{bib_journal_names_long,bib_myname,bib}

\label{lastpage}
\end{document}